\documentclass[english,onecolumn,11p]{IEEEtran}

\usepackage[latin9]{inputenc}
\usepackage{geometry}
\usepackage{array}
\usepackage{amsthm}
\usepackage{amsmath}
\usepackage{amssymb}
\usepackage{fixltx2e}
\usepackage{graphicx}
\usepackage{setspace}
\usepackage{esint}
\usepackage{xargs}[2008/03/08]
\makeatletter

\providecommand{\tabularnewline}{\\}

  \theoremstyle{plain}
  \newtheorem{thm}{\protect\theoremname}
  \theoremstyle{plain}
  \newtheorem{cor}{\protect\corollaryname}
  \theoremstyle{plain}
  \newtheorem{prop}{\protect\propositionname}
  \theoremstyle{definition}
  \newtheorem{defn}{\protect\definitionname}
  \theoremstyle{plain}
  \newtheorem{lem}{\protect\lemmaname}

\usepackage{amsmath}
\usepackage{amsthm}
\usepackage{amsfonts}
\usepackage{epsfig}
\usepackage{latexsym}
\usepackage{url}
\usepackage[hypertexnames=false]{hyperref}
\usepackage{ifthen}
\usepackage[noadjust]{cite}

\makeatother

\usepackage{babel}
\providecommand{\corollaryname}{Corollary}
\providecommand{\definitionname}{Definition}
\providecommand{\lemmaname}{Lemma}
\providecommand{\propositionname}{Proposition}
\providecommand{\theoremname}{Theorem}

\begin{document}

\title{Comparison of the Achievable Rates in\\OFDM and Single Carrier Modulation\\with
I.I.D. Inputs}

\author{Yair Carmon\IEEEauthorrefmark{1}, Shlomo Shamai\IEEEauthorrefmark{1}
and Tsachy Weissman\IEEEauthorrefmark{2}%
\thanks{\IEEEauthorrefmark{1} Technion, Israel Institute of Technology. Emails:
yairc@tx.technion.ac.il, sshlomo@ee.technion.ac.il%
}%
\thanks{\IEEEauthorrefmark{2} Stanford University. Email: tsachy@stanford.edu

This paper was presented at the 2014 International Zurich Seminar
on Communications.%
}}
\maketitle
\begin{abstract}
We compare the maximum achievable rates in single-carrier and OFDM
modulation schemes, under the practical assumptions of i.i.d. finite
alphabet inputs and linear ISI with additive Gaussian noise. We show
that the Shamai-Laroia approximation serves as a bridge between the
two rates: while it is well known that this approximation is often
a \emph{lower bound} on the single-carrier achievable rate, it is
revealed to also essentially \emph{upper bound} the OFDM achievable
rate. We apply Information-Estimation relations in order to rigorously
establish this result for both general input distributions and to
sharpen it for commonly used PAM and QAM constellations. To this end,
novel bounds on MMSE estimation of PAM inputs to a scalar Gaussian
channel are derived, which may be of general interest. Our results
show that, under reasonable assumptions, optimal single-carrier schemes
may offer spectral efficiency significantly superior to that of OFDM,
motivating further research of such systems.
\end{abstract}

\global\long\def\Ihat{{\hat{I}}}
\global\long\def\Immse{{I_{\mathrm{MMSE}}}}
\global\long\def\Isl{{I_{\textrm{SL}}}}
\global\long\def\Iofdm{{\mathcal{I}_{\textrm{OFDM}}}}
\global\long\def\Iach{{\mathcal{I}_{\textrm{SC}}}}
\global\long\def\Immi{{I_{\textrm{MMI}}}}
\newcommandx\Ilog[1][usedefault, addprefix=\global, 1=x]{{I_{#1}^{\log}}}
\newcommandx\dIlog[1][usedefault, addprefix=\global, 1=x]{{I_{#1}^{\log}}{'}}
\newcommandx\Ihatlog[1][usedefault, addprefix=\global, 1=x]{{\hat{I}_{#1}^{\log}}}
\newcommandx\dIhatlog[1][usedefault, addprefix=\global, 1=x]{{\Ihat_{#1}^{\log}}{'}}
\newcommandx\dIchecklog[1][usedefault, addprefix=\global, 1=x]{{\check{I}_{#1}^{\log}}{'}}
\newcommandx\SNR[1][usedefault, addprefix=\global, 1=x]{{\mathsf{SNR}_{\textrm{#1}}}}
\global\long\def\Ebrack#1#2{\mathrm{E}_{#2}\left[#1\right]}
\newcommandx\E[2][usedefault, addprefix=\global, 1=x, 2=]{\mathrm{E}_{#2}#1}
\newcommandx\Ichecklog[1][usedefault, addprefix=\global, 1=x]{{\check{I}_{#1}^{\log}}}

\section{Introduction}

Intersymbol interference (ISI) is a ubiquitous impairment in communication
and data storage media \cite{proakis1987digital}. Techniques of information
transmission over ISI channels can be roughly divided into two types:
single-carrier (SC) modulation and multi-carrier modulation. 

In SC modulation, symbols are transmitted at a rate approximately
equal to the available bandwidth, and are distorted by ISI. The distortion
can be compensated for at the receiver using different equalization
techniques such as maximum-likelihood sequence estimation \cite{forney1972maximum}
and linear MMSE estimation, possibly with decision-feedback \cite{cioffi1995mmse}.
Equalization can also be combined with decoding, resulting in various
iterative schemes \cite{tuchler2002turbo}. 

In multi-carrier modulation, the available bandwidth is divided among
several lower symbol-rate signals, each with a different carrier frequency,
often referred to as subcarriers. Orthogonal frequency-division multiplexing
(OFDM) \cite{hwang2009ofdm,li2013ofdma} is the most important multi-carrier
modulation technique. In OFDM, the frequency spacing between subcarriers
is chosen so that they are orthogonal on every signaling interval.
Additionally, a portion of the end of each interval is copied to its
beginning and is commonly referred to as the cyclic prefix. These
two modifications allow for low-complexity optimal equalization using
the FFT algorithm, which is the primary advantage of OFDM.

Today, OFDM is the predominant modulation technique in high-bandwidth
communications over channels with significant ISI, and is featured
in a large number of standards, including DSL \cite{bingham2000adsl},
WiFi \cite{van2006wifi}, WiMAX \cite{ghosh2005wimax}, DVB-T \cite{reimers1996dvbt}
and the LTE downlink \cite{ghosh2010lte}. However, OFDM waveforms
suffer from a higher peak to average power ratio (PAPR) than SC waveforms.
Due to their lower PAPR, and due to the introduction of efficient
frequency-domain decision feedback equalization techniques \cite{benvenuto2010single},
SC schemes have become a viable alternative to OFDM in certain settings.
In particular, small, cheap and power-efficient amplifiers require
low PAPR input due to their limited dynamic range, making SC desirable
when they are used. Two such examples are the uplink air-interface
of LTE, where a scheme called SC-FDMA is used \cite{myung2008single},
and the new 802.11ad specification, which includes a SC option \cite{perahia2011gigabit}.

The purpose of this paper is to compare the maximum achievable rates
of reliable communication in OFDM and SC modulations, when optimal
equalization and channel coding is assumed. Since optimal OFDM equalization
is trivial, modern OFDM systems are able to approach this maximum
theoretical rate using advanced coding schemes such as turbo codes
or LDPC \cite{ryan2009channel}. In contrast, optimal equalization
and decoding cannot be decoupled in SC schemes, and instead must be
approximated by iterative turbo equalization techniques \cite{tuchler2002turbo}.
Such techniques incur a high computational cost which currently renders
them infeasible in practice. Nonetheless, with ever-growing computation
resources and steadily improving iterative receivers (c.f. \cite{nguyen2011spatially}),
SC schemes able to approach the maximum achievable rate might soon
become feasible. Therefore, comparison of the two maximum achievable
rates is of practical as well as theoretical interest.

When only average power constraints apply on the channel input, it
is well known that Gaussian signaling achieves the same maximum rate
in both OFDM and SC schemes \cite{hirt1988capacity}. However, we
impose two practical restrictions, which rule out the classical Gaussian
solution. The first restriction is that the channel inputs take values
in a certain fixed finite alphabet (constellation), as is always the
case in practice. The second restriction is that the inputs are i.i.d.,
or more precisely that an i.i.d. random coding distribution is used.
This restriction is justified as long as we limit ourselves to channel
coding schemes that were designed for memoryless channels, as is common
in practice. For OFDM the latter restriction is taken to mean that
all subcarriers use the same distribution --- this will simplify our
calculations and it also accurately models wireless ISI channels,
which most often change too rapidly for constellation loading to become
practical.

Under these constraints, there is no closed-form expression for the
achievable rate of SC modulation. However, an expression introduced
by Shamai and Laroia \cite{shamai1996intersymbol} is known to tightly
approximate the SC achievable rate in virtually all scenarios. The
Shamai-Laroia approximation is intimately connected to the performance
of decision-feedback equalization of SC signals.

In this paper, we show that in the sense of achievable rate and under
the above-mentioned restrictions, SC modulation will often offer performance
superior or equal to that of OFDM. In particular, we prove that for
BPSK and QPSK inputs, the OFDM achievable rate is lower than the Shamai-Laroia
approximation regardless of the ISI channel. A numerical study indicates
this holds also for 4-PAM, 8-PSK, 16-QAM and 32-QAM inputs. For general
finite-alphabet input distributions, we prove the same result in the
low- and high-SNR regimes. For PAM and square QAM inputs, we find
an SNR threshold above which the high-SNR regime is in effect. For
square QAM inputs of order 256 and above, this threshold corresponds
to a symbol error rate of more than 50\% and is therefore reasonably
low. We provide an exact characterization of the maximum advantage
OFDM may offer over the Shamai-Laroia approximation. Numerical evaluation
shows that this advantage is very small for QAM constellations of
orders up to 4096 (less than 0.1 bit). Complementing this result,
we show that the advantage of the Shamai-Laroia approximation over
the OFDM rate becomes arbitrarily large for some of ISI channels.
Thus, SC schemes may offer a significant performance gain over OFDM
in certain cases, but not vice versa. Finally, we study continuous
uniform input, which is limiting case of increasingly high-order QAM
inputs. We show that unlike the finite-alphabet cases, under uniform
input the Shamai-Laroia approximation cannot significantly exceed
the OFDM achievable rate. This indicates that the advantage of SC
over OFDM will become small if a sufficiently dense input constellation
is chosen. However, such inputs are not necessarily feasible. We provide
a detailed discussion on practical scenarios in which SC is expected
to offer a considerable performance gain over OFDM.

Our results stem from the concavity properties of the input-output
mutual information in a scalar Gaussian channel, as a function of
a rescaled SNR variable that we call the ``log-SNR''. In order to
investigate these properties, we make extensive use of Information-Estimation
results that link derivatives of the mutual information function and
estimation-theoretic quantities \cite{guo2005mutual,guo2011estimation,payaro2009hessian}.
We derive new bounds on MMSE estimation of PAM inputs to an additive
Gaussian channel in the high-SNR regime. Besides their use in proving
some of our main results, we believe them to be of general interest.

There is a considerable amount of literature that deals with comparison
between SC and OFDM modulations. It is mostly concerned with comparison
of specific schemes and quantities such as PAPR and bit or frame error
rates (c.f. \cite{wilson1995comparison,czylwik1997comparison,tubbax2001ofdm,wang2004ofdm,lin2003ber,de2011uncoded}).
Some works also compare fundamental limits. In \cite{fischer1997equivalence,zervos1989optimized,benvenuto2002comparison}
achievable rates are considered, but Gaussian inputs are assumed in
order to model adaptive constellation loading, leading to the expected
conclusion that both methods offer the same rates. In \cite{zhang2012comparison}
achievable rates are compared via simulation for binary input and
two-taps ISI channels. The authors report that in these settings SC
is superior, but make no general or theoretically-backed claims. In
\cite{franceschini2008information} it is conjectured that the SC
achievable rate is always higher than the OFDM rate. This conjecture
is supported by numerical evidence, but no theoretical analysis is
performed. In \cite{aue1998comparison} and \cite{de2011comparison}
the cut-off rate is compared analytically, and the SC rate is shown
to exceed the OFDM rate in several scenarios. 

In a recent report \cite{de2013comparison} the authors independently
found that concavity with respect to log-SNR yields an inequality
between the OFDM achievable rate and the Shamai-Laroia approximation.\textbf{
}Based on numerical evidence, they argue that concavity holds for
QPSK and 16-QAM inputs and that while it does not always hold for
higher order constellations, the maximum difference in favor of OFDM
is small. However, the majority of our results remain exclusive to
this work, including the analytic proofs of concavity, the study of
the concave envelope and the application of Information-Estimation
tools.

The rest of this paper is organized as follows. Section \ref{sec:preliminaries}
formulates the problem and presents the main results, briefly discussing
their extension when linear precoding is allowed. Section \ref{sec:concavity results}
introduces the log-SNR scale and studies the concavity of the mutual
information function with respect to it, culminating in a proof to
our main result for general finite-alphabet inputs. Section \ref{sec:mmse bounds}
focuses on the nonlinear MMSE estimation of PAM random variables corrupted
by additive Gaussian noise. In this section, a novel ``pointwise''
result is presented, bounding the conditional variance of the channel
input given an observation of the channel output. This bound is then
applied to derive a tight high-SNR characterization of the MMSE function
and its derivative.  Section \ref{sec:pam qam proofs} uses insights
from the two previous sections in order to prove our results pertaining
PAM and square QAM inputs, as well as continuous uniform input. Section
\ref{sec:discussion} contains an in-depth discussion on the differences
between the SC and OFDM achievable rates likely to occur in practice,
and the implications of increasing the constellation order. Section
\ref{sec:conclusion} concludes this paper.

\section{\label{sec:preliminaries}Preliminaries and main results}

We consider a complex-valued, discrete-time ISI channel model,
\begin{equation}
y_{k}=\sum_{i=0}^{L-1}h_{i}x_{k-i}+n_{k},
\end{equation}
where $x_{-\infty}^{\infty}$ is the channel input sequence%
\footnote{We use the standard notation $a_{N_{1}}^{N_{2}}$ for the sequence
$[a_{N_{1}},a_{N_{1}+1},...,a_{N_{2}}]$ with the natural interpretation
when $N_{1}=-\infty$ and/or $N_{2}=\infty$.%
}, $h_{0}^{L-1}$ are arbitrary complex-valued ISI taps and $n_{-\infty}^{\infty}$
is an i.i.d. standard complex Gaussian%
\footnote{A standard complex Gaussian variable is of the form $n^{I}+jn^{Q}$,
where $n^{I},n^{Q}\sim\mathcal{N}\left(0,1/2\right)$ and $n^{I}\perp n^{Q}$.%
} sequence independent of the input. Here $L$ denotes the length of
the channel impulse response. Let $H\left(\theta\right)=\sum_{k=0}^{L-1}h_{k}e^{-jk\theta}$
be the ISI channel transfer function. We assume throughout the paper
that the input sequence has zero mean and unit average power, \emph{i.e.}
$\E[\left|x_{i}\right|^{2}]=1$. Since the input and noise are both
normalized to unit variance, the quantity $\sum_{k=0}^{L-1}\left|h_{k}\right|^{2}=\frac{1}{2\pi}\int_{-\pi}^{\pi}|H(\theta)|^{2}d\theta$,
which is not normalized to unity, expresses the input signal-to-noise
ratio (SNR).

\subsection{Single carrier modulation model}

In our model for single-carrier modulation, the channel input sequence
is assumed to be i.i.d., zero-mean and unit power, with every input
symbol drawn from a finite complex-valued alphabet also known as a
signal constellation. Conventional constellations are composed of
$2^{m}$ uniformly spaced symbols, each representing $m$ data bits.
Commonly used constellations include BPSK, QPSK, 8-PSK, 16-QAM and
64-QAM \cite{proakis1987digital}. Unless specifically mentioned otherwise,
our results apply for any (finite alphabet) input distribution. However,
when we refer to a certain input distribution by its constellation
name (\emph{e.g.} ``BPSK input'' or ``256-QAM'' input), a uniform
distribution over the constellation points will be assumed. The maximum
achievable rate for reliable communication under the assumptions of
this model is given by the input-output Average Mutual Information
\cite{gray2010entropy}:
\begin{equation}
\Iach\triangleq\lim_{K\rightarrow\infty}\frac{1}{2K+1}I\left(x_{-K}^{K}\:;\: y_{-K}^{K}\right)=I\left(x_{0}\:;\: y_{-\infty}^{\infty}\:|\: x_{-\infty}^{-1}\right)
\end{equation}
where $I\left(A\:;\: B\right)$ denotes the mutual information between
$A$ and $B$, and $I\left(A\:;\: B\:|\: C\right)$ denotes the mutual
information between $A$ and $B$, conditioned on $C$ (c.f. \cite{cover2012elements}).

When the input distribution is symmetric Gaussian, a closed-form expression
for $\Iach$ can be easily derived (cf. \cite{hirt1988capacity,shamai1996intersymbol}),
\begin{equation}
\mathcal{I}_{\textrm{SC,Gaussian}}=\frac{1}{2\pi}\int_{-\pi}^{\pi}\log\left(1+|H(\theta)|^{2}\right)d\theta
\end{equation}

Let
\begin{equation}
I_{x}\left(\gamma\right)\triangleq I\left(x\:;\:\sqrt{\gamma}x+n\right)\label{eq:Ix_def}
\end{equation}
stand for the input-output mutual information in a scalar complex-valued
Gaussian channel with unit-power input%
\footnote{By ``Gaussian channel'' we mean a channel with additive Gaussian
noise independent of the input, which is not necessarily Gaussian.
The subscript $x$ will commonly be used to indicate that a general
input distribution $x$ is discussed, and subscripts with indicative
names will be employed when referring to specific input distributions,
\emph{e.g.} $I_{\mathrm{BPSK}}\left(\gamma\right)$ and $I_{\mathrm{Gaussian}}\left(\gamma\right)$.%
} $x$, SNR $\gamma$ and standard complex Gaussian noise $n$, independent
of $x$. The Gaussian achievable rate can also be expressed as
\begin{equation}
\mathcal{I}_{\textrm{SC,Gaussian}}=I_{\mathrm{Gaussian}}\left(\SNR[MMSE-DFE-U]\right)
\end{equation}
where 
\begin{equation}
\SNR[MMSE-DFE-U]=\exp\left\{ \frac{1}{2\pi}\int_{-\pi}^{\pi}\log\left(1+|H(\theta)|^{2}\right)d\theta\right\} -1\label{eq:SNR-dfe-def}
\end{equation}
is the output SNR of the unbiased MMSE linear estimator of $x_{0}$
given $x_{-\infty}^{-1}$ and $y_{-\infty}^{\infty}$, known as the
MMSE decision-feedback equalizer (DFE), and $I_{\mathrm{Gaussian}}\left(\gamma\right)=\log\left(1+\gamma\right)$
is the input-ouput mutual information for a complex-valued Gaussian
channel with standard complex Gaussian input. A concise presentation
of the MMSE descision-feedback equalizer, as well as a derivation
of its output SNR can be found in \cite{cioffi1995mmse}.

When the input distribution in not Gaussian, no closed-form expression
for $\Iach$ is known and it must be approximated either analytically
\cite{shamai1996intersymbol,jeong2012easily,carmon2013slc} or by
Monte-Carlo simulations \cite{arnold2006simulation,pfister2001achievable,radosevic2011bounds}.
A simple and often-used approximation for $\Iach$ was first proposed
by Shamai and Laroia \cite{shamai1996intersymbol},
\begin{equation}
\Iach\approx\Isl\triangleq I_{x}\left(\SNR[MMSE-DFE-U]\right)
\end{equation}
where $x$ is a random variable distributed as one of the input symbols
to the ISI channel. This approximation can be derived by applying
the MMSE DFE on the channel output sequence and replacing the residual
ISI by independent Gaussian variables with equal power --- however,
as explained in \cite{shamai1996intersymbol}, the central limit theorem
cannot be used to rigorously justify this approximation, as the residual
ISI coefficients do not meet its conditions. 

The Shamai-Laroia approximation was originally conjectured to be a
lower bound on $\Iach$. However, in \cite{carmon2013slc} a counterexample
based on highly skewed binary input is constructed, showing that this
conjecture does not hold for all input distributions. Nonetheless,
extensive experimentation has shown that when conventional input distributions
are used, $\Isl$ is an extremely tight lower bound for $\Iach$ for
any ISI channel and SNR \cite{shamai1996intersymbol,arnold2006simulation,jeong2012easily}.
In particular, there is no known counterexample to $\Iach\geq\Isl$
that involves symmetrically distributed inputs (as all conventional
inputs are). Moreover, in \cite{carmon2013slc} the lower bound $\Iach\geq\Isl$
is proven to hold for sufficiently high SNR, further establishing
its validity. Whether $\Iach\geq\Isl$ can be proven to always hold
for specific input distributions is a question open to future research.

\subsection{OFDM modulation model}

In OFDM, information is transmitted in blocks of $N+N_{CP}$ channel
inputs, where the first $N_{CP}$ elements of each block are identical
to its last $N_{CP}$ elements and thus constitute a cyclic prefix
(CP). With the CP discarded at the receiver, the ISI channel is transformed
into a vector channel,
\begin{equation}
\mathbf{y}=\mathbf{H}\mathbf{x}+\mathbf{n}
\end{equation}
The vectors $\mathbf{y}$ and $\mathbf{x}$ represent blocks of $N$
channel outputs and inputs, respectively, $\mathbf{n}$ is a standard
complex Gaussian noise vector, and \textbf{$\mathbf{H}$ }is a matrix
representing the ISI. Practical OFDM schemes are designed so that
the cyclic prefix is longer than the channel memory ($N_{CP}>L$).
Moreover, we consider a sequence of schemes in which the block size
$N$ and the cyclic prefix size $N_{CP}$ grow to infinity, so that
$N_{CP}$ is guaranteed to exceed $L$ eventually. Assuming $N_{CP}>L$,
$\mathbf{H}$ is a circulant matrix, with first row equal to $\left[h_{0},0\cdots0,h_{L-1},\cdots,h_{1}\right]$.
Therefore, \textbf{$\mathbf{H}$} is diagonalized by the the DFT matrix
of order $N$,
\begin{equation}
\mathbf{H^{d}}=\mathbf{W}\mathbf{H}\mathbf{W}^{-1}
\end{equation}
with $\mathbf{H^{d}}$ a diagonal matrix and $W_{m,k}=\frac{1}{\sqrt{N}}e^{-2\pi jmk/N}$
the DFT matrix. Applying the input precoding $\mathbf{x}=\mathbf{W}^{-1}\mathbf{\tilde{x}}$
and output transformation $\tilde{\mathbf{y}}=\mathbf{W}\mathbf{y}$
thus yields an equivalent diagonal vector channel,
\begin{equation}
\tilde{\mathbf{y}}=\mathbf{H^{d}}\mathbf{\tilde{x}}+\tilde{\mathbf{n}}\label{eq:ofdm diagonalized channel model}
\end{equation}

We assume that the elements of $\mathbf{\tilde{x}}$ are i.i.d.%
\footnote{This is usually the case in wireless links, where the communication
overhead of coordinating different powers and constellations for different
subcarriers often makes doing so undesirable.%
}, zero mean and have unit average power. Since the channel model in
(\ref{eq:ofdm diagonalized channel model}) is simply $N$ parallel
channels, the maximum achievable rate per channel input is given by
\begin{equation}
\Iofdm^{\left(N\right)}\triangleq\frac{1}{N+N_{CP}}I\left(\tilde{\mathbf{x}};\tilde{\mathbf{y}}\right)=\frac{1}{N+N_{CP}}\sum_{i=1}^{N}I_{x}\left(|H_{i,i}^{d}|^{2}\right)
\end{equation}
with $I_{x}\left(\cdot\right)$ as defined in (\ref{eq:Ix_def}) and
$x$ distributed as one of the elements of $\tilde{\mathbf{x}}$.
The Toeplitz Distribution Theorem \cite{gray2006toeplitz} allows
us to take the limit of the large block size,
\begin{equation}
\Iofdm\triangleq\lim_{N\rightarrow\infty}\Iofdm^{\left(N\right)}=\frac{1}{2\pi}\int_{-\pi}^{\pi}I_{x}\left(|H(\theta)|^{2}\right)d\theta
\end{equation}
where it is assumed that $N_{CP}$ grows as $o(N),$ so that the rate
overhead of the cyclic prefix vanishes.

\subsection{MMSE estimation in a scalar Gaussian channel}

Consider once more the scalar complex-valued Gaussian channel $y=\sqrt{\gamma}x+n$
with input $x$ and standard complex Gaussian noise $n$ independent
of $x$. The minimum mean square error (MMSE) in estimating $x$ from
$y$ is given by
\begin{equation}
\mathrm{mmse}_{x}\left(\gamma\right)=\E[\left|x-\Ebrack{x|y}{}\right|^{2}]\label{eq:mmsex_def}
\end{equation}
For a standard complex Gaussian input we have $\mathrm{mmse}_{\textrm{Gaussian}}\left(\gamma\right)=1/\left(1+\gamma\right)\geq\mathrm{mmse}_{x}\left(\gamma\right)$,
for any other unit-power input $x$.

Assuming $x$ has zero mean and unit variance (\emph{i.e.} $\E[\left|x\right|^{2}]=1)$,
the connection between the mutual information (\ref{eq:Ix_def}) and
the MMSE (\ref{eq:mmsex_def}) is given by:
\begin{equation}
I_{x}'\left(\gamma\right)=\mathrm{mmse}_{x}\left(\gamma\right)\label{eq:Immse relation}
\end{equation}
Note that the above equation differs from the familiar Guo-Shamai-Verd{\'u}
formula \cite{guo2005mutual} by a factor of 2 on the right-hand side.
This is due to the fact that our channel model is complex-valued.
The relation (\ref{eq:Immse relation}) can be easily derived from
the vector version of the GSV theorem (eq. (22) in \cite{guo2005mutual}),
by considering a two-dimensional scalar channel matrix.

\subsection{Statement of results}

Our main result provides a connection between $\Iofdm$ and $\Isl$
for finite-alphabet inputs. First, we show that an inequality of the
form $\Iofdm\leq\Isl+\Delta_{x}$ always holds, where $\Delta_{x}\geq0$
depends only on the input distribution (and not on the ISI channel).
Second, we characterize low and high SNR regions in which the strengthened
inequality $\Iofdm\leq\Isl$ holds, even when $\Delta_{x}\neq0$.
To this end, we introduce two pairs of SNR thresholds. The first pair,
denoted $\underline{\gamma}_{1}$ and $\bar{\gamma}_{2}$, constitutes
low and high SNR thresholds with respect to $\SNR[MMSE-DFE-U]$, defined
in (\ref{eq:SNR-dfe-def}) --- when $\SNR[MMSE-DFE-U]$ is below $\underline{\gamma}_{1}$
or above $\bar{\gamma}_{2}$, we have $\Iofdm\leq\Isl$. The second
threshold pair is denoted $\underline{\gamma}_{0}$ and $\bar{\gamma}_{0}$,
and relates to the channel's frequency response --- when $|H(\theta)|^{2}$
is bounded by $\underline{\gamma}_{0}$ from above or by $\bar{\gamma}_{0}$
from below, we are guaranteed once more to have $\Iofdm\leq\Isl$.
Like $\Delta_{x}$, these thresholds depend only on the input distribution. 

The explicit construction of $\underline{\gamma}_{0}$, $\bar{\gamma}_{0}$,
$\underline{\gamma}_{1}$, $\bar{\gamma}_{2}$ and $\Delta_{x}$ is
deferred to Section \ref{sec:concavity results} and Definitions \ref{def:zeta0},
\ref{def:zeta12} and \ref{def:delta_x} therein, as it relies on
concepts and results developed there. These quantities are defined
in terms of concavity properties of a certain function, and are straightforward
to evaluate numerically. In particular, we provide an expression for
$\Delta_{x}$ in terms of a three-variable optimization problem which
is easily solved numerically --- see (\ref{eq:dmax-def pre}) below.

Formally stated, our main result is as follows,
\begin{thm}
\label{thm:Iofdm_sl_bound}For any ISI channel and any finite alphabet
distribution $x$,\textup{
\begin{equation}
\Iofdm\leq\Isl+\Delta_{x}\label{eq:Iofdm log-snr concave envelope bound}
\end{equation}
}Where $\Delta_{x}\geq0$ is given by
\begin{equation}
\Delta_{x}\triangleq\sup_{\substack{\gamma_{1},\gamma_{2},\gamma\,\mathrm{s.t.}\\
\gamma_{1}\leq\gamma\leq\gamma_{2}
}
}\frac{\log\left(\frac{1+\gamma}{1+\gamma_{1}}\right)\left[I_{x}\left(\gamma_{2}\right)-I_{x}\left(\gamma\right)\right]-\log\left(\frac{1+\gamma_{2}}{1+\gamma}\right)\left[I_{x}\left(\gamma\right)-I_{x}\left(\gamma_{1}\right)\right]}{\log\left(\left[1+\gamma_{2}\right]/\left[1+\gamma_{1}\right]\right)}\label{eq:dmax-def pre}
\end{equation}
Additionally, if $\Delta_{x}>0$\textup{,} there exit $0<\underline{\gamma}_{1}\leq\underline{\gamma}_{0}\leq\bar{\gamma}_{0}\leq\bar{\gamma}_{2}<\infty$
that depend only on the input distribution, such that \textup{$\Iofdm\leq\Isl$}\textup{
}holds whenever the channel transfer function $H\left(\theta\right)$
satisfies at least one of the following conditions:
\begin{enumerate}
\item \textup{$\SNR[MMSE-DFE-U]\in[0,\underline{\gamma}_{1}]\cup[\bar{\gamma}_{2},\infty)$}
\item $|H(\theta)|^{2}\leq\underline{\gamma}_{0}$ for every $\theta\in\left(-\pi,\pi\right)$
\item $|H(\theta)|^{2}\geq\bar{\gamma}_{0}$ for every $\theta\in\left(-\pi,\pi\right)$
\end{enumerate}
\end{thm}
Next, we provide the following results, which sharpen Theorem \ref{thm:Iofdm_sl_bound}
for specific input distributions,
\begin{thm}
\label{thm:low_order_QAM}For BPSK and QPSK inputs, $\Delta_{x}=0$
and so \textup{$\Iofdm\leq\Isl$} for every ISI channel\textup{.}
\end{thm}

\begin{thm}
\label{thm:MQAM_high_snr}For $M$-PAM and square $M^{2}$-QAM inputs,
\textup{$\left(d_{\min}/2\right)^{2}\bar{\gamma}_{0}\leq1$,} where
$d_{\min}$ is the minimum distance between input symbols, assuming
unit input power.
\end{thm}
Combined with Theorem \ref{thm:Iofdm_sl_bound}, Theorem \ref{thm:MQAM_high_snr}
implies that for $M$-PAM and square $M^{2}$-QAM inputs, $\Iofdm\leq\Isl$
whenever the ISI channel is such that $\left(d_{\min}/2\right)^{2}|H(\theta)|^{2}\geq1$
for every $\theta\in\left(-\pi,\pi\right)$. Since the uncoded symbol
error rate in OFDM subcarrier frequency $\theta_{0}$ is a function
of $\left(d_{\min}/2\right)^{2}|H(\theta_{0})|^{2}$, the following
corollary is immediate,
\begin{cor}
\label{cor:MQAM SER implication}For a given ISI channel and square
$M^{2}$-QAM inputs with $M\geq16$, if the uncoded symbol error rate
is below 50\% in all OFDM subcarriers, \textup{$\Iofdm\leq\Isl$.}
\end{cor}
Our last result deals with uniformly distributed input. This input
distribution represents the limit of infinitely high-order QAM, and
is therefore referred to also as $\infty$-QAM. Since this input has
an infinite alphabet, Theorem \ref{thm:Iofdm_sl_bound} does not apply
to it. Instead, we have

\begin{thm}
\label{thm:uniform_sl_ofdm}For uniformly distributed complex input
and any ISI channel,\textup{
\begin{equation}
-\tilde{\Delta}_{\infty\text{-QAM}}\leq\Iofdm-\Isl\leq\bar{\Delta}_{\infty\text{-QAM}}\left(\max_{\theta\in\left(-\pi,\pi\right)}\left|H\left(\theta\right)\right|^{2}\right)
\end{equation}
}where \textup{$\tilde{\Delta}_{\infty\text{-QAM}}\approx0.0608\text{ [bit]}$},
and \textup{$\bar{\Delta}_{\infty\text{-QAM}}\left(\gamma\right)$},
is a non-decreasing function that satisfies \textup{$\bar{\Delta}_{\infty\text{-QAM}}\left(\gamma\right)=0$}
for every \textup{$\gamma\leq\underline{\gamma}_{0}^{\left(\infty\text{-QAM}\right)}\approx8.76\text{ [dB]}$},
and\textup{
\begin{equation}
\lim_{\gamma\to\infty}\bar{\Delta}_{\infty\text{-QAM}}\left(\gamma\right)\triangleq\Delta_{\infty\text{-QAM}}=\log\left(\pi e/6\right)\approx0.509\text{ [bit]}
\end{equation}
}is the uniform input shaping loss with respect to Gaussian input.

Moreover, \textup{$\Iofdm\geq\Isl$} when $H\left(\theta\right)$
satisfies at least one of the following conditions:
\begin{enumerate}
\item \textup{$\SNR[MMSE-DFE-U]\geq\tilde{\gamma}_{2}^{\left(\infty\text{-QAM}\right)}\approx16.5\text{ [dB]}$}
\item \textup{$|H(\theta)|^{2}\geq\underline{\gamma}_{0}^{\left(\infty\text{-QAM}\right)}\approx8.76\text{ [dB]}$}
for every $\theta\in\left(-\pi,\pi\right)$
\end{enumerate}
\end{thm}

\begin{figure}
\begin{centering}
\includegraphics[width=8cm]{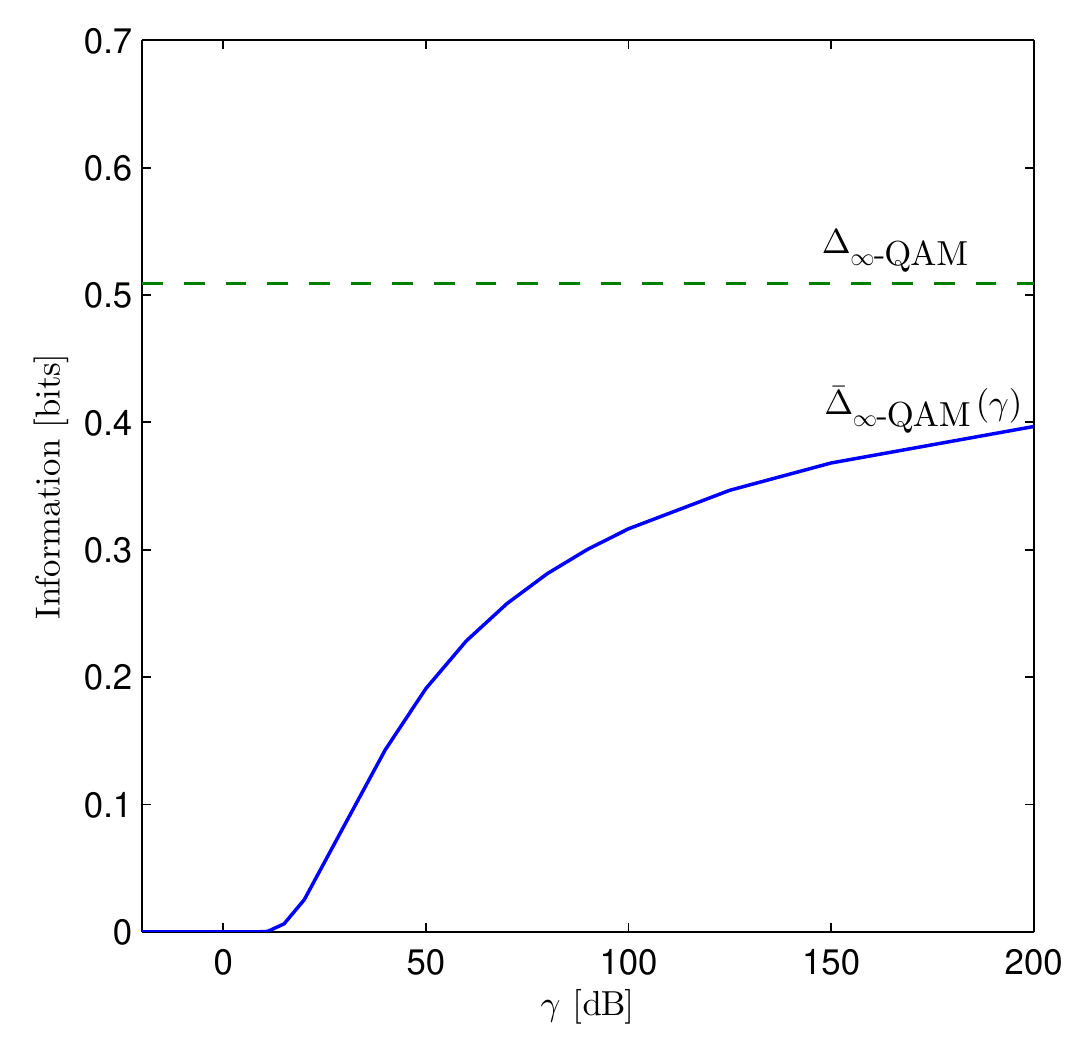}
\par\end{centering}

\protect\caption{\label{fig:deltabar}Numerical evaluation of $\bar{\Delta}_{\infty\text{-QAM}}\left(\gamma\right)$.}
\end{figure}

Note that Theorem \ref{thm:uniform_sl_ofdm} provides some of the
guarantees of Theorem \ref{thm:Iofdm_sl_bound}. In particular, we
have $\Iofdm\leq\Isl+\Delta_{\infty\text{-QAM}}$ for any ISI channel,
as well as $\Iofdm\leq\Isl$ for every channel that satisfies $|H(\theta)|^{2}\leq\underline{\gamma}_{0}^{\left(\infty\text{-QAM}\right)}$.
Figure \ref{fig:deltabar} graphs $\bar{\Delta}_{\infty\text{-QAM}}\left(\gamma\right)$,
which is evaluated numerically based on the analysis carried out in
subsection \ref{sub:uniform analysis}. As seen in the figure, the
convergence of $\bar{\Delta}_{\infty\text{-QAM}}\left(\gamma\right)$
to $\Delta_{\infty\text{-QAM}}$ is extremely slow. Thus, for any
practical purpose we may select an SNR level $\bar{\gamma}$ that
is much higher than any plausible value of $|H(\theta)|^{2}$, and
use $\bar{\Delta}_{\infty\text{-QAM}}(\bar{\gamma})$ instead of $\Delta_{\infty\text{-QAM}}$.
Depending on the application, appropriate choices of $\bar{\gamma}$
are likely to yield values between $\bar{\Delta}_{\infty\text{-QAM}}\left(30\text{ dB}\right)\approx0.0841$
and $\bar{\Delta}_{\infty\text{-QAM}}\left(60\text{ dB}\right)\approx0.228$.

Table \ref{tab:log concave summary} summarizes a numerical study
of the quantities that appear in Theorem \ref{thm:Iofdm_sl_bound}.
The table is consistent with Theorem \ref{thm:low_order_QAM} and
indicates it also extends to 4-PAM, 8-PSK, 16-QAM and 32-QAM inputs.
It also reveals that while nonzero, $\Delta_{\textrm{64-QAM}}$ is
negligible, being of the order of a millionth of a bit. For higher
order constellations $\Delta_{x}$ is more significant, but remains
quite small even in very high-order constellations such as 4096-QAM.
The limiting case of $\infty$-QAM input is also included in the table.
It is seen that 4096-QAM has a value of $\underline{\gamma}_{0}$
quite close to that of $\infty$-QAM, but that $\Delta_{\textrm{4096-QAM}}$
is still far from converging to $\Delta_{\infty\text{-QAM}}$. Examining
Figure \ref{fig:deltabar}, it is seen that for the various QAM inputs
considered, $\Delta_{x}$ is well-approximated by $\bar{\Delta}_{\infty\text{-QAM}}\left(\bar{\gamma}_{2}\right)$.
Finally, the table shows that the general bound provided by Theorem
\ref{thm:MQAM_high_snr} is slack by an approximate factor of 2 ---
\emph{i.e.}, for higher-order QAM, $\left(d_{\min}/2\right)^{2}\bar{\gamma}_{0}\approx1/2$.

Our results show that $\Iofdm$ may only exceed $\Isl$ by a small
amount, but the opposite is not true. Indeed, in subsection \ref{sub:extremal diff}
we construct a family of channels for which $\Iofdm$ tends to zero
while $\Isl$ tends to the input entropy and in subsection \ref{sub:practical-difference}
we discuss practical scenarios in which the SC achievable rate is
significantly higher than the OFDM rate. However, Theorem \ref{thm:uniform_sl_ofdm}
indicates that this difference can be made small by increasing the
constellation order. Subsection \ref{sub:implications} further discusses
this course of action.

\begin{table}
\begin{centering}
\protect\caption{\label{tab:log concave summary}Numerical evaluation of the quantities
appearing in Theorem \ref{thm:Iofdm_sl_bound}}

\par\end{centering}

\begin{centering}
\begin{tabular}{|>{\centering}m{0.1\paperwidth}|c|c|c|c|c|c|}
\hline 
Input & $\left(\frac{d_{\min}}{2}\right)^{2}$ {[}dB{]} & $\underline{\gamma}_{1}$ {[}dB{]} & $\underline{\gamma}_{0}$ {[}dB{]} & $\bar{\gamma}_{0}$ {[}dB{]} & $\bar{\gamma}_{2}$ {[}dB{]} & $\Delta_{x}$ {[}bits{]}\tabularnewline
\hline 
\hline 
BPSK, 4-PAM, QPSK, 8-PSK, 16-QAM, 32-QAM & (varies) & - & - & - & - & 0\tabularnewline
\hline 
64-QAM & $-16.2$ & 10.7 & 11.0 & 11.7 & 12.0 & 1.86$\cdot10^{-6}$\tabularnewline
\hline 
256-QAM & $-22.3$ & 5.29 & 9.01 & 19.2 & 21.0 & 0.0202\tabularnewline
\hline 
1024-QAM & $-28.3$ & 3.63 & 8.81 & 25.2 & 27.5 & 0.0585\tabularnewline
\hline 
4096-QAM & $-34.4$ & 2.60 & 8.77 & 31.0 & 33.6 & 0.0987\tabularnewline
\hline 
$\infty$-QAM & $-\infty$ & - & 8.76 & - & - & 0.509\tabularnewline
\hline 
\end{tabular}
\par\end{centering}

\medskip{}

\centering{}\textbf{\small{}Note:}{\small{} All values are rounded
to three significant digits.}
\end{table}

\subsection{A note on linear precoding}

Our results extend straightforwardly to the following generalized
problem setting. In the SC case, we add a linear precoding filter
that is applied on the i.i.d. inputs prior to their transmission.
In the OFDM case, we allow a different power allocation for each subcarrier.
More concretely, the SC precoded transmitted symbols are
\begin{equation}
x_{n}^{\text{precoded}}=\sum_{i}c_{i}x_{n-i}
\end{equation}
where $x_{-\infty}^{\infty}$ is an i.i.d. input sequence and the
precoder taps satisfy $\sum_{i}\left|c_{i}\right|^{2}=1$. The OFDM
transmitted symbols are
\begin{equation}
\tilde{x}_{i}^{\text{precoded}}=\sqrt{P_{i}}\tilde{x}_{i}
\end{equation}
where $\tilde{x}_{1}^{N}$ are the i.i.d. OFDM block inputs as in
(\ref{eq:ofdm diagonalized channel model}), and the power allocations
satisfy $\sum_{i}P_{i}=1$.

Clearly, precoded SC modulation with taps $c_{-\infty}^{\infty}$
and ISI channel $\left|H\left(\theta\right)\right|^{2}$ is equivalent
to normal SC with channel $\left|C\left(\theta\right)H\left(\theta\right)\right|^{2}$
where $C\left(\theta\right)=\sum_{k}c_{k}e^{-ik\theta}$. Moreover,
OFDM with non-uniform power allocation is equivalent to normal OFDM
with channel $P\left(\theta\right)\left|H\left(\theta\right)\right|^{2}$
where $P\left(2\pi i/N\right)=P_{i}$. Equating $\left|C\left(\theta\right)\right|^{2}$
with $P\left(\theta\right)$ we conclude that for any SC linear precoder
there exist an OFDM power allocation such that both yield the same
equivalent ISI channel. Hence, given a degree of freedom in choosing
any SC linear precoder and any OFDM power allocation policy, our results
are still applicable, revealing that SC has significant advantages
over OFDM in this case as well.

The introduction of linear precoding lends additional viability to
our assumption of i.i.d. input, as the capacity-achieving SC scheme
can be viewed as i.i.d. Gaussian inputs linearly precoded with an
optimal Waterfilling filter. Thus, it is reasonable to assume that
when combined with a suitably chosen linear precoder, statistically
independent symbols will be close to optimal even when input alphabet
constraints prohibit Gaussian signaling. For OFDM with independent
non-Gaussian inputs, an optimal power allocation policy called Mercury/Waterfilling
was proposed in \cite{lozano2006optimum}. However, the Mercury/Waterfilling
spectrum does not necessarily describe the optimal linear precoder
for the i.i.d. non-Gaussian single carrier case. Using the methods
described in \cite{xiao2011globally}, it should be possible to find
the taps of this optimal precoder.

\section{\label{sec:concavity results}Concavity of mutual information with
respect to log-SNR}

\subsection{Log-SNR scale}

\begin{figure}
\begin{centering}
\includegraphics[width=10cm]{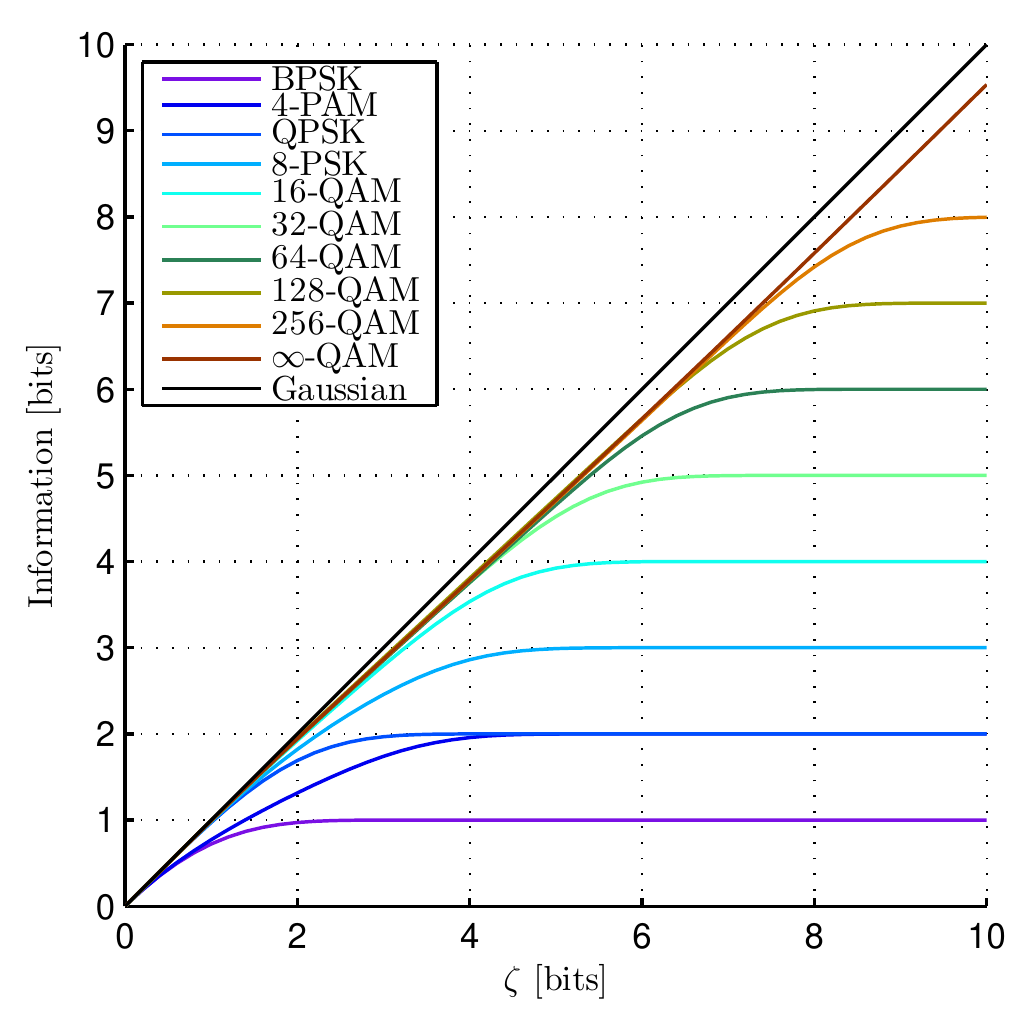}
\par\end{centering}

\protect\caption{\label{fig:Ilog}$\protect\Ilog\left(\zeta\right)$ for some common
input distributions.}
\end{figure}

For a given SNR $\gamma$, define the log-SNR as $\zeta=\log\left(1+\gamma\right)$,
and let
\begin{equation}
\Ilog\left(\zeta\right)\triangleq I_{x}(e^{\zeta}-1)\label{eq:ilog def}
\end{equation}
be the input output mutual information, as a function of log-SNR,
for a scalar complex Gaussian channel with zero-mean, unit-variance
input $x$. Since $\zeta$ is identical to $\Ilog[\mathrm{Gaussian}]\left(\zeta\right)$,
it is naturally measured in units of information. Moreover, we have
$\Ilog\left(\zeta\right)\leq\Ilog[\mathrm{Gaussian}]\left(\zeta\right)=\zeta$
for all inputs. Figure \ref{fig:Ilog} shows $\Ilog\left(\zeta\right)$
for some common input distributions. As can be seen in the figure,
$\Ilog\left(\zeta\right)$ is nearly linear for low $\zeta$ and,
for finite alphabet inputs, it is nearly constant for high $\zeta$,
with the shoulder occurring at around the input entropy.

The main results of this paper hinge on the concavity properties of
$\Ilog[x]\left(\zeta\right)$. In this section we study these properties
for a general input distribution. We begin by showing that $\Ilog\left(\zeta\right)$
is concave for sufficiently low and sufficiently high $\zeta$. Next,
we consider the concave envelope of $\Ilog\left(\zeta\right)$ and
show that it must equal $\Ilog\left(\zeta\right)$ for sufficiently
low and sufficiently high $\zeta$. Finally, we apply these conclusions
to prove Theorem \ref{thm:Iofdm_sl_bound}.

\subsection{Asymptotic concavity results}
\begin{prop}
\label{prop:concavity-low-SNR}For every input distribution $x$,
there exists $0<\zeta_{0}$ such that \textup{$\Ilog\left(\zeta\right)$}
is concave for every \textup{$\zeta\in[0,\zeta_{0}]$.}\end{prop}
\begin{IEEEproof}
Setting $\gamma=e^{\zeta}-1$ and differentiating $\Ilog[x]$ twice,
we find that
\begin{eqnarray}
\Ilog''(\zeta) & = & e^{2\zeta}I_{x}''(e^{\zeta}-1)+e^{\zeta}I_{x}'(e^{\zeta}-1)\nonumber \\
 & = & \left(1+\gamma\right)^{2}I_{x}''(\gamma)+\left(1+\gamma\right)I_{x}'(\gamma)\nonumber \\
 & = & \left(1+\gamma\right)\left[\mathrm{\mathrm{mmse}}_{x}(\gamma)+\left(1+\gamma\right)\mathrm{\mathrm{mmse}}_{x}'(\gamma)\right]\label{eq:ddIlog}\\
 & = & \left(1+\gamma\right)\frac{d}{d\gamma}\left[\left(1+\gamma\right)\mathrm{mmse}_{x}\left(\gamma\right)\right]=\left(1+\gamma\right)r_{x}'\left(\gamma\right)\label{eq:ddIlog as drx}
\end{eqnarray}
where the transition to (\ref{eq:ddIlog}) is due to the I-MMSE relation
(\ref{eq:Immse relation}). The function $r_{x}\left(\gamma\right)\triangleq\left(1+\gamma\right)\mathrm{mmse}_{x}\left(\gamma\right)$
denotes the ratio between the MMSE's of the non-linear and linear
optimal estimators of $x$ in the scalar complex Gaussian channel
with SNR $\gamma$. Clearly, $r_{x}\left(\gamma\right)\leq1$, and
$r_{x}\left(0\right)=1$. Therefore, by continuity there must be a
neighborhood of $0$, denoted by $[0,\gamma_{0}]$, in which $r_{x}$
is decreasing. Hence, by (\ref{eq:ddIlog as drx}) we find that $\Ilog\left(\zeta\right)$
is concave in $[0,\zeta_{0}]$, with $\zeta_{0}=\log(1+\gamma_{0})$.\end{IEEEproof}
\begin{prop}
\label{prop:Ilog eventually concave high SNR}For every input distribution
$x$ over a finite alphabet, there exists $\zeta_{0}<\infty$ such
that \textup{$\Ilog\left(\zeta\right)$} is concave for every \textup{$\zeta\in[\zeta_{0},\infty]$.}\end{prop}
\begin{IEEEproof}
Let $d_{\min}$ denote the minimum distance between any two symbols
in the input alphabet. By the standard probability of error upper
bound (c.f. Appendix C in \cite{lozano2006optimum}), we have
\begin{equation}
\mathrm{mmse}_{x}(\gamma)\leq D^{2}e^{-\left(d_{\min}/2\right)^{2}\gamma}
\end{equation}
for some $D>0$. Moreover in Appendix \ref{app:dmmse_bound_general_constel}
it is shown that
\begin{equation}
\mathrm{mmse}_{x}'(\gamma)\leq-C\frac{e^{-\left(d_{\min}/2\right)^{2}\gamma}}{\sqrt{\gamma}}
\end{equation}
for sufficiently large $\gamma$ and some $C>0$ . Therefore, denoting
again $\gamma=e^{\zeta}-1$ and substituting the above bounds in (\ref{eq:ddIlog}),
we find that
\begin{alignat}{1}
\Ilog''(\zeta) & \leq\left(1+\gamma\right)\left(D^{2}-\frac{1+\gamma}{\sqrt{\gamma}}C\right)e^{-\left(d_{\min}/2\right)^{2}\gamma}
\end{alignat}
for some $C>0$ and sufficiently large $\gamma$. Clearly, this implies
$\Ilog''(\zeta)<0$ for sufficiently large $\zeta$.\end{IEEEproof}
\begin{defn}
\label{def:zeta0}Let $0<\underline{\zeta}_{0}\leq\infty$ be the
\emph{maximal} $\zeta_{0}$ for which $\Ilog\left(\zeta\right)$ is
concave for every $\zeta\in[0,\zeta_{0}]$ and similarly let $0\leq\bar{\zeta}_{0}<\infty$
be the \emph{minimal} $\zeta_{0}$ for which $\Ilog\left(\zeta\right)$
is concave for every $\zeta\in[\zeta_{0},\infty)$. Let $\underline{\gamma}_{0}=e^{\underline{\zeta}_{0}}-1$
and $\bar{\gamma}_{0}=e^{\bar{\zeta}_{0}}-1$ denote the SNR's corresponding
to $\underline{\zeta}_{0}$ and $\bar{\zeta}_{0}$, respectively.
\end{defn}
Notice that $\underline{\zeta}_{0}<\bar{\zeta}_{0}$ if and only if
$\Ilog\left(\zeta\right)$ is not a concave function of $\zeta$,
in which case 
\begin{equation}
\Ilog''(\underline{\zeta}_{0})=\Ilog''(\bar{\zeta}_{0})=0
\end{equation}

\subsection{\label{sub:concave-envelope}Concave envelope}

Let $\Ihatlog\left(\zeta\right)$ denote the concave envelope \cite{boyd2009convex}
of $\Ilog$, \emph{i.e. }the smallest concave function that upper
bounds $\Ilog$, also given by, 
\begin{equation}
\Ihatlog\left(\zeta\right)=\sup_{\substack{\zeta_{1},\zeta_{2}\,\mathrm{s.t.}\\
\zeta_{1}\leq\zeta\leq\zeta_{2}
}
}\left[\frac{\left(\zeta-\zeta_{1}\right)\Ilog\left(\zeta_{2}\right)+\left(\zeta_{2}-\zeta\right)\Ilog\left(\zeta_{1}\right)}{\zeta_{2}-\zeta_{1}}\right]\label{eq:def concave envelope log}
\end{equation}
Clearly, $\Ihatlog$ exists and is upper bounded by $\zeta$, which
is a concave functions that upper bounds $\Ilog\left(\zeta\right)$.
Since $\Ilog$ is real-analytic, $\Ihatlog\left(\zeta\right)$ is
continuous and has a continuous derivative. Moreover, our previous
results allow the following,
\begin{prop}
For every input distribution with finite alphabet, There exist $\zeta_{1}>0$
and $\zeta_{2}<\infty$ such that $\Ihatlog\left(\zeta\right)=\Ilog\left(\zeta\right)$
for every $\zeta\in[0,\zeta_{1}]\cup[\zeta_{2},\infty)$.\end{prop}
\begin{IEEEproof}
At any point $\zeta$, the concave envelope $\Ihatlog\left(\zeta\right)$
is either equal to $\Ilog\left(\zeta\right)$ or linear in an interval
containing $\zeta$, such that the concave envelope is equal to $\Ilog$
at the edges of the intervals. Put in other words, there exists a
set of disjoint intervals $\left\{ [\zeta_{1,i},\zeta_{2,i}]\right\} _{i\in S}$
such that
\begin{equation}
\Ihatlog\left(\zeta\right)=\begin{cases}
\frac{\zeta-\zeta_{1,i}}{\zeta_{2,i}-\zeta_{1,i}}\Ilog\left(\zeta_{2,i}\right)+\frac{\zeta_{2,i}-\zeta}{\zeta_{2,i}-\zeta_{1,i}}\Ilog\left(\zeta_{1,i}\right) & \zeta\in[\zeta_{1,i},\zeta_{2,i}]\\
\Ilog\left(\zeta\right) & \mbox{otherwise}
\end{cases}\label{eq:def concave envelope piecewise}
\end{equation}
Moreover, since $\dIhatlog\left(\zeta\right)$ is also continuous,
the above statement implies that $\dIhatlog\left(\zeta\right)=\dIlog\left(\zeta_{1,i}\right)=\dIlog\left(\zeta_{2,i}\right)$
for every $\zeta\in[\zeta_{1,i},\zeta_{2,i}]$. 

Suppose by contradiction that there exists $i_{0}$ such that $\zeta_{1,i_{0}}=0$.
Denoting $\gamma=e^{\zeta}-1$, By the I-MMSE relationship we have
$\dIhatlog\left(\zeta\right)=r_{x}\left(\zeta\right)=\left(1+\gamma\right)\mathrm{mmse}_{x}\left(\gamma\right)$.
Since the input is normalized to unit power, we have $\dIlog\left(0\right)=r_{x}\left(0\right)=1$,
and so there must exist $\zeta_{2,i_{i}}>0$ such that $\dIlog\left(\zeta_{2,i}\right)=\dIlog\left(\zeta_{1,i}\right)=1$.
However, since the input is not Gaussian (it has finite alphabet),
and since $\mathrm{mmse}_{x}\left(0\right)=1$, the single-crossing
property \cite{guo2011estimation} implies that $\mathrm{mmse}_{x}\left(\gamma\right)<1/\left(1+\gamma\right)$
for every $\gamma>0$ and therefore $\dIlog\left(\zeta\right)<1$
for every $\zeta>0$, contradicting $\dIlog\left(\zeta_{2,i}\right)=1$.
Hence, setting $\zeta_{1}=\min_{i\in S}\zeta_{i,1}$, we find that
$\Ilog\left(\zeta\right)=\Ihatlog\left(\zeta\right)$ for any $\zeta\in[0,\zeta_{1}]$.

Since the input has finite alphabet, $\Ihatlog\left(\zeta\right)$
converges to a finite value (the input entropy). By the data-processing
inequality, mutual information is an increasing function of SNR, and
since the mapping $\zeta=\log\left(1+\gamma\right)$ is monotonic
and increasing, it follows that $\Ilog\left(\zeta\right)$ is also
increasing in $\zeta$. Therefore, $\dIlog\left(\zeta\right)>0$ for
any $0\leq\zeta<\infty$ and $\lim_{\zeta\to\infty}\dIlog\left(\zeta\right)=0$.
Moreover, by Proposition \ref{prop:Ilog eventually concave high SNR}
we know that there exist $\zeta_{0}<\infty$ such that $\dIlog\left(\zeta\right)$
is monotonically decreasing for every $\zeta>\zeta_{0}$. By the above
observations, there must exist $\zeta_{0}\leq\zeta_{2}<\infty$ such
that $\zeta_{a}\leq\zeta_{2}\leq\zeta_{b}$ implies $\dIlog\left(\zeta_{a}\right)>\dIlog\left(\zeta_{2}\right)\geq\dIlog\left(\zeta_{b}\right)$.
Clearly $\zeta_{2,i}\leq\zeta_{2}$ for any $i\in S$, as otherwise
the equality $\dIlog\left(\zeta_{2,i}\right)=\dIlog\left(\zeta_{1,i}\right)$
contradicts $\dIlog\left(\zeta_{1,i}\right)>\dIlog\left(\zeta_{2}\right)\geq\dIlog\left(\zeta_{2,i}\right)$.
Therefore, we have $\Ilog\left(\zeta\right)=\Ihatlog\left(\zeta\right)$
for any $\zeta\in[\zeta_{2},\infty)$, concluding our proof.\end{IEEEproof}
\begin{defn}
\label{def:zeta12}Let $0<\underline{\zeta}_{1}\leq\infty$ be the
\emph{maximal} $\zeta_{1}$ for which $\Ihatlog\left(\zeta\right)=\Ilog\left(\zeta\right)$
for every $\zeta\in[0,\zeta_{1}]$ and similarly let $0\leq\bar{\zeta}_{2}<\infty$
be the \emph{minimal} $\zeta_{2}$ for which $\Ihatlog\left(\zeta\right)=\Ilog\left(\zeta\right)$
for every $\zeta\in[\zeta_{2},\infty)$. Let $\underline{\gamma}_{1}=e^{\underline{\zeta}_{1}}-1$
and $\bar{\gamma}_{2}=e^{\bar{\zeta}_{2}}-1$ denote the SNR's corresponding
to $\underline{\zeta}_{1}$ and $\bar{\zeta}_{2}$, respectively. 
\end{defn}
Notice that in light of the above definition, the optimization in
the definition of the concave envelope (\ref{eq:def concave envelope log})
can be limited to $\zeta_{1}$ and $\zeta_{2}$ in the interval $[\underline{\zeta}_{1},\bar{\zeta}_{2}]$.
\begin{defn}
\label{def:delta_x}Let $\Delta_{x}$ denote that maximum difference
between $\Ilog\left(\zeta\right)$ and its concave envelope.
\end{defn}
By the above definition and (\ref{eq:def concave envelope log}) we
have,
\begin{eqnarray}
\Delta_{x} & = & \max_{\zeta}\left[\Ihatlog\left(\zeta\right)-\Ilog\left(\zeta\right)\right]\\
 & = & \sup_{\substack{\gamma_{1},\gamma_{2},\gamma\,\mathrm{s.t.}\\
\gamma_{1}\leq\gamma\leq\gamma_{2}
}
}\frac{\log\left(\frac{1+\gamma}{1+\gamma_{1}}\right)\left[I_{x}\left(\gamma_{2}\right)-I_{x}\left(\gamma\right)\right]-\log\left(\frac{1+\gamma_{2}}{1+\gamma}\right)\left[I_{x}\left(\gamma\right)-I_{x}\left(\gamma_{1}\right)\right]}{\log\left(\left[1+\gamma_{2}\right]/\left[1+\gamma_{1}\right]\right)}\label{eq:dmax def}
\end{eqnarray}
Using (\ref{eq:def concave envelope piecewise}), it is seen that
\begin{eqnarray}
\Delta_{x} & = & \Ihatlog\left(\zeta_{m}\right)-\Ilog\left(\zeta_{m}\right)=\left(\zeta_{m}-\zeta_{1,i}\right)\dIlog\left(\zeta_{m}\right)-\left[\Ilog\left(\zeta_{m}\right)-\Ilog\left(\zeta_{1,i}\right)\right]\nonumber \\
 & = & \left[\Ilog\left(\zeta_{2,i}\right)-\Ilog\left(\zeta_{m}\right)\right]-\left(\zeta_{2,i}-\zeta_{m}\right)\dIlog\left(\zeta_{m}\right)\label{eq:Delta_x zeta_m}
\end{eqnarray}
for some $i\in S$ and $\zeta_{m}\in[\zeta_{1,i},\zeta_{2,i}]$ that
satisfies $\dIlog\left(\zeta_{m}\right)=\dIlog\left(\zeta_{1,i}\right)=\dIlog\left(\zeta_{2,i}\right)$.

\begin{figure}
\centering{}\includegraphics[width=16cm]{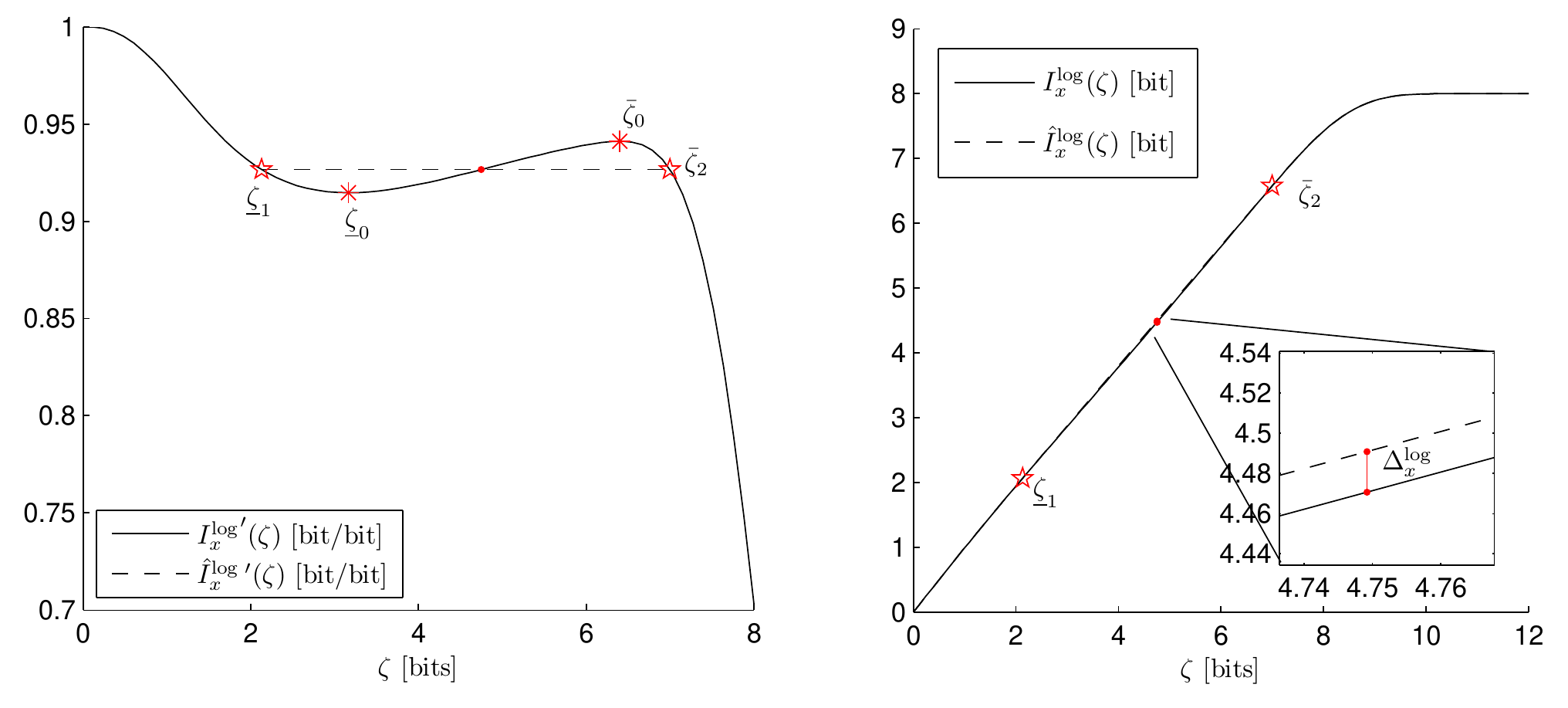}\protect\caption{\label{fig:Iloghat demo}$\protect\Ilog$ and $\protect\Ihatlog$
(right), and their derivatives with respect to $\zeta$ (left), with
$\underline{\zeta}_{0},\bar{\zeta}_{0},\underline{\zeta}_{1},\bar{\zeta}_{2}$
and $\Delta_{x}$ highlighted, for 256-QAM input.}
\end{figure}

Clearly, $\Delta_{x}=0$ if and only if $\Ilog$ is concave in $\mathbb{R}^{+}$.
As seen from Table \ref{tab:log concave summary}, $\Delta_{x}$ is
quite small, even when $\Ilog$ is not concave. Figure \ref{fig:Iloghat demo}
illustrates $\Ihatlog,\underline{\zeta}_{0},\bar{\zeta}_{0},\underline{\zeta}_{1},\bar{\zeta}_{2}$
and $\Delta_{x}$ as defined above, for an input uniformly distributed
on a 256-QAM constellation (two 16-PAM constellations in quadrature). 

\subsection{Proof of Theorem \ref{thm:Iofdm_sl_bound}}
\begin{IEEEproof}
The inequality (\ref{eq:Iofdm log-snr concave envelope bound}) is
immediate from the definitions of $\Isl$, $\Iofdm$, $\Ilog[x]$,
$\Ihatlog$ and $\Delta_{x}$:
\begin{alignat}{1}
\Iofdm & =\frac{1}{2\pi}\int_{-\pi}^{\pi}I_{x}\left(|H(\theta)|^{2}\right)d\theta\\
 & =\frac{1}{2\pi}\int_{-\pi}^{\pi}\Ilog\left(\log\left(1+|H(\theta)|^{2}\right)\right)d\theta\\
 & \leq\frac{1}{2\pi}\int_{-\pi}^{\pi}\Ihatlog\left(\log\left(1+|H(\theta)|^{2}\right)\right)d\theta\\
 & \leq\Ihatlog\left(\frac{1}{2\pi}\int_{-\pi}^{\pi}\log\left(1+|H(\theta)|^{2}\right)d\theta\right)\label{eq:Iofdm jensen application}\\
 & =\Ihatlog\left(\log\left(1+\SNR[MMSE-DFE-U]\right)\right)\\
 & \leq I_{x}\left(\SNR[MMSE-DFE-U]\right)+\Delta_{x}=\Isl+\Delta_{x}\label{eq:Iofdm definition application}
\end{alignat}
where in (\ref{eq:Iofdm jensen application}) the concavity of $\Ihatlog$
was used to invoke Jensen's inequality. Choosing $\underline{\gamma}_{1}$
and $\bar{\gamma}_{2}$ to be as defined in Definition \ref{def:zeta12},
it is clear that if condition 1 holds, then
\begin{equation}
\Ihatlog\left(\log\left(1+\SNR[MMSE-DFE-U]\right)\right)=\Ilog\left(\log\left(1+\SNR[MMSE-DFE-U]\right)\right)=\Isl
\end{equation}
and $\Iofdm\leq\Isl$ is therefore valid. Choosing $\underline{\gamma}_{0}$
and $\bar{\gamma}_{0}$ according to Definition \ref{def:zeta0},
if either condition 2 or conditions 3 hold, then $\Ilog$ is a concave
function for every value of $|H(\theta)|^{2}$, and we may therefore
exchange $\Ihatlog$ with $\Ilog$ in (\ref{eq:Iofdm jensen application}),
yielding $\Iofdm\leq\Isl$ once more.
\end{IEEEproof}

\section{\label{sec:mmse bounds}Interlude --- MMSE bounds for PAM signaling}

In order to prove Theorems \ref{thm:low_order_QAM} and \ref{thm:MQAM_high_snr},
we first need to derive a tight high-SNR characterization of the MMSE
function and its derivative, for PAM inputs and Gaussian noise. In
this section, we first present a novel ``pointwise'' bound on the
MMSE of PAM signaling conditioned on the channel output. The bound
is then applied to derive inequalities that characterize the MMSE
and its derivative in the high-SNR regime, as required. Finally, some
bounds on the MMSE and its derivative are presented for the special
case of BPSK input. Besides their use in proving Theorems \ref{thm:low_order_QAM}
and \ref{thm:MQAM_high_snr}, the results presented in this section
may be of general interest.

\subsection{A pointwise MMSE inequality}

Let $X$ be a real-valued random variable and define%
\footnote{We have chosen here to let $\gamma$ scale $N$ and not $X$, as opposed
to the convention in the I-MMSE literature. This definition ensures
that $Y_{\gamma}$ and $X$ are on the same scale, which simplifies
many of the derivations that follow. We have also set the noise variance
to be $1/2$ when $\gamma=1$, in the purpose of making these results
easily applicable in the complex setting of the rest of the paper.%
} $Y_{\gamma}=X+\frac{1}{\sqrt{\gamma}}N$ with $N\sim\mathcal{N}(0,\frac{1}{2})$
and independent of $X$. $Y_{\gamma}$ is the Gaussian-noise contaminated
version of $X$, at SNR $\gamma$. Let
\begin{equation}
\phi_{X}\left(y;\gamma\right)=\Ebrack{\left|X-{\Ebrack{X|Y_{\gamma}=y}{}}\right|^{2}\,|\: Y_{\gamma}=y}X
\end{equation}
denote the ``point-wise'' conditional variance of $X$ given a noisy
channel observation. Clearly, $\mathrm{mmse}_{X}(\gamma)=\E[\phi_{X}\left(Y_{\gamma};\gamma\right)][Y_{\gamma}]$.
Moreover, exploration of Information-Estimation relations \cite{guo2011estimation}
has revealed that\emph{,}
\begin{equation}
\mathrm{mmse}_{X}'(\gamma)=-2\E[\phi_{X}\left(Y_{\gamma};\gamma\right)][Y_{\gamma}]
\end{equation}
Hence, intimate understanding of $\phi_{X}\left(y;\gamma\right)$
is expected to provide insights on both the MMSE function and its
derivative. For the case of PAM input distribution, this understanding
is presented in the form of the following,
\begin{thm}
\label{thm:pointwise PAM bounds}Let $\mathcal{X}=\left\{ x_{1},...,x_{M}\right\} $
be the alphabet of an $M$-ary PAM constellation such that $x_{m+1}-x_{m}=d$
for every $1\leq m<M$ . Let $X$ be uniformly distributed in $\mathcal{X}$.
Fix $y\in\mathbb{R}$ and choose $1\leq J<M$ such that $x_{J},x_{J+1}$
are the nearest values to $y$ in $\mathcal{X}$. Let $B_{J}$ be
uniformly distributed on $\left\{ x_{J},x_{J+1}\right\} $. Then,
\begin{equation}
0\leq\phi_{X}\left(y;\gamma\right)-\phi_{B_{J}}\left(y;\gamma\right)\leq\left(\frac{d}{2}\right)^{2}\bar{D}\left(\left(\frac{d}{2}\right)^{2}\gamma\right)\label{eq:pointwise PAM bounds}
\end{equation}
with
\begin{equation}
\bar{D}\left(\gamma\right)=4\sum_{k=1}^{\infty}\left(k+1\right)^{2}e^{-4\gamma k^{2}}\leq\frac{16e^{-4\gamma}}{\left(1-e^{-4\gamma}\right)^{3}}
\end{equation}
\end{thm}
\begin{IEEEproof}
Appendix \ref{sec:PAM pointwise}.
\end{IEEEproof}
Note that 
\begin{equation}
\phi_{B_{J}}\left(y;\gamma\right)=\left(\frac{d}{2}\right)^{2}\phi_{\mathrm{BPSK}}\left(\left(\frac{d}{2}\right)^{-1}\left[y-\frac{x_{J}+x_{J+1}}{2}\right];\left(\frac{d}{2}\right)^{2}\gamma\right)
\end{equation}
where $\phi_{\mathrm{BPSK}}\left(y;\gamma\right)=1-\tanh^{2}\left(2\gamma y\right)$
is the conditional variance function for BPSK input. 

Put in words, Theorem \ref{thm:pointwise PAM bounds} means that
for PAM input and given channel output, the expected value of the
MMSE is lower-bounded by the expected MMSE given the same channel
output and assuming an input equally distributed on the two PAM symbols
nearest to it.

Figure \ref{fig:mpam_pointwise_bounds_demo} illustrates the bounds
in (\ref{eq:pointwise PAM bounds}) for 4-PAM input. As the figure
indicates, the lower bound is reasonably tight for $\left(d/2\right)^{2}\gamma$
as low as 0 dB, and both bounds are very tight for $\left(d/2\right)^{2}\gamma=3$
dB and above.

\begin{center}
\begin{figure}
\centering{}\includegraphics[width=12cm]{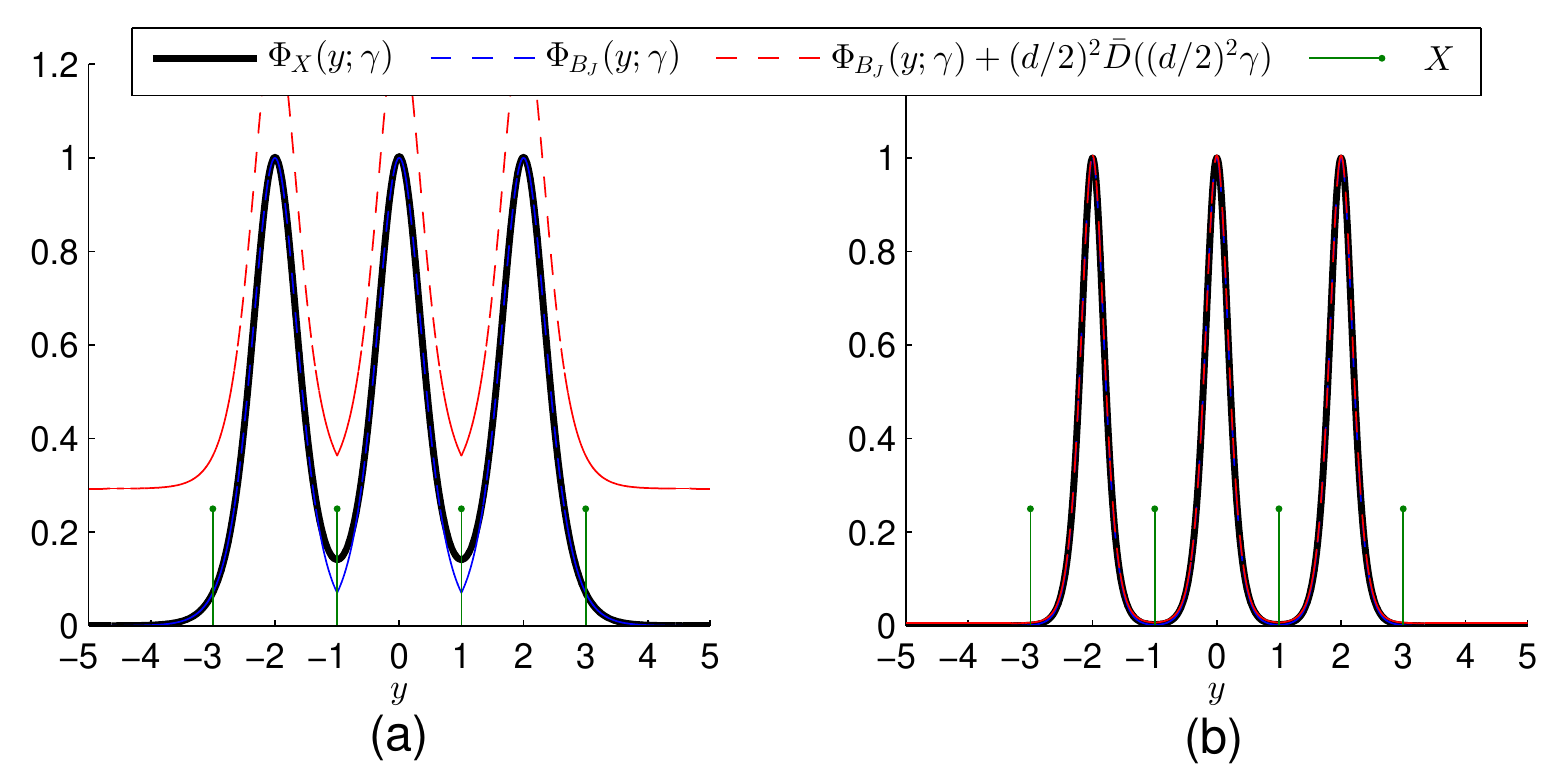}\protect\caption{\label{fig:mpam_pointwise_bounds_demo}Illustration of Theorem \ref{thm:pointwise PAM bounds}
for 4-PAM input with $d=2$ and two different SNR's: (a) $\left(d/2\right)^{2}\gamma=1$
and (b) $\left(d/2\right)^{2}\gamma=2$.}
\end{figure}

\par\end{center}

\subsection{\label{sub:mpam high SNR}High-SNR characterization of the MMSE}

Let $\mathrm{mmse}_{d,M\textrm{-PAM}}\left(\gamma\right)$ stand for
the $\mathrm{mmse}_{X}\left(\gamma\right)$, with $X$ uniformly distributed
on an $M$-PAM constellation with distance $d$ between adjacent points.
We show that 
\begin{equation}
\mathrm{mmse}_{d,M\textrm{-PAM}}\left(\gamma\right)\approx2\frac{M-1}{M}\left(\frac{d}{2}\right)^{2}\mathrm{mmse}_{\mathrm{BPSK}}\left(\left(\frac{d}{2}\right)^{2}\gamma\right)\label{eq:PAM mmse characterization approx}
\end{equation}
in the sense that the difference between the terms tends to zero with
a faster exponential rate than $\mathrm{mmse}_{M-PAM}\left(\gamma\right)$,
where $\mathrm{mmse}_{\mathrm{BPSK}}\left(\gamma\right)\equiv\mathrm{mmse}_{2,2-PAM}\left(\gamma\right)$.
Moreover, we use similar techniques in order to show that this approximation
also applies to the derivative of the MMSE with respect to $\gamma$,
\emph{i.e.,}
\begin{equation}
\mathrm{mmse}_{d,M\textrm{-PAM}}'\left(\gamma\right)\approx2\frac{M-1}{M}\left(\frac{d}{2}\right)^{4}\mathrm{mmse}_{\mathrm{BPSK}}'\left(\left(\frac{d}{2}\right)^{2}\gamma\right)\label{eq:PAM dmmse characterization approx}
\end{equation}
It should be noted that (\ref{eq:PAM mmse characterization approx})
can be seen as a special case of the high-SNR MMSE characterization
for general discrete inputs that was recently presented in \cite{alvarado2013high}.
However, the following analysis provides two important advantages.
First, it enables us to also establish (\ref{eq:PAM dmmse characterization approx})
--- a characterization of the derivative of the MMSE. Second, it allows
for explicit and tight bounds on the difference between the exact
$M$-PAM quantities and their BPSK approximations. Both of these features
are crucial for establishing the bound in Theorem \ref{thm:MQAM_high_snr}.

Letting
\begin{equation}
Q\left(x\right)\triangleq\int_{x}^{\infty}\frac{1}{\sqrt{2\pi}}e^{-t^{2}/2}dt\label{eq:Qfunc_def}
\end{equation}
denote the standard error function, the result (\ref{eq:PAM mmse characterization approx})
is stated formally as follows,
\begin{thm}
\label{thm:PAM bounds}The following bounds hold for every $M\geq2$,
$d\geq0$ and $\gamma\geq0$:\textup{
\begin{equation}
\mathrm{mmse}_{d,M\textrm{-PAM}}\left(\gamma\right)\leq2\frac{M-1}{M}\left(\frac{d}{2}\right)^{2}\left[\mathrm{mmse}_{\mathrm{BPSK}}\left(\left(\frac{d}{2}\right)^{2}\gamma\right)+\bar{B}\left(\left(\frac{d}{2}\right)^{2}\gamma\right)\right]
\end{equation}
}with
\begin{equation}
\bar{B}\left(\gamma\right)=16Q\left(\sqrt{8\gamma}\right)+4\sum_{k=2}^{\infty}\left(2k+1\right)Q\left(k\sqrt{8\gamma}\right)\label{eq:Bup}
\end{equation}
and\textup{
\begin{equation}
\mathrm{mmse}_{d,M\textrm{-PAM}}\left(\gamma\right)\geq2\frac{M-1}{M}\left(\frac{d}{2}\right)^{2}\left[\mathrm{mmse}_{\mathrm{BPSK}}\left(\left(\frac{d}{2}\right)^{2}\gamma\right)-\underline{B}\left(\left(\frac{d}{2}\right)^{2}\gamma\right)\right]
\end{equation}
}with
\begin{equation}
\underline{B}\left(\gamma\right)=4Q\left(\sqrt{8\gamma}\right)\leq\frac{1}{\sqrt{\pi\gamma}}e^{-4\gamma}
\end{equation}
\end{thm}
\begin{IEEEproof}
Appendices \ref{sub:PAM mmse low proof} and \ref{sub:PAM mmse up proof}.
\end{IEEEproof}
Similarly, (\ref{eq:PAM dmmse characterization approx}) has the following
formal form,
\begin{thm}
\label{thm:PAM d bounds}The following bounds hold for every $M\geq2$,
$d\geq0$ and $\gamma\geq0$:\textup{
\begin{equation}
\mathrm{mmse}_{d,M\textrm{-PAM}}'\left(\gamma\right)\leq2\frac{M-1}{M}\left(\frac{d}{2}\right)^{4}\left[\mathrm{mmse}_{\mathrm{BPSK}}'\left(\left(\frac{d}{2}\right)^{2}\gamma\right)+\bar{C}\left(\left(\frac{d}{2}\right)^{2}\gamma\right)\right]
\end{equation}
}with
\begin{equation}
\bar{C}\left(\gamma\right)=32e^{8\gamma}Q\left(\sqrt{32\gamma}\right)\leq\frac{4}{\sqrt{\pi\gamma}}e^{-8\gamma}\label{eq:Cup}
\end{equation}
and\textup{
\begin{equation}
\mathrm{mmse}_{d,M\textrm{-PAM}}'\left(\gamma\right)\geq2\frac{M-1}{M}\left(\frac{d}{2}\right)^{4}\left[\mathrm{mmse}_{\mathrm{BPSK}}'\left(\left(\frac{d}{2}\right)^{2}\gamma\right)-\underline{C}\left(\left(\frac{d}{2}\right)^{2}\gamma\right)\right]
\end{equation}
}with
\begin{equation}
\underline{C}\left(\gamma\right)=2\left[4\left(\sum_{k=1}^{\infty}\left(k+1\right)^{2}e^{-4\gamma k^{2}}\right)\left(8\sum_{k=1}^{\infty}\left(k+1\right)^{2}e^{-4\gamma k^{2}}+1\right)+Q\left(\sqrt{8\gamma}\right)\right]
\end{equation}
\end{thm}
\begin{IEEEproof}
Appendices \ref{sub:PAM dmmse low proof} and \ref{sub:PAM dmmse up proof}.
\end{IEEEproof}
Figure \ref{fig:mpam_mmse_dmmse_bounds_demo} illustrates the high
SNR behavior of $\mathrm{mmse}_{d,M\textrm{-PAM}}\left(\gamma\right)/\left[2\frac{M-1}{M}\left(\frac{d}{2}\right)^{2}\right]$
and $\mathrm{mmse}_{d,M\textrm{-PAM}}'\left(\gamma\right)/\left[2\frac{M-1}{M}\left(\frac{d}{2}\right)^{4}\right]$
for different values of $M$. It is seen that the above bounds become
tight at $\left(d/2\right)^{2}\gamma$ values of around 3 dB.

\begin{center}
\begin{figure}
\centering{}\includegraphics[width=16cm]{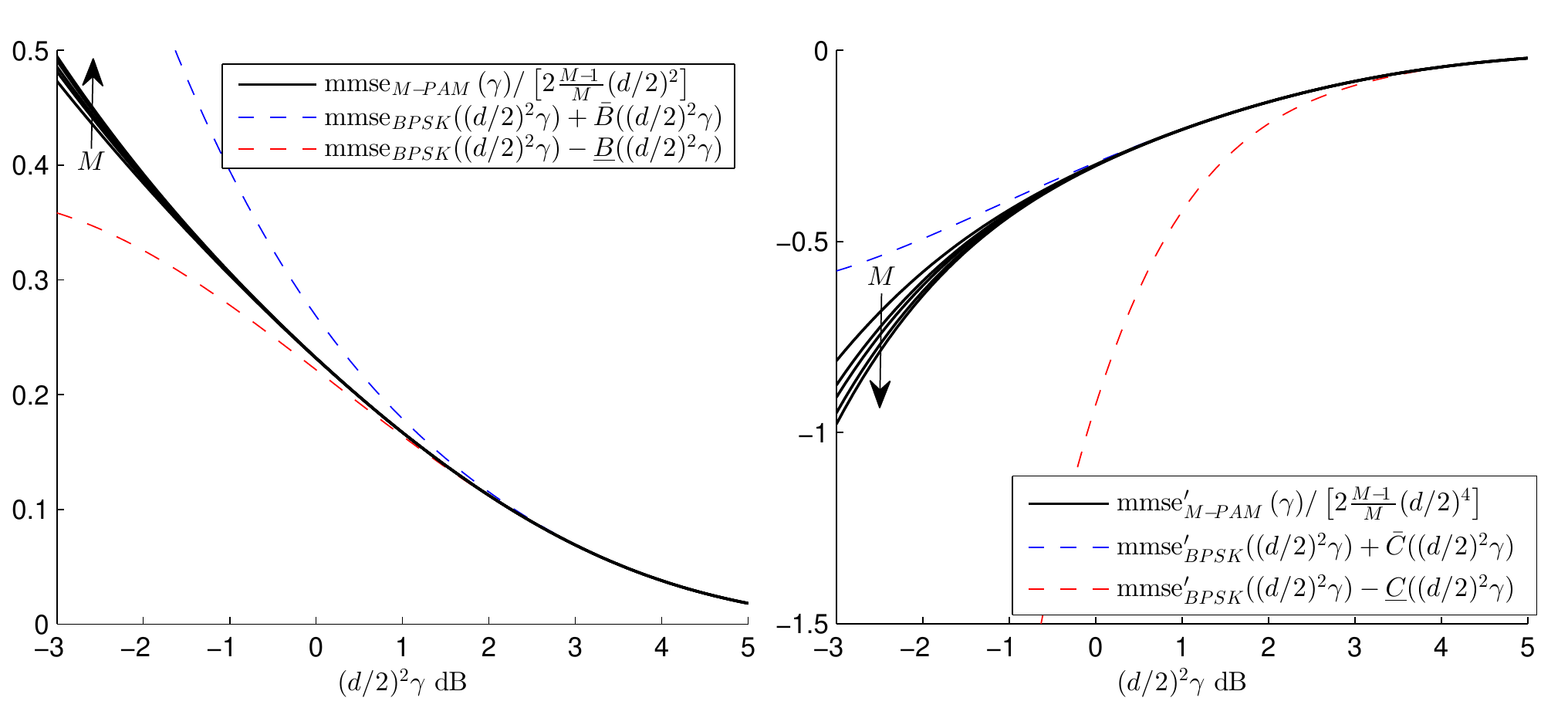}\protect\caption{\label{fig:mpam_mmse_dmmse_bounds_demo}Illustration of Theorems \ref{thm:PAM bounds}
and \ref{thm:PAM d bounds} for $M$ values of 3, 4, 5, 8 and 16.}
\end{figure}

\par\end{center}

\subsection{Bounds for BPSK inputs}

We present some upper and lower bound on the MMSE function and its
derivative for the case of BPSK inputs. These bounds will be of use
in proving analytically that $\Delta_{x}=0$ for BPSK and QPSK inputs,
as claimed in Theorem \ref{thm:low_order_QAM}.
\begin{thm}
\label{thm:BPSK bounds}The following bounds on $\mathrm{mmse}_{\mathrm{BPSK}}\left(\gamma\right)$
hold
\begin{gather}
\left(1-\frac{1}{2\gamma}\frac{\pi^{2}}{8}\right)\frac{\sqrt{\pi}}{2}\frac{1}{\sqrt{\gamma}}e^{-\gamma}\leq\mathrm{mmse}_{\mathrm{BPSK}}\left(\gamma\right)\leq\frac{\sqrt{\pi}}{2}\frac{1}{\sqrt{\gamma}}e^{-\gamma}\label{eq:BPSK MMSE asymptotic bound}\\
\frac{e^{-\gamma}}{\sqrt{1+2\gamma}}\leq\mathrm{mmse}_{\mathrm{BPSK}}\left(\gamma\right)\leq e^{-\gamma}\label{eq:BPSK MMSE lowsnr bound}
\end{gather}

similarly, the following bounds on\textup{ $\mathrm{mmse}_{\mathrm{BPSK}}'\left(\gamma\right)$}
hold
\begin{gather}
\left(1-\frac{1}{2\gamma}\left(\frac{\pi^{2}}{8}-1\right)\right)\frac{\sqrt{\pi}}{2}\frac{1}{\sqrt{\gamma}}e^{-\gamma}\leq-\mathrm{mmse}_{\mathrm{BPSK}}'\left(\gamma\right)\leq\frac{\sqrt{\pi}}{2}\frac{1}{\sqrt{\gamma}}e^{-\gamma}\\
\frac{2e^{-\gamma}}{\sqrt{1+6\gamma}}\leq-\mathrm{mmse}_{\mathrm{BPSK}}'\left(\gamma\right)\leq2e^{-\gamma}
\end{gather}
\end{thm}
\begin{IEEEproof}
Appendix \ref{sec:BPSK bounds proof}.
\end{IEEEproof}
We note that the bounds in (\ref{eq:BPSK MMSE asymptotic bound})
are composed of the first two terms in the asymptotic high SNR expansion
of $\mathrm{mmse}_{\mathrm{BPSK}}\left(\gamma\right)$ derived in
\cite{lozano2006optimum}. Our contribution in this case is the proof
that these approximations are upper and lower bounds. Our proof of
the bounds also allows for a simpler derivation of the series expansion
than the one in \cite{lozano2006optimum}, as well as extension of
these bounds to $\mathrm{mmse}_{\mathrm{BPSK}}'\left(\gamma\right)$.
It is worth noting that while not asymptotically tight as $\gamma\to\infty$,
the lower bound $e^{-\gamma}/\sqrt{1+2\gamma}$ is a good approximation
for $\mathrm{mmse}_{\mathrm{BPSK}}\left(\gamma\right)$ for all values
of $\gamma$, with a maximum slackness of less than $0.022$.

\section{\label{sec:pam qam proofs}Results for PAM and square QAM inputs}

In this section we utilize the results from Sections \ref{sec:concavity results}
and \ref{sec:mmse bounds} in order to prove Theorems \ref{thm:MQAM_high_snr},
\ref{thm:low_order_QAM} and \ref{thm:uniform_sl_ofdm}.

\subsection{Proof of Theorem \ref{thm:low_order_QAM}}

We recall the definition of $\Ilog$ (\ref{eq:ilog def}) and prove
the following,
\begin{lem}
\textup{\label{lem:Ilog cocave BPSK QPSK}$\Ilog[\textrm{BPSK}]\left(\zeta\right)$
and $\Ilog[\textrm{QPSK}]\left(\zeta\right)$ are concave in $\zeta$.}\end{lem}
\begin{IEEEproof}
Let $\gamma=e^{\zeta}-1$. As shown in (\ref{eq:ddIlog}), $\Ilog''(\zeta)=\left(1+\gamma\right)\left[\mathrm{\mathrm{mmse}}_{x}(\gamma)+\left(1+\gamma\right)\mathrm{\mathrm{mmse}}_{x}'(\gamma)\right]$.
By Theorem \ref{thm:BPSK bounds},
\begin{equation}
\mathrm{mmse}_{\mathrm{BPSK}}(\gamma)+\left(1+\gamma\right)\mathrm{mmse}_{\mathrm{BPSK}}'(\gamma)\leq e^{-\gamma}\left[1-\frac{2\left(1+\gamma\right)}{\sqrt{1+6\gamma}}\right]<0
\end{equation}
and so $\Ilog[\mathrm{BPSK}]''\left(\zeta\right)<0$ for every $\zeta$,
meaning $\Ilog[\mathrm{BPSK}]$ is concave in $\zeta$. Since,
\begin{equation}
\mathrm{\mathrm{mmse}}_{\mathrm{QPSK}}(\gamma)=\mathrm{mmse}_{\mathrm{BPSK}}\left(\frac{\gamma}{2}\right)
\end{equation}
we have,
\begin{equation}
\mathrm{\mathrm{mmse}}_{\mathrm{QPSK}}(\gamma)+\left(1+\gamma\right)\mathrm{\mathrm{mmse}}_{\mathrm{QPSK}}'(\gamma)\leq e^{-\gamma/2}\left[1-\frac{1+\gamma}{\sqrt{1+3\gamma}}\right]
\end{equation}
and it is easily verified that $\left(1+\gamma\right)/\sqrt{1+3\gamma}\geq1$
for every $\gamma\geq1$. For lower SNR's, we use the Gaussian upper
bound $\mathrm{\mathrm{mmse}}_{\mathrm{QPSK}}(\gamma)\leq\left(1+\gamma\right)^{-1}$
as well as $e^{-\gamma/2}\geq1-\gamma/2$ to show that
\begin{equation}
\mathrm{\mathrm{mmse}}_{\mathrm{QPSK}}(\gamma)+\left(1+\gamma\right)\mathrm{\mathrm{mmse}}_{\mathrm{QPSK}}'(\gamma)\leq\frac{1}{1+\gamma}\left[1-\frac{\left(1+\gamma\right)^{2}\left(1-\gamma/2\right)}{\sqrt{1+3\gamma}}\right]
\end{equation}
and once more it is easily verified that $\left(1+\gamma\right)^{2}\left(1-\gamma/2\right)/\sqrt{1+3\gamma}\geq1$
for every $\gamma\leq1$. We thus conclude that $\Ilog[\mathrm{QPSK}]''\left(\zeta\right)<0$
for every $\zeta$, and so $I_{\mathrm{QPSK}}^{\log}$ is concave
in $\zeta$.
\end{IEEEproof}
Clearly, if $\Ilog[x]$ is concave then $\Ilog[x]\left(\zeta\right)=\Ihatlog[x]\left(\zeta\right)$
for every $\zeta$ and by its definition (\ref{eq:dmax def}), $\Delta_{x}=0$.
Thus, Lemma \ref{lem:Ilog cocave BPSK QPSK} proves Theorem \ref{thm:low_order_QAM}.
While numeric investigation shows that $\Ilog[4\textrm{-PAM}]$, $\Ilog[8\textrm{-PSK}]$,
$\Ilog[16\textrm{-QAM}]$ and $\Ilog[32\textrm{-QAM}]$ are also concave,
no analytical proof of concavity has been found for these cases.

\subsection{Proof of Theorem \ref{thm:MQAM_high_snr}}
\begin{IEEEproof}
Let $\gamma=e^{\zeta}-1$ and let $\rho=\left(d_{\min}^{\textrm{PAM}}/2\right)^{2}\gamma$
where $d_{\min}^{\textrm{PAM}}=\sqrt{\frac{12}{M^{2}-1}}$ is the
minimum distance between symbols of a unit-power $M$-PAM input. Using
(\ref{eq:ddIlog}) and bounds provided in Theorems \ref{thm:PAM bounds}
and \ref{thm:PAM d bounds}, we find that
\begin{eqnarray}
\Ilog[M\textrm{-PAM}]''(\zeta) & = & \left(1+\gamma\right)\left[\mathrm{\mathrm{mmse}}_{M\textrm{-PAM}}(\gamma)+\left(1+\gamma\right)\mathrm{\mathrm{mmse}}_{M\textrm{-PAM}}'(\gamma)\right]\nonumber \\
 & \leq & K\left[\rho\mathrm{mmse}_{\mathrm{BPSK}}'\left(\rho\right)+\mathrm{mmse}_{\mathrm{BPSK}}\left(\rho\right)+\bar{B}\left(\rho\right)+\rho\bar{C}\left(\rho\right)\right]\label{eq:M-PAM dd}
\end{eqnarray}
with $\bar{B}\left(\rho\right)$ and $\bar{C}\left(\rho\right)$ given
in (\ref{eq:Bup}) and (\ref{eq:Cup}), respectively, and $K=2\left(1+\gamma\right)\frac{M-1}{M}\left(d_{\min}^{\textrm{PAM}}/2\right)^{2}$.
Evaluating numerically the expression in square brackets, it is found
that $\Ilog[M\textrm{-PAM}]''(\zeta)<0$ for $\rho\geq1$, see Figure
\ref{fig:mpam concave bound}. Consequently, $\left(d_{\min}^{\textrm{PAM}}/2\right)^{2}\bar{\gamma}_{0}\leq1$
for $M$-PAM inputs, as required.

\begin{figure}
\begin{centering}
\includegraphics[width=12cm]{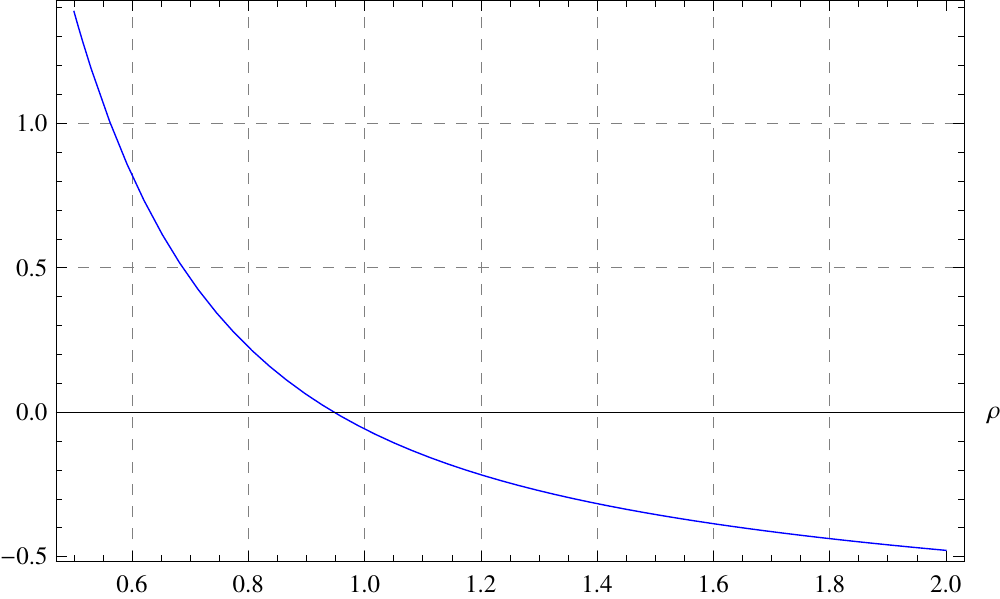}
\par\end{centering}

\protect\caption{\label{fig:mpam concave bound}Evaluation of $\left(\sqrt{\rho}e^{-\rho}\right)^{-1}\left[\rho\mathrm{mmse}_{\mathrm{BPSK}}'\left(\rho\right)+\mathrm{mmse}_{\mathrm{BPSK}}\left(\rho\right)+\bar{B}\left(\rho\right)+\rho\bar{C}\left(\rho\right)\right]$.}
\end{figure}

An $M^{2}$-ary square QAM constellation is composed of two $M$-ary
PAM constellations in quadrature. Letting $d_{\min}^{\textrm{QAM}}=\sqrt{\frac{6}{M^{2}-1}}$
denote the minimum distance between symbols of a unit-power $M^{2}$-QAM
input, we have
\begin{equation}
\mathrm{mmse}_{M^{2}\textrm{-QAM}}\left(\gamma\right)=2\mathrm{mmse}_{d_{\min}^{\textrm{QAM}},M\textrm{-PAM}}\left(\gamma\right)
\end{equation}
and
\begin{equation}
\mathrm{mmse}_{M^{2}\textrm{-QAM}}'\left(\gamma\right)=2\mathrm{mmse}_{d_{\min}^{\textrm{QAM}},M\textrm{-PAM}}'\left(\gamma\right)
\end{equation}
where $\mathrm{mmse}_{d,M\textrm{-PAM}}\left(\gamma\right)$ denotes
the MMSE for an $M$-PAM input with distance $d$ between adjacent
symbols and complex-valued additive Gaussian noise with power $1/\gamma$,
as defined in subsection \ref{sub:mpam high SNR}. Applying Theorems
\ref{thm:PAM bounds} and \ref{thm:PAM d bounds} to the above equations,
we find that $\Ilog[M^{2}\textrm{-QAM}]''(\zeta)$ is also bounded
from above by (\ref{eq:M-PAM dd}), with $\rho=\left(d_{\min}^{\textrm{QAM}}/2\right)^{2}\gamma$.
Therefore, we have $\left(d_{\min}^{\textrm{QAM}}/2\right)^{2}\bar{\gamma}_{0}\leq1$
for $M^{2}$-QAM inputs, and the proof is complete.
\end{IEEEproof}
Examining Figure \ref{fig:mpam concave bound} more closely, we find
that the term (\ref{eq:M-PAM dd}) becomes negative for $\rho$ values
around 0.95, and the bound on $\left(d_{\min}/2\right)^{2}\bar{\gamma}_{0}$
might be slightly tightened accordingly. However, as remarked on Table
\ref{tab:log concave summary}, numerical evaluation of $\bar{\gamma}_{0}$
for square QAM inputs indicate that $\left(d_{\min}/2\right)^{2}\bar{\gamma}_{0}$
is closer to 0.5. Therefore, reducing the bound by $0.05$ does not
significantly improve its tightness.

\subsection{Proof of Corollary \ref{cor:MQAM SER implication}}
\begin{IEEEproof}
In the large block size limit, the input SNR at the $k$'th OFDM subcarrier
is given by $\gamma_{k}=\left|H\left(\theta_{k}\right)\right|^{2}$
where $\theta_{k}=2\pi k/N$ is the subcarrier frequency and $k$
is its index spanning from 0 to $N-1$. Consider a unit-power square
$M^{2}$-QAM input and Gaussian noise at SNR $\gamma$, and set $q=Q\left(\sqrt{\left(d_{\min}/2\right)^{2}\gamma}\right)$,
with the error function $Q\left(\cdot\right)$ as given in (\ref{eq:Qfunc_def}).
For $M$-PAM input with spacing $d_{\min}$, the probability of a
symbol error is $2q$ for the $M-2$ inner constellation points, and
$q$ for the 2 outer points, \emph{i.e.}
\begin{equation}
P_{err}^{M\text{-PAM}}=\frac{M-2}{M}2q+\frac{2}{M}q=2\frac{M-1}{M}q
\end{equation}

For $M^{2}$-QAM input, a symbol error event is the union of two independent
error events along the in-phase and quadrature directions, each being
$M$-ary PAM error events. Therefore,
\begin{equation}
P_{err}^{M^{2}\text{-QAM}}=2P_{err}^{M\text{-PAM}}-\left(P_{err}^{M\text{-PAM}}\right)^{2}=4\frac{M-1}{M}q\left(1-\frac{M-1}{M}q\right)
\end{equation}
It can thus be seen that for $M\geq16$, $P_{err}^{M^{2}\text{-QAM}}<50\%$
implies $\left(d_{\min}/2\right)^{2}\gamma>1$ and hence $\gamma>\bar{\gamma}_{0}$.
Assuming this holds for all subcarriers and assuming large enough
OFDM block size, we find that $\left|H\left(\theta\right)\right|^{2}>\bar{\gamma}_{0}$
for all values of $\theta$, thus satisfying condition 3 of Theorem
\ref{thm:Iofdm_sl_bound} and proving the corollary.
\end{IEEEproof}

\subsection{\label{sub:uniform analysis}Analysis of uniform input}

\begin{figure}
\begin{centering}
\includegraphics[width=16cm]{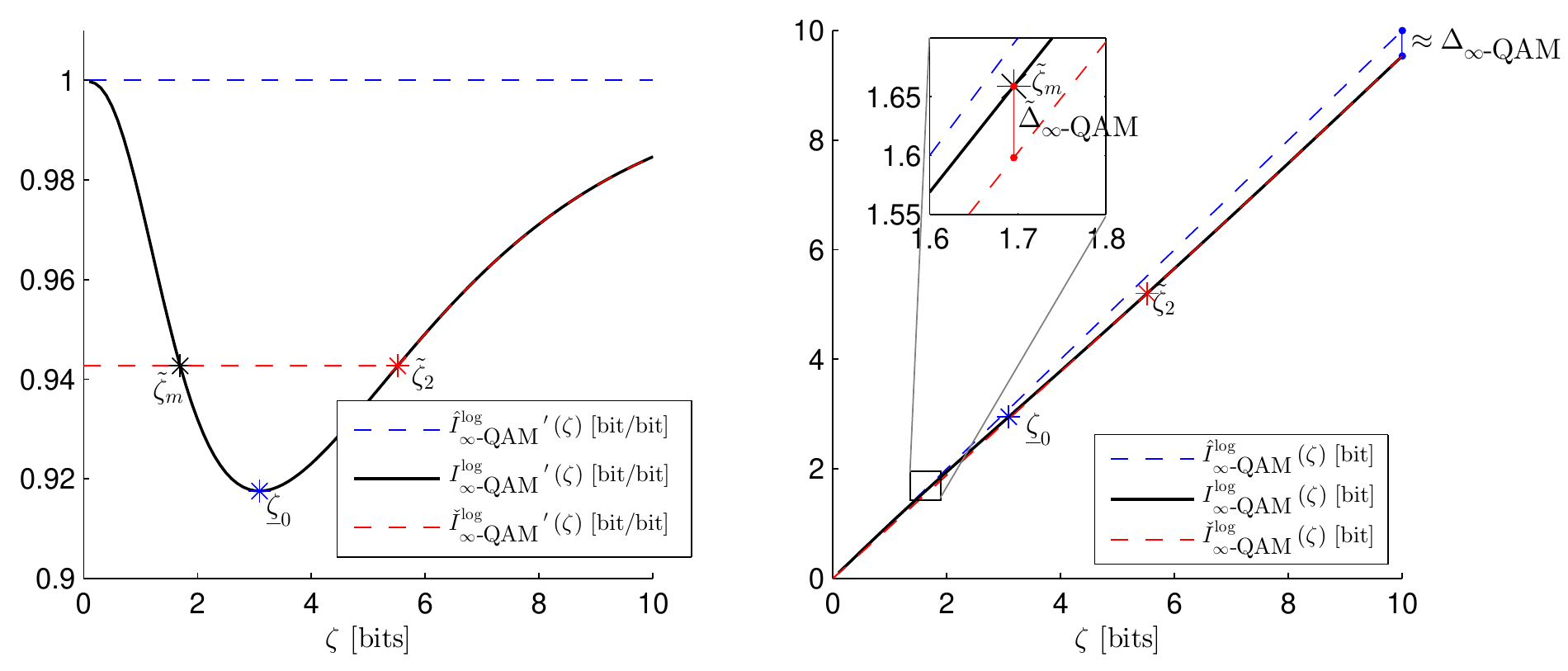}
\par\end{centering}

\protect\caption{\label{fig:unif-ilog}$\protect\Ilog$  with its concave and convex
envelopes (right), and their derivatives with respect to $\zeta$
(left), with $\underline{\zeta}_{0},\tilde{\zeta}_{2}$, $\Delta{}_{x}$
and $\tilde{\Delta}_{x}$ highlighted, for $\infty$-QAM input.}
\end{figure}

We consider a unit-power input distributed uniformly on the square
$\left[-\sqrt{\frac{3}{2}},\sqrt{\frac{3}{2}}\right]\times\left[-\sqrt{\frac{3}{2}},\sqrt{\frac{3}{2}}\right]$,
also referred to as $\infty$-QAM input. As usual we let $I_{\infty\textrm{-QAM }}\left(\gamma\right)$
denote the mutual information between such input and its complex Gaussian
noise corrupted version at SNR $\gamma$, and we let $\Ilog[\infty\textrm{-QAM}]\left(\zeta\right)=I_{\infty\textrm{-QAM }}\left(e^{\zeta}-1\right)$
be the mutual information with respect to log SNR. Figure \ref{fig:unif-ilog}
illustrates $\Ilog[\infty\text{-QAM}]$ and its derivative, as well
as the quantities to be defined in the following paragraphs.

For high SNR, it well known \cite{smith1971information} that
\begin{equation}
I_{\infty\textrm{-QAM }}\left(\gamma\right)\approx\log\left(\frac{6}{\pi e}\gamma\right)
\end{equation}
and therefore
\begin{equation}
\lim_{\zeta\to\infty}\left(\zeta-\Ilog[\infty\textrm{-QAM}]\left(\zeta\right)\right)=\log\left(\frac{\pi e}{6}\right)\approx0.509\text{ [bit]}\label{eq:unif_Ilog_lim}
\end{equation}
as $\zeta$ is also the mutual information for Gaussian input at log-SNR
$\zeta$, the above limit represent the loss of using uniform input
rather than Gaussian input at high SNR, and is commonly referred to
as the shaping gain. 

Proposition \ref{prop:concavity-low-SNR} applies to the case of uniform
input, and so we know there must a constant $\underline{\zeta}_{0}$
such $\Ilog[\infty\textrm{-QAM}]\left(\zeta\right)$ is concave for
every $\zeta\leq\underline{\zeta}_{0}$. However, the high-SNR behavior
of $\Ilog[\infty\textrm{-QAM}]\left(\zeta\right)$ is quite different
than the finite-alphabet case, and is characterized as follows,
\begin{prop}
\textup{\label{prop:uniform-convex}}\textup{$\Ilog[\infty\textrm{-QAM}]\left(\zeta\right)$}
is concave for every $\zeta\leq\zeta_{0}$ and convex for every $\zeta\geq\underline{\zeta}_{0}$,
where $\underline{\zeta}_{0}\approx3.09$ bits.\end{prop}
\begin{IEEEproof}
Appendix \ref{sec:uniform-convex-proof}.
\end{IEEEproof}
Even though it becomes convex in high SNR's, $\Ilog[\infty\textrm{-QAM}]$
still has a concave envelope. Additionally, it is of interest to study
the \emph{convex envelope} of $\Ilog[\infty\textrm{-QAM}]$, \emph{i.e.
}the maximum convex function that lower bounds $\Ilog[\infty\textrm{-QAM}]$,
which we denote by $\Ichecklog[\infty\text{-QAM}]$. Moreover, we
are interested in the concave envelope of $\Ilog[\infty\textrm{-QAM}]$
when limited to the interval $\left[0,\bar{\zeta}\right]$, \emph{i.e.
}the minimum concave function that upper bounds $\Ilog[\infty\textrm{-QAM}]$
for every $\zeta\in\left[0,\bar{\zeta}\right]$. We will denote this
function by $\hat{I}_{\infty\text{-QAM}}^{\log;\left[0,\bar{\zeta}\right]}\left(\zeta\right)$.
These concave and convex envelopes are characterized as follows,
\begin{prop}
\label{prop:unif-envelopes}The concave envelope of \textup{$\Ilog[\infty\textrm{-QAM}]$}
is given by\textup{
\begin{equation}
\Ihatlog[\infty\textrm{-QAM }]\left(\zeta\right)=\lim_{\bar{\zeta}\to\infty}\hat{I}_{\infty\text{-QAM}}^{\log;\left[0,\bar{\zeta}\right]}\left(\zeta\right)=\zeta
\end{equation}
}and satisfies\textup{
\begin{equation}
\Delta_{\infty\text{-QAM}}=\sup_{\zeta}\left(\Ihatlog[\infty\textrm{-QAM }]\left(\zeta\right)-\Ilog[\infty\textrm{-QAM}]\left(\zeta\right)\right)=\log\left(\frac{\pi e}{6}\right)
\end{equation}
}Limited to the interval $\left[0,\bar{\zeta}\right]$, where $\bar{\zeta}>\underline{\zeta}_{0}$,
the concave envelope of \textup{$\Ilog[\infty\textrm{-QAM}]$} is
given by\textup{\textup{
\begin{equation}
\hat{I}_{\infty\text{-QAM}}^{\log;\left[0,\bar{\zeta}\right]}\left(\zeta\right)=\begin{cases}
\Ilog[\infty\text{-QAM}](\underline{\zeta}_{1})+(\zeta-\underline{\zeta}_{1})\dIlog[\infty\text{-QAM}](\underline{\zeta}_{1}) & \zeta\geq\underline{\zeta}_{1}\\
\Ilog[\infty\text{-QAM}]\left(\zeta\right) & \text{otherwise}
\end{cases}\label{eq:unif_concave_envelope_interval}
\end{equation}
}}where $\underline{\zeta}_{1}<\underline{\zeta}_{0}$ depends on
$\bar{\zeta}$ and is determined by the condition \textup{$\hat{I}_{\infty\text{-QAM}}^{\log;\left[0,\bar{\zeta}\right]}(\bar{\zeta})=\Ilog[\infty\text{-QAM}](\bar{\zeta})$}.
The function \textup{$\bar{\Delta}_{\infty\text{-QAM}}\left(\bar{\gamma}\right)$}
is given by,\textup{
\begin{equation}
\bar{\Delta}_{\infty\text{-QAM}}\left(e^{\bar{\zeta}}-1\right)\triangleq\sup_{\zeta\leq\bar{\zeta}}\left(\hat{I}_{\infty\text{-QAM}}^{\log;\left[0,\bar{\zeta}\right]}\left(\zeta\right)-\Ilog[\infty\textrm{-QAM}]\left(\zeta\right)\right)=\hat{I}_{\infty\text{-QAM}}^{\log;\left[0,\bar{\zeta}\right]}(\zeta_{m})-\Ilog[\infty\textrm{-QAM}](\zeta_{m})
\end{equation}
}where $\zeta_{m}>\underline{\zeta}_{1}$ depends on $\bar{\zeta}$
and satisfies \textup{$\dIlog[\infty\text{-QAM}](\zeta_{m})=\dIlog[\infty\text{-QAM}](\underline{\zeta}_{1})$}.
For $\bar{\zeta}\leq\underline{\zeta}_{0}$, \textup{$\Ilog[\infty\textrm{-QAM}]$}
is concave on the interval $\left[0,\bar{\zeta}\right]$, so that
\textup{$\hat{I}_{\infty\text{-QAM}}^{\log;\left[0,\bar{\zeta}\right]}\left(\zeta\right)=\Ilog[\infty\text{-QAM}]\left(\zeta\right)$}
and \textup{$\bar{\Delta}_{\infty\text{-QAM}}\left(e^{\bar{\zeta}}-1\right)=0$}.

The convex envelope of \textup{$\Ilog[\infty\textrm{-QAM}]$} is given
by\textup{ 
\begin{equation}
\Ichecklog[\infty\text{-QAM}]\left(\zeta\right)=\begin{cases}
\zeta\dIlog[\infty\text{-QAM}](\tilde{\zeta}_{2}) & \zeta\leq\tilde{\zeta}_{2}\\
\Ilog[\infty\text{-QAM}]\left(\zeta\right) & \text{otherwise}
\end{cases}\label{eq:unif_convex_envelope}
\end{equation}
}with \textup{$\tilde{\zeta}_{2}\approx5.52\text{ [bits]}$} determined
by the continuity condition \textup{$\Ichecklog[\infty\text{-QAM}](\tilde{\zeta}_{2})=\Ilog[\infty\text{-QAM}](\tilde{\zeta}_{2})$}.
The constant \textup{$\tilde{\Delta}_{\infty\text{-QAM}}$} is given
by\textup{
\begin{equation}
\tilde{\Delta}_{\infty\text{-QAM}}\triangleq\sup_{\zeta}\left[\Ilog[\infty\text{-QAM}]\left(\zeta\right)-\Ichecklog[\infty\text{-QAM}]\left(\zeta\right)\right]\approx0.0608\text{ [bit]}
\end{equation}
}\end{prop}
\begin{IEEEproof}
Appendix \ref{sec:unif-envelopes-proof}.
\end{IEEEproof}
We let $\underline{\gamma}_{0}=e^{\underline{\zeta}_{0}}-1\approx8.76$
dB and $\tilde{\gamma}_{2}=e^{\tilde{\zeta}_{2}}-1\approx16.5$ dB.
Armed with the above results, the proof of Theorem \ref{thm:uniform_sl_ofdm}
is straightforward,
\begin{IEEEproof}[Proof of Theorem \ref{thm:uniform_sl_ofdm}]
Letting $\bar{\gamma}=\max_{\theta\in\left(-\pi,\pi\right)}\left|H\left(\theta\right)\right|^{2}$
and $\bar{\zeta}=\log\left(1+\bar{\gamma}\right),$ the proof that
$\Iofdm\leq\Isl+\bar{\Delta}_{\infty\text{-QAM}}\left(\bar{\gamma}\right)$
is identical to the proof of Theorem \ref{thm:Iofdm_sl_bound}, with
$\hat{I}_{\infty\text{-QAM}}^{\log;\left[0,\bar{\zeta}\right]}\left(\zeta\right)$
replacing $\Ihatlog[x]\left(\zeta\right)$ and $\bar{\Delta}_{\infty\text{-QAM}}\left(\bar{\gamma}\right)$
replacing $\Delta_{x}$. To show that $\Isl\leq\Iofdm+\tilde{\Delta}_{\infty\textrm{-QAM}}$
we reverse the direction of the derivation:
\begin{alignat}{1}
\Iofdm & =\frac{1}{2\pi}\int_{-\pi}^{\pi}I_{\infty\textrm{-QAM}}\left(|H(\theta)|^{2}\right)d\theta\\
 & =\frac{1}{2\pi}\int_{-\pi}^{\pi}\Ilog[\infty\textrm{-QAM}]\left(\log\left(1+|H(\theta)|^{2}\right)\right)d\theta\label{eq:unif_int_form}\\
 & \geq\frac{1}{2\pi}\int_{-\pi}^{\pi}\Ichecklog[\infty\textrm{-QAM}]\left(\log\left(1+|H(\theta)|^{2}\right)\right)d\theta\\
 & \geq\Ichecklog[\infty\textrm{-QAM}]\left(\frac{1}{2\pi}\int_{-\pi}^{\pi}\log\left(1+|H(\theta)|^{2}\right)d\theta\right)\label{eq:unif_jensen_step}\\
 & =\Ichecklog[\infty\textrm{-QAM}]\left(\log\left(1+\SNR[MMSE-DFE-U]\right)\right)\\
 & \geq I_{\infty\textrm{-QAM}}\left(\SNR[MMSE-DFE-U]\right)-\tilde{\Delta}_{\infty\textrm{-QAM}}=\Isl-\tilde{\Delta}_{\infty\textrm{-QAM}}
\end{alignat}
where in (\ref{eq:unif_jensen_step}) the convexity of $\Ichecklog[x]$
was used to invoke Jensen's inequality. When $|H(\theta)|^{2}\geq\underline{\gamma}_{0}$
for all $\theta\in(-\pi,\pi)$, the function $\Ilog$ is convex for
all values of $\log\left(1+|H(\theta)|^{2}\right)$, and we may therefore
replace $\Ichecklog[\infty\textrm{-QAM}]$ with $\Ilog[\infty\textrm{-QAM}]$
and obtain $\Iofdm\geq\Isl.$ When $\SNR[MMSE-DFE-U]\geq\tilde{\gamma}_{2}$,
\begin{equation}
\Ichecklog[\infty\textrm{-QAM}]\left(\log\left(1+\SNR[MMSE-DFE-U]\right)\right)=I_{\infty\textrm{-QAM}}\left(\SNR[MMSE-DFE-U]\right)
\end{equation}
and hence the introduction of $\tilde{\Delta}_{\infty\textrm{-QAM}}$
is unnecessary, resulting once more in $\Iofdm\geq\Isl$.
\end{IEEEproof}

\section{\label{sec:discussion}Discussion}

In this section we use the results obtained in the paper to draw insight
on the differences between the maximum achievable rates of SC and
OFDM that expected in practical scenarios. Additionally, we consider
how increasing the constellation order affects these difference and
discuss the implications of doing so.

\subsection{\label{sub:extremal diff}Maximum and minimum difference $\protect\Iofdm$
and $\protect\Isl$}

Our analysis enables us to characterize the ISI channels for which
$\Isl-\Iofdm$ will be smallest, and the channels for which it will
be the largest. If the input distribution is such that $\Delta_{x}=0$
then clearly every memoryless (flat fading) channel achieves the minimum
difference $\Isl-\Iofdm=0$. If $\Delta_{x}>0$, the minimum difference
is obtained by a channel with two-level transfer function,
\begin{equation}
\left|H\left(\theta\right)\right|^{2}=\begin{cases}
\underline{\gamma}_{1} & \left|\frac{\theta}{2\pi}\right|\leq\log\left(\frac{1+\gamma_{2}}{1+\gamma_{m}}\right)/\log\left(\frac{1+\gamma_{2}}{1+\gamma_{1}}\right)\\
\bar{\gamma}_{2} & \mbox{otherwise}
\end{cases}
\end{equation}
with $\gamma_{m}=e^{\zeta_{m}}-1$ and $\zeta_{m}$ as defined in
(\ref{eq:Delta_x zeta_m}). Clearly for this channel $\Isl-\Iofdm=-\Delta_{x}$
, which is the minimum possible difference according to Theorem \ref{thm:Iofdm_sl_bound}.

Consider the channel,
\begin{equation}
\left|H\left(\theta\right)\right|^{2}=\begin{cases}
e^{\Gamma^{2}}-1 & \left|\theta\right|\leq\pi/\Gamma\\
0 & \mbox{otherwise}
\end{cases}
\end{equation}
where $\Gamma>0$ is an arbitrary sharpness parameter. As $\Gamma$
increases, the channel's frequency response becomes narrower and steeper.
For this channel, $\Iofdm=I_{x}\left(e^{\Gamma^{2}}-1\right)/\Gamma$
and $\Isl=I_{x}\left(e^{\Gamma}-1\right)$. For any finite-alphabet
input, we therefore have $\Isl-\Iofdm\to H\left(x_{0}\right)$ as
$\Gamma\to\infty$, where $H\left(x_{0}\right)$ denotes the input
entropy. Clearly, $H\left(x_{0}\right)$ is the maximum possible difference
between $\Isl$ and $\Iofdm$, as it upper bounds both quantities.

Now consider uniform input, characterized in Theorem \ref{thm:uniform_sl_ofdm}
and analyzed in subsection \ref{sub:uniform analysis}. Since $I_{\infty\textrm{-QAM }}\left(\gamma\right)\approx\log\left(\gamma\right)-\log\left(\pi e/6\right)$
at high SNR, for the above channel we will have $\Iofdm-\Isl\to\log\left(\pi e/6\right)=\Delta_{\infty\text{-QAM}}$
as $\Gamma\to\infty$. Thus, the extreme case that maximizes $\Isl-\Iofdm$
for finite-alphabet inputs also minimizes it for uniform inputs. However,
$\Delta_{\infty\text{-QAM}}$ is approached only for highly impractical
channels. For example, in order for $\Iofdm-\Isl$ to reach 90\% of
$\Delta_{\infty\text{-QAM}}$, we need $\Gamma\approx10$, which yields
$\left|H\left(\theta\right)\right|^{2}\approx430$ dB! Indeed, $\Delta_{\infty\text{-QAM}}$
can only be approached as $\left|H\left(\theta\right)\right|^{2}$
becomes exceedingly large --- this is proven by the very slow rate
of convergence of $\bar{\Delta}_{\infty\text{-QAM}}\left(\gamma\right)$
(Figure \ref{fig:deltabar}).

\subsection{\label{sub:practical-difference}Difference between $\protect\Iofdm$
and $\protect\Isl$in practical settings}

In OFDM wireless communication systems, the constellation and error
correcting code will usually be chosen so that the code rate is between
$1/2$ and $5/6$ \cite{van2006wifi,ghosh2005wimax,reimers1996dvbt,ghosh2010lte}.
Assuming the system is efficient enough to have performance close
to the maximum achievable rate, this means we would have 
\begin{equation}
\frac{1}{2}H\left(x_{0}\right)\leq\Iofdm=\frac{1}{2\pi}\int_{-\pi}^{\pi}\Ilog[x]\left(\zeta_{\theta}\right)d\theta\leq\frac{5}{6}H\left(x_{0}\right)\label{eq:practical-Iofdm-val}
\end{equation}
where $H\left(x_{0}\right)$ is the input entropy which equals the
number of uncoded bits per input symbol for equiprobably inputs, and
$\zeta_{\theta}=\log\left(1+\left|H\left(\theta\right)\right|^{2}\right)$
is the log-SNR at subcarrier frequency $\theta$. Let $\zeta_{\text{OFDM }}$
be the log-SNR in a memoryless channel with achievable rate of $\Iofdm$,\emph{
i.e.} $\Ilog\left(\zeta_{\text{OFDM }}\right)=\Iofdm$. Under this
notation, the single carrier achievable rate satisfies,
\begin{equation}
\Iach\approx\Isl=I_{x}\left(\SNR[MMSE-DFE-U]\right)=\Ilog\left(\frac{1}{2\pi}\int_{-\pi}^{\pi}\zeta_{\theta}d\theta\right)
\end{equation}

Examining Figures \ref{fig:Ilog} and \ref{fig:Iloghat demo} while
keeping (\ref{eq:practical-Iofdm-val}) in mind, we are able to estimate
the performance gain of SC over OFDM for different ISI channels. Clearly,
for channels with little ISI (nearly constant $\left|H\left(\theta\right)\right|^{2}$),
there will be little difference between $\Iofdm$ and $\Iach$, as
we will have $\zeta_{\theta}\approx\zeta_{\text{OFDM }}$ and $\Ilog$
will typically be nearly linear around $\zeta_{\text{OFDM }}$. For
higher-order constellations, $\Iofdm$ might even be slightly larger
than $\Isl$ as $\zeta_{\theta}$ will take values in the interval
where $\Ilog$ is convex. Little performance gain is also expected
when the overall code rate is low so that $\max_{\theta}\Ilog\left(\zeta_{\theta}\right)\leq\frac{5}{6}H\left(x_{0}\right)$,
since in such scenario $\Ilog\left(\zeta_{\theta}\right)$ will be
nearly linear. Conversely, a large difference between the SC and OFDM
achievable rates is expected whenever there exists a significant bandwidth
of sub-carriers for which $\Ilog\left(\zeta_{\theta}\right)$ is close
to the input entropy. This event is likely for channels with significant
ISI, and the difference will become more pronounced as the code rate
grows.

\begin{figure}
\begin{centering}
\includegraphics[width=7cm]{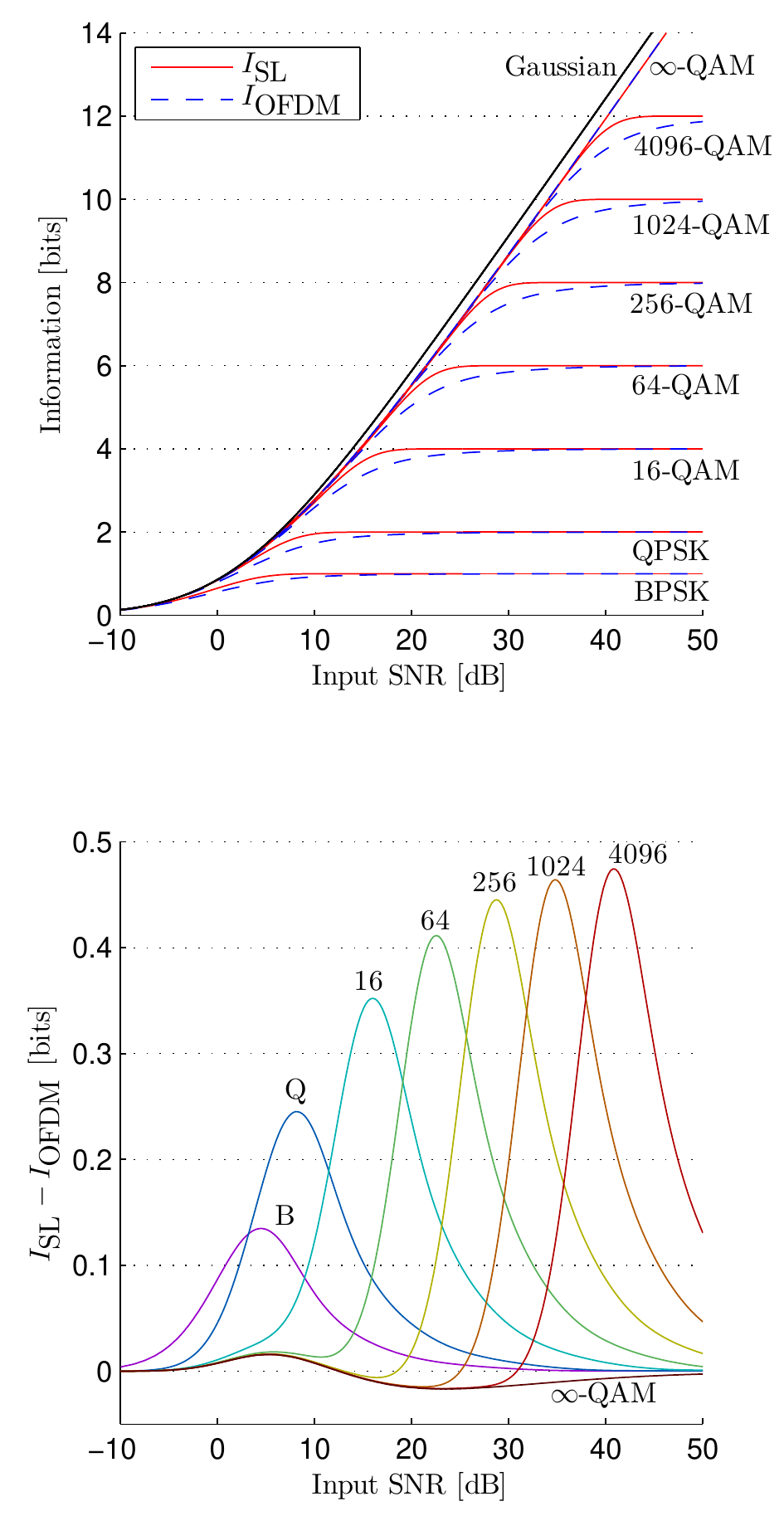}
\par\end{centering}

\protect\caption{\label{fig:80211n-sim-compare}$\protect\Isl$ and $\protect\Iofdm$
(up) and $\protect\Isl-\protect\Iofdm$ (down) as a function of input
SNR, for different input distributions.}
\end{figure}

To demonstrate our conclusions we a present a numerical experiment
using a 9-tap ISI channel randomly drawn from a distribution defined
by the 802.11n NLOS channel model B \cite{erceg2004ieee}. The input
SNR is given by $\frac{1}{2\pi}\int_{-\pi}^{\pi}\left|H\left(\theta\right)\right|^{2}d\theta$,
and is varied by scaling $\left|H\left(\theta\right)\right|^{2}$.
At unit input SNR and rounded to 2 significant digits, the taps of
the specific channel used are given by,
\[
h=\left[0.62e^{1.3j},\,0.42e^{2.8j},\,0.33e^{-1.3j},\,0.091e^{2.5j},\,0.51e^{0.66j},\,0.25e^{2j},\,0.039e^{-0.087j},\,0.028e^{-0.28j},\,0.039e^{1.7j}\right]
\]
\begin{figure}
\begin{centering}
\includegraphics[width=9cm]{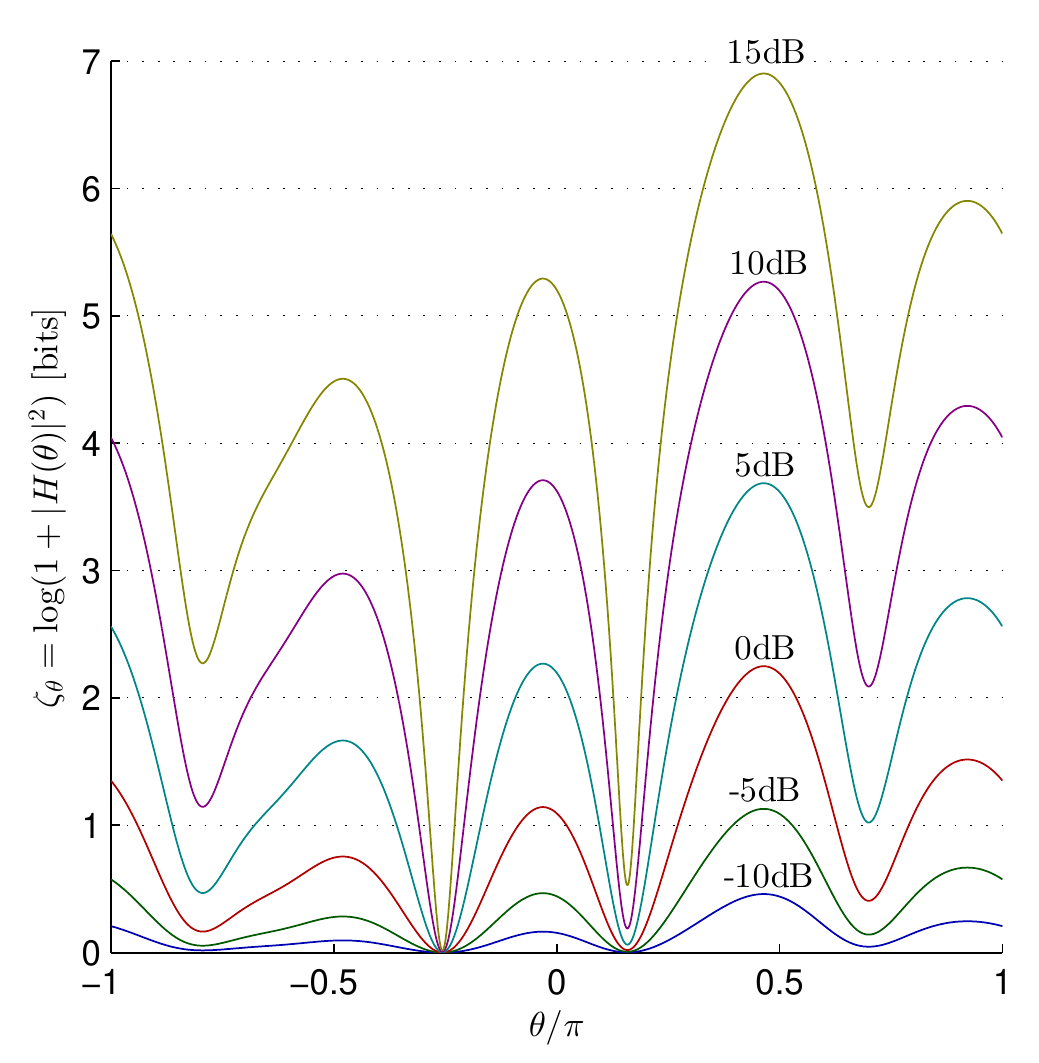}
\par\end{centering}

\protect\caption{\label{fig:80211n-sim-channel}The channel transfer function expressed
in terms of log-SNR $\zeta_{\theta}=\log\left(1+\left|H\left(\theta\right)\right|^{2}\right),$
different values of input SNR. }
\end{figure}
Figure \ref{fig:80211n-sim-channel} shows $\zeta_{\theta}$ as a
function of $\theta$ for different input SNR's. Clearly, this channel
shows considerable variation in $\zeta_{\theta}$, and so observable
differences between the achievable rates are expected. Figure \ref{fig:80211n-sim-compare}
shows $\Isl$, $\Iofdm$ and the difference between them for different
input distributions. As expected, differences between the achievable
rates are very small for low code rates, but become significant as
the code rates grow. In some cases, $\Isl$ is seen to exceed $\Iofdm$
by over 15\%, and for code rate 5/6 the differences between SC and
OFDM in terms of required SNR reach up to 3dB. The very low rate in
which $\Iofdm$ converges to the input entropy as the SNR grows can
be explained by the strong notches in the ISI transfer functions,
where $\zeta_{\theta}$ approaches the input entropy only for very
high SNR's. For uniform input, $\Isl-\Iofdm$ is positive for low
SNR's and negative for high SNR's, as Theorem \ref{thm:uniform_sl_ofdm}
predicts. Moreover, the difference is always very small, never exceeding
0.02 bits in magnitude. In particular, the maximum theoretical difference
of about $\Delta_{\infty\text{-QAM}}\approx0.509$ bits in favor of
OFDM is never attained, and ever the tighter bound $\bar{\Delta}_{\infty\text{-QAM}}\left(60\text{ dB}\right)\approx0.228$
is quite loose for this channel.

Qualitatively, when the log-SNR at a given frequency grows, the contribution
of that frequency to the overall OFDM achievable rate saturates, while
its contribution to the single-carrier achievable rate continues to
grow, resulting in a growing difference between the two rates. We
note that this saturation effect is due to the finite-entropy nature
of the input, and does not occur in the Gaussian and uniform cases.
This behavior echoes the Mercury/Water-filling results of \cite{lozano2006optimum},
where it is seen that the optimal OFDM power allocation policy for
finite-entropy inputs differs significantly from classical Waterfilling
in the high-SNR regime.

\subsection{\label{sub:implications}The implications of increasing the constellation
order}

From our analysis of uniform QAM input in Theorem \ref{thm:uniform_sl_ofdm}
and from the discussion above, it is clear that for a given ISI channel
and SNR, the performance of OFDM can be made close to that of SC,
by sufficiently increasing the constellation order. This is due to
the fact that if the input alphabet is chosen to be sufficiently large,
no saturation of $\Ilog$ will occur at any subcarrier frequency,
and therefore no significant difference between achievable rates is
to be expected.

Thus, the potential performance gain of moving from OFDM to SC, and
maintaining the same constellation order, may also be realized by
using OFDM with a higher-order constellation. However, there are two
system design considerations that may not allow for arbitrary increase
in constellation order. First, increasing the number of bits per symbol
necessitates using lower code rates and perhaps more sophisticated
coded modulation schemes. For example, in the setting depicted in
Figure \ref{fig:80211n-sim-compare}, for QPSK to be used with code
rate 1/2 there is a difference of about 0.5dB, or 12\%, between SC
and OFDM in terms of SNR. Changing the constellation to 16-QAM will
essentially eliminate this difference, but require an unconventional
code rate of 1/4.

The second consideration is channel estimation. As the constellation
order grows, the requirements on estimation accuracy of the channel
gain become more stringent. Conversely, for BPSK and QPSK inputs amplitude
estimation in not necessary at all. Hence, when increasing the constellation
order in an OFDM system, the overhead of pilot subcarriers might have
to grow as well. 

In light of the issues above, as well as the sate-of-the-art OFDM
wireless communication technology, where code rates below 1/2 and
constellations above 256-QAM are uncommon, we conclude that using
higher order constellations at low SNR's is not trivial. Therefore,
we maintain that fixing the constellation order and using SC in lieu
of OFDM is an option well-worth investigating.

\section{\label{sec:conclusion}Conclustion}

In this paper a comparison of the achievable rates of OFDM and single-carrier
modulations was performed, under the assumption of a fixed i.i.d.
input distribution. In lieu of a tractable expression for the achievable
rate of single-carrier modulation, the Shamai-Laroia approximation
was used, since it is well known to essentially reflect tight lower
bounds on the achievable rate. We revealed an intimate relation between
the Shamai-Laroia approximation and the OFDM achievable rate, that
stems from the concavity properties of the input-output mutual information
in a scalar Gaussian channel with respect to a modified SNR variable
--- namely, that the Shamai-Laroia approximation is essentially an
upper bound on the OFDM achievable rate.

In particular, the upper bound always holds for conventional low order
input distributions including BPSK, QAM and 16-QAM. It also holds
for all PAM and square QAM inputs, when the SNR exceeds a certain
relatively modest threshold. Moreover, we quantified the amount by
which the OFDM achievable rate might exceed the Shamai-Laroia approximation,
and found it to be very small for any input distribution of interest.
In contrast, we demonstrated that the Shamai-Laroia approximation
may be arbitrarily larger than the OFDM rate for some ISI channels
and any finite-alphabet input distribution and may provide significant
improvement in practical scenarios. By similar analysis of a continuous
uniform input distribution, it is shown that the difference between
achievable rates can be made small by selecting a sufficiently dense
input distribution. However, such choice of input might not be practical.
Our conclusions extend to the case when linear precoding is allowed,
giving additional validity to our assumption of i.i.d. input. 

Estimation-theoretic bounds along with Information-Estimation identities
were the primary tools used in our analysis. In order to establish
our results regarding PAM and square QAM inputs, novel bounds on nonlinear
MMSE estimation of PAM inputs in an additive Gaussian channel were
developed. They include a ``pointwise'' bound on the conditional
variance of the channel input given the channel output, as well as
a tight high-SNR characterization of the MMSE. These bounds might
be more widely useful.

We conclude that single-carrier modulation offers a fundamental, possibly
large, improvement in spectral efficiency over OFDM when the input
alphabet is constrained. However, virtually all state-of-the-art high-performance
communication systems over ISI use OFDM. This is mainly due to the
fact that implementing optimal joint equalization and decoding is
straightforward in OFDM, but difficult in single-carrier modulation.
However, practical iterative schemes that approach the SC achievable
may be within reach. We believe that this work provides motivation
for research and development of such schemes.

\section*{Acknowledgment}

We are grateful to Uri Erez and the anonymous reviewers for their
comments and suggestions, which have enhanced the scope of this paper
and its clarity.

This research has been supported by the S. and N. Grand research fund,
by the Israel Science Foundation (ISF) and by the European Commission
in the framework of the FP7 Network of Excellence in Wireless COMmunications
NEWCOM\#.

\appendices{}

\section{\label{app:dmmse_bound_general_constel}High-SNR upper bound on MMSE
derivative}
\begin{figure}[h]
\centering{}\includegraphics{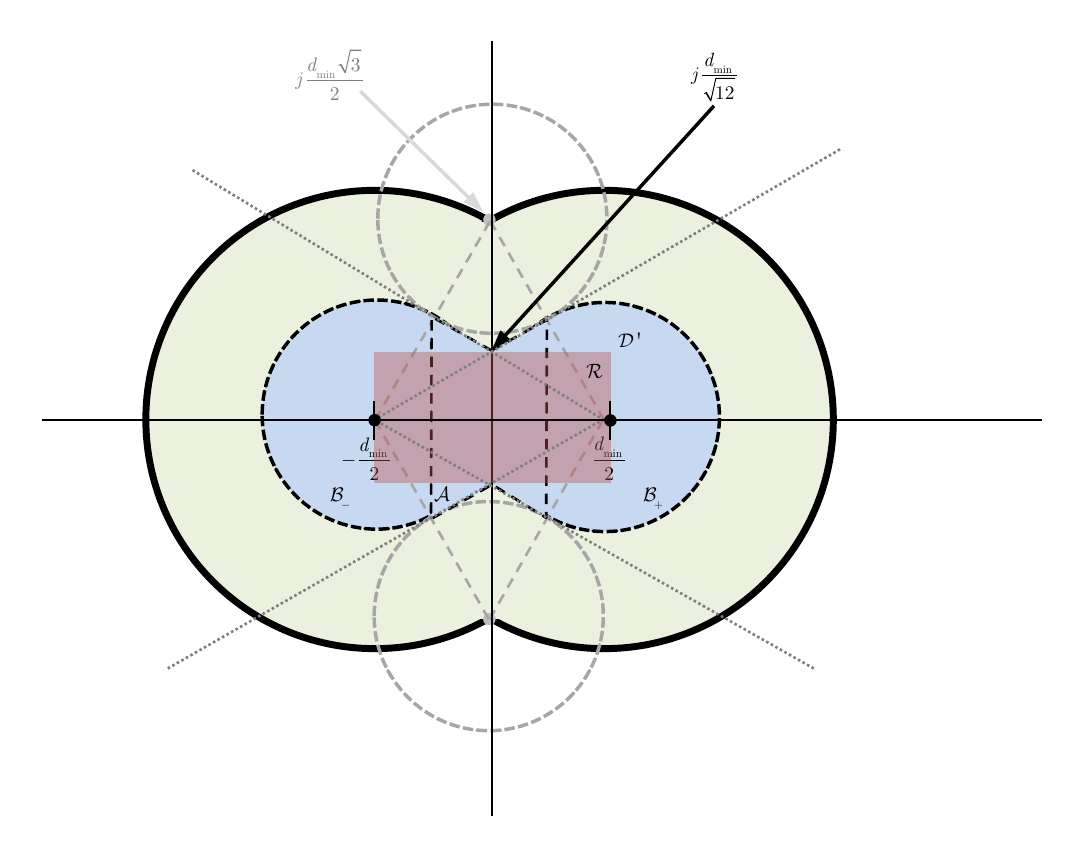}\protect\caption{\label{fig:Dtag R illustration}Illustration of the sets $\mathcal{D}'$
(blue) and $\mathcal{R}$ (red). The black dots indicate the location
of the points $x_{\pm}=\pm d_{\min}/2.$ The region $\mathcal{D}'$
is formed by assuming the existence of other input points on the curve
formed by points that have distance $d_{\min}$ from either $x_{+}$
or $x_{-}$ and distance greater than $d_{\min}$ from the other point
(the edge of the green region in the figure).}
\end{figure}

\begin{lem}
For any finite-alphabet unit-power input distribution $x$, there
exists $C>0$ such that
\begin{equation}
\mathrm{mmse}_{X}'(\gamma)\leq-C\frac{e^{-\left(d_{\min}/2\right)^{2}\gamma}}{\sqrt{\gamma}}
\end{equation}
for sufficiently large $\gamma$, where $d_{\min}$ is the minimum
distance between any two input values.\end{lem}
\begin{IEEEproof}
Let $\mathcal{X}$ be the input alphabet and let $K=\left|\mathcal{X}\right|$.
The derivative of the MMSE function in the complex scalar channels
can be read from the results of \cite{payaro2009hessian},
\begin{equation}
\textrm{mmse}_{X}'(\gamma)=-\Ebrack{\phi_{X}(Y_{\gamma};\gamma)+|\psi_{X}(Y_{\gamma};\gamma)|^{2}}{Y_{\gamma}}
\end{equation}
where $Y_{\gamma}=X+\frac{1}{\sqrt{\gamma}}N$ with $N$ standard
complex Gaussian and independent of $X$, and
\begin{eqnarray}
\phi_{X}(y;\gamma) & = & \Ebrack{\left|X-{\Ebrack{X|Y_{\gamma}=y}{}}\right|^{2}\,|\, Y_{\gamma}=y}X\\
\psi_{X}(y;\gamma) & = & \Ebrack{\left(X-{\Ebrack{X|Y_{\gamma}=y}{}}\right)^{2}\,|\, Y_{\gamma}=y}X
\end{eqnarray}
$\phi_{X}(y;\gamma)$ can be thought of as a point-wise MMSE function,
given channel outcome $y$, but $\psi_{X}(y;\gamma)$ is complex and
does not posses much intuitive meaning. Let $x_{+}$ and $x_{-}$
be two input values such that $\left|x_{+}-x_{-}\right|=d_{\min}$.
We may assume without loss of generality that
\begin{equation}
x_{\pm}=\pm d_{\min}/2
\end{equation}
because the input alphabet can always be shifted and rotated so
that the above relation holds. Let $p_{+}$ and $p_{-}$ denote the
probabilities of $x_{+}$ and $x_{-}$ respectively and assume without
loss of generality that $p_{+}\leq p_{-}$. Let $U$ be random variable
independent of $X$ and distributed on $\left\{ 0,1\right\} $ with
$\Pr\left(U=1\right)=p_{+}/p_{-}$. Define the random variable $I=1_{\left\{ X=x_{+}\right\} }+1_{\left\{ X=x_{-}\wedge U=1\right\} }$,
so that given $I=1$, $X$ is distributed equiprobably on $\left\{ x_{+},x_{-}\right\} $.
We have,
\begin{eqnarray}
\phi_{X}(y;\gamma) & = & \Pr(I=1|Y_{\gamma}=y)\Ebrack{\left|X-{\Ebrack{X|Y_{\gamma}=y}{}}\right|^{2}\,|\, Y_{\gamma}=y\ ,\ I=1}X\\
 &  & +\Pr(I=0|Y_{\gamma}=y)\Ebrack{\left|X-{\Ebrack{X|Y_{\gamma}=y}{}}\right|^{2}\,|\, Y_{\gamma}=y\ ,\ I=0}X
\end{eqnarray}
Notice that 
\begin{equation}
\Ebrack{\left|X-{\Ebrack{X|Y_{\gamma}=y}{}}\right|^{2}\,|\, Y_{\gamma}=y\ ,\ I=1}X\geq\Ebrack{\left|X-{\Ebrack{X|Y_{\gamma}=y\ ,\ I=1}{}}\right|^{2}\,|\, Y_{\gamma}=y\ ,\ I=1}X
\end{equation}
since we add the information $I=1$ to the MMSE estimator. Since the
input is binary and symmetric given $I=1$, the RHS of the above inequality
is the pointwise MMSE for symmetric binary input with variance $\left(d_{\min}/2\right)^{2}$
at SNR $\rho\triangleq\left(d_{\min}/2\right)^{2}\gamma$:
\begin{equation}
\Ebrack{\left|X-{\Ebrack{X|Y_{\gamma}=y\ ,\ I=1}{}}\right|^{2}\,|\, Y_{\gamma}=y\ ,\ I=1}X=\left(\frac{d_{\min}}{2}\right)^{2}\phi_{\mathrm{BPSK}}\left(\left(\frac{d_{\min}}{2}\right)^{-1}y;\rho\right)
\end{equation}
with
\begin{equation}
\phi_{\mathrm{BPSK}}(z;\gamma)=1-\tanh^{2}(2\gamma\mathrm{Re}\left\{ z\right\} )=\frac{1}{\cosh^{2}\left(2\gamma\mathrm{Re}\left\{ z\right\} \right)}\label{eq:phi BPSK}
\end{equation}
therefore
\begin{equation}
\phi_{X}(y;\gamma)\geq\Pr(I=1|Y_{\gamma}=y)\left(\frac{d_{\min}}{2}\right)^{2}\phi_{\mathrm{BPSK}}\left(\left(\frac{d_{\min}}{2}\right)^{-1}y;\rho\right)
\end{equation}
and so,
\begin{equation}
\Ebrack{\phi_{X}^{2}(Y_{\gamma};\gamma)}{Y_{\gamma}}\geq\left(\frac{d_{\min}}{2}\right)^{4}\int_{\mathbb{C}}{f_{Y_{\gamma}}(y)\left[\Pr(I=1|Y_{\gamma}=y)\phi_{\mathrm{BPSK}}\left(\left(\frac{d_{\min}}{2}\right)^{-1}y;\rho\right)\right]^{2}dy}\label{eq:phi integral bound 1}
\end{equation}

We have
\begin{equation}
\Pr(I=1|Y_{\gamma}=y)f_{Y_{\gamma}}(y)=\Pr(I=1)f_{Y_{\gamma}|I}(y|I=1)=p_{+}\frac{\gamma}{\pi}\left(e^{-\gamma\left|y-\frac{d_{\min}}{2}\right|^{2}}+e^{-\gamma\left|y+\frac{d_{\min}}{2}\right|^{2}}\right)\label{eq:PrI fy}
\end{equation}
and also
\begin{equation}
\Pr(I=1|Y_{\gamma}=y)=\Pr(X=x_{+}|Y_{\gamma}=y)+\left(p_{+}/p_{-}\right)\Pr(X=x_{-}|Y_{\gamma}=y)
\end{equation}
with
\begin{equation}
\Pr(X=x|Y_{\gamma}=y)=\frac{\Pr\left(X=x\right)e^{-\gamma\left|y-x\right|^{2}}}{\sum_{x'\in\mathcal{X}}\Pr\left(X=x'\right)e^{-\gamma\left|y-x'\right|^{2}}}
\end{equation}

Let $\mathcal{D}\subseteq\mathbb{C}$ denote the set of points for
which $\arg\min_{x\in\mathcal{X}}{\left|y-x\right|}$ is either $x_{+}$
or $x_{-}$. Clearly, for every $y\in\mathcal{D}$, either $\Pr(X=x_{-}|Y_{\gamma}=y)>p_{-}$
or $\Pr(X=x_{+}|Y_{\gamma}=y)>p_{+}$ and so
\begin{equation}
\Pr(I=1|Y_{\gamma}=y)>p_{+}\ \forall y\in\mathcal{D}\label{eq:prI bound}
\end{equation}

The set $\mathcal{D}$ depends on other points in $\mathcal{X}$,
but can be lower bounded by $\mathcal{D}'\subseteq\mathcal{D}$ which
is formed by adding to $\mathcal{X}$ all the points with distance
greater than $d_{\min}$ from both $x_{+}$ and $x_{-}$. Figure \ref{fig:Dtag R illustration}
illustrates the construction of $\mathcal{D}'$, which is given by
\begin{equation}
\mathcal{D}'=\mathcal{A}\cup\mathcal{B}_{+}\cup\mathcal{B}_{-}
\end{equation}
where
\begin{equation}
\mathcal{B}_{\pm}=\left\{ y\in\mathbb{C}\,|\:\left|y\mp\frac{d_{\min}}{2}\right|^{2}<\left(\frac{d_{\min}}{2}\right)^{2}\right\} 
\end{equation}
and
\begin{equation}
\mathcal{A}=\left\{ y\in\mathbb{C}\,|\:\left|\mathrm{Im}\left\{ y\right\} \right|<\frac{\left|\mathrm{Re}\left\{ y\right\} \right|+d_{\min}/2}{\sqrt{3}}\wedge\left|\mathrm{Re}\left\{ y\right\} \right|<\frac{d_{\min}}{4}\right\} 
\end{equation}
Finally, the set $\mathcal{D}'$ contains the rectangular subset $\mathcal{R}\subset\mathcal{D}'$
given by%
\footnote{The real axis border of $\mathcal{R}$ can be extended to $d_{\min}\sqrt{11/12}$,
but this doesn't change the leading exponent in the bound nor does
it change its coefficient. It only changes the faster decreasing exponents.%
}
\begin{equation}
\mathcal{R}=\left\{ y\in\mathbb{C}\,|\:\left|\mathrm{Im}\left\{ y\right\} \right|<\frac{d_{\min}}{\sqrt{12}}\wedge\left|\mathrm{Re}\left\{ y\right\} \right|<\frac{d_{\min}}{2}\right\} 
\end{equation}

Limiting the integration in (\ref{eq:phi integral bound 1}) to $\mathcal{R}$
and substituting (\ref{eq:PrI fy}) and (\ref{eq:prI bound}) we obtain,
\begin{eqnarray}
\E[\phi_{X}^{2}(Y_{\gamma};\gamma)][Y_{\gamma}] & \geq & p_{+}^{2}\left(\frac{d_{\min}}{2}\right)^{4}\int_{\mathcal{R}}{\frac{\gamma}{\pi}\left(e^{-\gamma\left|y-\frac{d_{\min}}{2}\right|^{2}}+e^{-\gamma\left|y+\frac{d_{\min}}{2}\right|^{2}}\right)\phi_{\mathrm{BPSK}}^{2}\left(\frac{y}{d_{\min}/2};\rho\right)dy}\\
 & = & p_{+}^{2}\left(1-2Q\left(\sqrt{2\rho/3}\right)\right)\left(\frac{d_{\min}}{2}\right)^{4}2M(\rho)
\end{eqnarray}
with
\begin{eqnarray}
M(\rho) & = & \frac{1}{\sqrt{\pi}}\int_{-\sqrt{\rho}}^{\sqrt{\rho}}{dz\left(\frac{1}{2}e^{-\left(z-\sqrt{\rho}\right)^{2}}+\frac{1}{2}e^{-\left(z+\sqrt{\rho}\right)^{2}}\right)\phi_{\mathrm{BPSK}}^{2}\left(\frac{z}{\sqrt{\rho}};\rho\right)}\\
 & = & \frac{e^{-\rho}}{\sqrt{\pi}}\int_{-\sqrt{\rho}}^{\sqrt{\rho}}{dze^{-z^{2}}\left[\cosh\left(2\sqrt{\rho}z\right)\right]^{-3}}\\
 & \geq & \frac{e^{-\rho}}{\sqrt{\pi}}\int_{-\sqrt{\rho}}^{\sqrt{\rho}}{dze^{-z^{2}\left(1+6\rho\right)}}=\frac{e^{-\rho}}{\sqrt{1+6\rho}}\left(1-2Q\left(\sqrt{2\rho\left(1+6\rho\right)}\right)\right)
\end{eqnarray}
where we have used the expression (\ref{eq:phi BPSK}) for $\phi_{\mathrm{BPSK}}$
along with $\cosh x\leq e^{x^{2}/2}$ to establish the above bound.
Using $\sqrt{2\pi}xQ(x)\leq e^{-x^{2}/2}$, we find that $Q\left(\sqrt{2\rho\left(1+6\rho\right)}\right)=o(e^{-6\rho^{2}})$
and so,
\begin{equation}
M(\rho)\geq C'\frac{e^{-\rho}}{\sqrt{\rho}}
\end{equation}
for some $C'>0$ and for sufficiently large $\rho$. Similarly, noticing
that $Q\left(\sqrt{2\rho/3}\right)=o(e^{-\rho/3})$ we have also
\begin{equation}
\E[\phi_{X}^{2}(Y_{\gamma};\gamma)][Y_{\gamma}]\geq C\frac{e^{-\left(d_{\min}/2\right)^{2}\gamma}}{\sqrt{\gamma}}
\end{equation}
for some $C>0$ and for sufficiently large $\gamma$, where we have
substituted back $\rho=\left(d_{\min}/2\right)^{2}\gamma$. Finally,
\begin{equation}
\textrm{mmse}_{x}'(\gamma)\geq-\E[\phi_{X}^{2}(Y_{\gamma};\gamma)][Y_{\gamma}]\geq-C\frac{e^{-\left(d_{\min}/2\right)^{2}\gamma}}{\sqrt{\gamma}}
\end{equation}
under the same conditions.
\end{IEEEproof}

\section{\label{sec:PAM pointwise}Proof of Theorem \ref{thm:pointwise PAM bounds}}

We begin by establishing some notation. The input alphabet will be
denoted by $\mathcal{X}=\left\{ x_{1},...,x_{M}\right\} $ and we
assume that $x_{m+1}-x_{m}=d$ for every $1\leq m<M$. Let 
\begin{equation}
p_{m|y}=\Pr\left(X=x_{m}|Y_{\gamma}=y\right)=\frac{e^{-\gamma\left(y-x_{m}\right)^{2}}}{\sum_{m'=1}^{M}e^{-\gamma\left(y-x_{m'}\right)^{2}}}\label{eq:pmy-def}
\end{equation}
 denote the probability of symbol $x_{m}$ given observation $y$,
and let 
\begin{equation}
s\left(y\right)=\Ebrack{X|Y_{\gamma}=y}{}=\sum_{m=1}^{M}p_{m|y}x_{m}
\end{equation}
 be the expectation of $X$ conditioned on the observation $Y_{\gamma}=y$,
so that 
\begin{equation}
\phi_{X}\left(y;\gamma\right)=\sum_{m=1}^{M}p_{m|y}\left(x_{m}-s\left(y\right)\right)^{2}
\end{equation}
Notice that $\emph{s}\left(y\right)=\arg\min_{\emph{s}}\sum_{m=1}^{M}p_{m|y}\left(x_{m}-\emph{s}\right)^{2}$
--- \emph{i.e.} the conditional expectation is the point-wise optimal
estimator of $X$ given $Y_{\gamma}=y$. Finally, recall that $x_{J},x_{J+1}$
denote the two nearest neighbors to $y$ in $\mathcal{X}$.

The upper bound in Theorem \ref{thm:pointwise PAM bounds} is derived
by considering the sub-optimal estimator that assumes the input has
the same distribution as $B_{J}$ (uniform on $\left\{ x_{J},x_{J+1}\right\} $).
This estimator is given by 
\begin{equation}
\tilde{\emph{s}}\left(y\right)=\left(p_{J|y}x_{J}+p_{J+1|y}x_{J+1}\right)/\left(p_{J|y}+p_{J+1|y}\right)
\end{equation}
and the resulting bound reads:
\begin{eqnarray}
\phi_{X}\left(y;\gamma\right) & \leq & \sum_{m=1}^{M}p_{m|y}\left(x_{m}-\tilde{\emph{s}}\left(y\right)\right)^{2}\nonumber \\
 & \leq & \sum_{j=J,J+1}\frac{p_{j|y}}{p_{J|y}+p_{J+1|y}}\left(x_{j}-\tilde{\emph{s}}\left(y\right)\right)^{2}+\sum_{m\neq J,J+1}p_{m|y}\left(x_{m}-\tilde{\emph{s}}\left(y\right)\right)^{2}\label{eq:pointwise upper base}
\end{eqnarray}

The following bound is seen to hold,
\begin{alignat}{1}
\sum_{m=1}^{J-1} & \left(x_{m}-\tilde{\emph{s}}\left(y\right)\right)^{2}e^{-\gamma\left(y-x_{m}\right)^{2}}\leq d^{2}e^{-\gamma\left(y-x_{J}\right)^{2}}\sum_{m=1}^{J-1}\left(J-m+1\right)^{2}e^{-\gamma\left(x_{J}-x_{m}\right)^{2}}\nonumber \\
 & =d^{2}e^{-\gamma\left(y-x_{J}\right)^{2}}\sum_{k=1}^{J-1}\left(k+1\right)^{2}e^{-\gamma d^{2}k^{2}}\leq d^{2}e^{-\gamma\left(y-x_{J}\right)^{2}}\sum_{k=1}^{\infty}\left(k+1\right)^{2}e^{-\gamma d^{2}k^{2}}
\end{alignat}
where the first transition follows from $\left(y-x_{m}\right)^{2}\geq\left(y-x_{J}\right)^{2}+\left(x_{J}-x_{m}\right)^{2}$
which holds since $x_{m}<x_{J}\leq y$, and from $\left(x_{m}-\tilde{\emph{s}}\left(y\right)\right)^{2}\geq\left(x_{m}-x_{J}\right)^{2}=d^{2}\left(J-m+1\right)^{2}$,
which holds since $x_{m}<x_{J}\leq\tilde{s}\left(y\right)$. Similarly,
we have
\begin{equation}
\sum_{m=J+2}^{M}\left(x_{m}-\tilde{\emph{s}}\left(y\right)\right)^{2}e^{-\gamma\left(y-x_{m}\right)^{2}}\leq d^{2}e^{-\gamma\left(y-x_{J+1}\right)^{2}}\sum_{k=1}^{\infty}\left(k+1\right)^{2}e^{-\gamma d^{2}k^{2}}
\end{equation}

Using the above bounds and observing (\ref{eq:pmy-def}), we find
that
\begin{alignat}{1}
\sum_{m\neq J,J+1} & p_{m|y}\left(x_{m}-\tilde{\emph{s}}\left(y\right)\right)^{2}\leq\frac{e^{-\gamma\left(y-x_{J}\right)^{2}}+e^{-\gamma\left(y-x_{J+1}\right)^{2}}}{\sum_{m'=1}^{M}e^{-\gamma\left(y-x_{m'}\right)^{2}}}d^{2}\sum_{k=1}^{\infty}\left(k+1\right)^{2}e^{-\gamma d^{2}k^{2}}\nonumber \\
 & \leq d^{2}\sum_{k=1}^{\infty}\left(k+1\right)^{2}e^{-\gamma d^{2}k^{2}}\leq d^{2}\sum_{k=1}^{\infty}\left(k+1\right)^{2}e^{-\gamma d^{2}k}\leq\frac{4d^{2}e^{-\gamma d^{2}}}{\left(1-e^{-\gamma d^{2}}\right)^{3}}\label{eq:pointwise upper sum}
\end{alignat}
where the last inequality is due to,
\begin{equation}
\sum_{k=1}^{\infty}\left(k+1\right)^{2}x^{k}\leq2x\sum_{k=0}^{\infty}\left(k+2\right)\left(k+1\right)x^{k}=2x\left(\sum_{k=0}^{\infty}x^{k}\right)''=2x\left(\frac{1}{1-x}\right)''=\frac{4x}{\left(1-x\right)^{3}}
\end{equation}
Identifying $\sum_{j=J,J+1}\frac{p_{j|y}}{p_{J|y}+p_{J+1|y}}\left(x_{j}-\tilde{\emph{s}}\left(y\right)\right)^{2}$
with $\phi_{B_{J}}\left(y;\gamma\right)$, the upper bound follows
from (\ref{eq:pointwise upper base}) and (\ref{eq:pointwise upper sum}).

To prove the lower bound in Theorem \ref{thm:pointwise PAM bounds},
we first prove the following,
\begin{lem}
\label{lem:pointwise bound remove one} Let $X$ be uniformly distributed
on $\mathcal{X}=\left\{ x_{1},x_{2},...,x_{M}\right\} $ such that
$x_{m+1}-x_{m}=d$ for all $1\leq m<M$. For any $y\in\mathbb{R}$,
let $\tilde{x}_{y}$ be the point in $\mathcal{X}$ with maximum distance
from $y$. Let $\hat{X}$ be uniformly distributed on $\hat{\mathcal{X}}=\mathcal{X}\setminus\left\{ \tilde{x}_{y}\right\} $.
For every $\gamma>0$,
\begin{equation}
\phi_{X}\left(y;\gamma\right)\geq\phi_{\hat{X}}\left(y;\gamma\right)
\end{equation}
\end{lem}
\begin{IEEEproof}
Without loss of generality, assume $y\leq\left(x_{1}+x_{M}\right)/2$
so that $\tilde{x}_{y}\equiv x_{M}$ and $\hat{\mathcal{X}}=\left\{ x_{1},x_{2},...,x_{M-1}\right\} $.
Let
\begin{equation}
\hat{\emph{s}}\left(y\right)=\Ebrack{X|Y_{\gamma}=y,X\neq\tilde{x}_{y}}{}=\Ebrack{\hat{X}|\hat{Y}_{\gamma}=y}{}=\sum_{m=1}^{M-1}\frac{p_{m|y}}{1-p_{M|y}}x_{m}
\end{equation}
denote the expectation of $X$ given $Y=y$ and $X\neq\tilde{x}_{y}$
or equivalently the expectation of $\hat{X}$ given $\hat{Y}_{\gamma}=\hat{X}+\frac{1}{\sqrt{\gamma}}\hat{N}=y$,
with $\hat{N}\sim\mathcal{N}\left(0,1/2\right)$ and independent of
$\hat{X}$. Notice that
\begin{equation}
\emph{s}\left(y\right)-\hat{\emph{s}}\left(y\right)=p_{M|y}\left(x_{M}-\hat{\emph{s}}\left(y\right)\right)
\end{equation}

Using the orthogonality principle, we may therefore write,
\begin{eqnarray}
\phi_{X}(y;\gamma) & = & \sum_{m=1}^{M}p_{m|y}\left(x_{m}-\emph{s}\left(y\right)\right)^{2}=\sum_{m=1}^{M}p_{m|y}\left(x_{m}-\hat{\emph{s}}\left(y\right)\right)^{2}-\left(\emph{s}\left(y\right)-\hat{\emph{s}}\left(y\right)\right)^{2}\nonumber \\
 & = & \left(1-p_{M|y}\right)\sum_{m=1}^{M-1}\frac{p_{m|y}}{1-p_{M|y}}\left(x_{m}-\hat{\emph{s}}\left(y\right)\right)^{2}+\left(p_{M|y}-p_{M|y}^{2}\right)\left(x_{M}-\hat{\emph{s}}\left(y\right)\right)^{2}\nonumber \\
 & = & \phi_{\hat{X}}\left(y;\gamma\right)+p_{M|y}\left[\left(1-p_{M|y}\right)\left(x_{M}-\hat{\emph{s}}\left(y\right)\right)^{2}-\phi_{\hat{X}}\left(y;\gamma\right)\right]\label{eq:phiX phiXhat relate}
\end{eqnarray}
By our assumption that $y\leq\left(x_{1}+x_{M}\right)/2$ we have
$p_{m|y}\geq p_{M-m+1|y}$ for every $1\leq m\leq M/2$ and therefore
$\hat{\emph{s}}\left(y\right)\leq\emph{s}\left(y\right)\leq\left(x_{1}+x_{M}\right)/2$.
Thus,
\begin{equation}
\left(x_{M}-\hat{\emph{s}}\left(y\right)\right)^{2}\geq\left(\frac{x_{M}-x_{1}}{2}\right)^{2}=\left(\frac{d}{2}\right)^{2}\left(M-1\right)^{2}\label{eq:xM-fy bound}
\end{equation}

We obtain the following crude upper bound for $\phi_{\hat{X}}\left(y;\gamma\right)$
by considering the suboptimal estimator $\left(x_{1}+x_{M-1}\right)/2$,
\begin{eqnarray}
\phi_{\hat{X}}\left(y;\gamma\right) & \leq & \sum_{m=1}^{M-1}\frac{p_{m|y}}{1-p_{M|y}}\left(x_{m}-\frac{x_{1}+x_{M-1}}{2}\right)^{2}\nonumber \\
 & \leq & \left(\frac{x_{M-1}-x_{1}}{2}\right)^{2}=\left(\frac{d}{2}\right)^{2}\left(M-2\right)^{2}\label{eq:phiXhat up bound}
\end{eqnarray}
The second inequality follows from the fact that $x_{1}$ is the farthest
point from $\left(x_{1}+x_{M-1}\right)/2$ in $\hat{\mathcal{X}}$
and therefore moving all probability mass to $m=1$ increases the
sum. Since $x_{M}$ is farthest from $y$ in $\mathcal{X}$, we have
$p_{M|y}\leq p_{m|y}$ for any $1\leq m<M$, and consequently 
\begin{equation}
p_{M|y}\leq\frac{1}{M}\label{eq:pMy up bound}
\end{equation}

Combining (\ref{eq:xM-fy bound}), (\ref{eq:phiXhat up bound}) and
(\ref{eq:pMy up bound}) we find that
\begin{equation}
\left(1-p_{M|y}\right)\left(x_{M}-\hat{\emph{s}}\left(y\right)\right)^{2}-\phi_{\hat{X}}\left(y;\gamma\right)\geq\left(\frac{d}{2}\right)^{2}\left(M-1-\frac{1}{M}\right)\geq0
\end{equation}
 for every $M\geq2$. We therefore conclude by (\ref{eq:phiX phiXhat relate})
that $\phi_{X}(y;\gamma)\geq\phi_{\hat{X}}\left(y;\gamma\right)$
for every $y$ and every $\gamma$.
\end{IEEEproof}
The lower bound in Theorem \ref{thm:pointwise PAM bounds} follows
immediately from Lemma \ref{lem:pointwise bound remove one} by applying
it $M-2$ times and obtaining a chain of inequalities, starting from
$\phi_{X}(y;\gamma)$ and ending in $\phi_{B_{J}}\left(y;\gamma\right)$.

\section{\label{sec:PAM high SNR char proofs}Proof of Theorems \ref{thm:PAM bounds}
and \ref{thm:PAM d bounds}}

\subsection{\label{sub:PAM mmse low proof}Lower bound on $\mathrm{mmse}_{d,M\textrm{-PAM}}\left(\gamma\right)$}

Using the notation of Section \ref{sec:mmse bounds}, we have
\begin{equation}
\mathrm{mmse}_{d,M\textrm{-PAM}}\left(\gamma\right)=\E[\phi_{X}\left(Y_{\gamma};\gamma\right)][Y_{\gamma}]=\sum_{m=1}^{M}\frac{1}{M}\int_{-\infty}^{\infty}\sqrt{\frac{\gamma}{\pi}}e^{-\gamma\nu^{2}}\phi_{X}\left(x_{m}+\nu;\gamma\right)d\nu\label{eq:mmse PAM as phi expectation}
\end{equation}
Using Theorem \ref{thm:pointwise PAM bounds}, we find that for $m<M$
and $\nu\geq0$
\begin{eqnarray}
\phi_{X}\left(x_{m}+\nu;\gamma\right) & \geq & \left(\frac{d}{2}\right)^{2}\phi_{\mathrm{BPSK}}\left(\left(\frac{d}{2}\right)^{-1}\left[\frac{x_{m}-x_{m+1}}{2}+\nu\right];\left(\frac{d}{2}\right)^{2}\gamma\right)\nonumber \\
 & = & \left(\frac{d}{2}\right)^{2}\phi_{\mathrm{BPSK}}\left(-1+\left(\frac{d}{2}\right)^{-1}\nu;\left(\frac{d}{2}\right)^{2}\gamma\right)\label{eq:phiX_bpsk_bound}
\end{eqnarray}
for every $\nu\in\mathbb{R}$. Writing $\rho=\left(d/2\right)^{2}\gamma$,
and integrating the above inequality, we have
\begin{alignat}{1}
\int_{0}^{\infty}\sqrt{\frac{\gamma}{\pi}}e^{-\gamma\nu^{2}} & \phi_{X}\left(x_{m}+\nu;\gamma\right)d\nu\geq\left(\frac{d}{2}\right)^{2}\int_{0}^{\infty}\sqrt{\frac{\rho}{\pi}}e^{-\rho\nu^{2}}\phi_{\mathrm{BPSK}}\left(-1+\nu;\rho\right)d\nu\nonumber \\
= & \left(\frac{d}{2}\right)^{2}\mathrm{mmse}_{\mathrm{BPSK}}\left(\rho\right)-\left(\frac{d}{2}\right)^{2}\int_{-\infty}^{0}\sqrt{\frac{\rho}{\pi}}e^{-\rho\nu^{2}}\phi_{\mathrm{BPSK}}\left(-1+\nu;\rho\right)d\nu\label{eq:phiX integral upper bound}
\end{alignat}
where the first transition follows from applying (\ref{eq:phiX_bpsk_bound}),
scaling the integration variable by $d/2$ and using $\rho=\left(d/2\right)^{2}\gamma$.

Since $\phi_{\mathrm{BPSK}}\left(y;\rho\right)=\left[\cosh\left(2y\rho\right)\right]^{-2}\leq4e^{-4\left|y\right|\rho}$,
we have the following upper bound
\begin{equation}
\int_{-\infty}^{0}\sqrt{\frac{\rho}{\pi}}e^{-\rho\nu^{2}}\phi_{\mathrm{BPSK}}\left(-1+\nu;\rho\right)d\nu\leq4\int_{-\infty}^{0}\sqrt{\frac{\rho}{\pi}}e^{-\rho\left(\nu-2\right)^{2}}d\nu=4Q\left(\sqrt{8\rho}\right)=\underline{B}\left(\rho\right)
\end{equation}
where $Q\left(\cdot\right)$ is the standard error function (\ref{eq:Qfunc_def}).
Therefore,
\begin{equation}
\int_{-\infty}^{\infty}\sqrt{\frac{\gamma}{\pi}}e^{-\gamma\nu^{2}}\phi_{X}\left(x_{m}+\nu;\gamma\right)d\nu\geq\left(\frac{d}{2}\right)^{2}\mathrm{mmse}_{\mathrm{BPSK}}\left(\left(\frac{d}{2}\right)^{2}\gamma\right)-\underline{B}\left(\left(\frac{d}{2}\right)^{2}\gamma\right)
\end{equation}
for every $m<M$. Similarly, for every $m>1$ and every $\nu\leq0$,
\begin{equation}
\phi_{X}\left(x_{m}+\nu;\gamma\right)\geq\left(\frac{d}{2}\right)^{2}\phi_{\mathrm{BPSK}}\left(\left(\frac{d}{2}\right)^{-1}\left[\frac{x_{m}-x_{m-1}}{2}+\nu\right];\left(\frac{d}{2}\right)^{2}\gamma\right)
\end{equation}
 and so
\begin{equation}
\int_{-\infty}^{0}\sqrt{\frac{\gamma}{\pi}}e^{-\gamma\nu^{2}}\phi_{X}\left(x_{m}+\nu;\gamma\right)d\nu\geq\left(\frac{d}{2}\right)^{2}\mathrm{mmse}_{\mathrm{BPSK}}\left(\left(\frac{d}{2}\right)^{2}\gamma\right)-\underline{B}\left(\left(\frac{d}{2}\right)^{2}\gamma\right)
\end{equation}

Consequently, we find that
\begin{equation}
\int_{-\infty}^{\infty}\sqrt{\frac{\gamma}{\pi}}e^{-\gamma\nu^{2}}\phi_{X}\left(x_{m}+\nu;\gamma\right)d\nu\geq2\left(\frac{d}{2}\right)^{2}\mathrm{mmse}_{\mathrm{BPSK}}\left(\left(\frac{d}{2}\right)^{2}\gamma\right)-2\underline{B}\left(\left(\frac{d}{2}\right)^{2}\gamma\right)\:\forall1<m<M\label{eq:PAM phix int lower bound}
\end{equation}
while for $m=1$ and $m=M$ it is easily seen that
\begin{equation}
\int_{-\infty}^{\infty}\sqrt{\frac{\gamma}{\pi}}e^{-\gamma\nu^{2}}\phi_{X}\left(x_{m}+\nu;\gamma\right)d\nu\geq\left(\frac{d}{2}\right)^{2}\mathrm{mmse}_{\mathrm{BPSK}}\left(\left(\frac{d}{2}\right)^{2}\gamma\right)\label{eq:PAM phix2 int lower bound}
\end{equation}
Substituting back to (\ref{eq:mmse PAM as phi expectation}), we find
that, 
\begin{equation}
\mathrm{mmse}_{d,M\textrm{-PAM}}\left(\gamma\right)\geq2\frac{M-1}{M}\left(\frac{d}{2}\right)^{2}\left[\mathrm{mmse}_{\mathrm{BPSK}}\left(\left(\frac{d}{2}\right)^{2}\gamma\right)-\underline{B}\left(\left(\frac{d}{2}\right)^{2}\gamma\right)\right]
\end{equation}
as required.

\subsection{\label{sub:PAM dmmse up proof}Upper bound on $\mathrm{mmse}_{d,M\textrm{-PAM}}'\left(\gamma\right)$}

Similarly to (\ref{eq:mmse PAM as phi expectation}), we have
\begin{equation}
\E[\phi_{X}^{2}\left(Y_{\gamma};\gamma\right)][Y_{\gamma}]=\sum_{m=1}^{M}\frac{1}{M}\int_{-\infty}^{\infty}\sqrt{\frac{\gamma}{\pi}}e^{-\gamma\nu^{2}}\phi_{X}^{2}\left(x_{m}+\nu;\gamma\right)d\nu
\end{equation}
Thus, the upper bound on $\mathrm{mmse}_{d,M\textrm{-PAM}}'\left(\gamma\right)=-2\E[\phi_{X}^{2}\left(Y_{\gamma};\gamma\right)][Y_{\gamma}]$
is obtained by applying the procedure of \ref{sub:PAM mmse low proof}
on $\phi_{X}^{2}$. In particular, similarly to (\ref{eq:PAM phix int lower bound}),
for $1<m<M$ we have
\begin{alignat}{1}
2\int_{-\infty}^{\infty} & \sqrt{\frac{\gamma}{\pi}}e^{-\gamma\nu^{2}}\phi_{X}^{2}\left(x_{m}+\nu;\gamma\right)\nu\nonumber \\
 & \geq-2\left(\frac{d}{2}\right)^{4}\mathrm{mmse}_{\mathrm{BPSK}}'\left(\left(\frac{d}{2}\right)^{2}\gamma\right)-2\left(\frac{d}{2}\right)^{4}\int_{-\infty}^{0}\sqrt{\frac{\rho}{\pi}}e^{-\rho\nu^{2}}\phi_{\mathrm{BPSK}}^{2}\left(-1+\nu;\rho\right)d\nu
\end{alignat}
with $\rho=\left(d/2\right)^{2}\gamma$. Using $\phi_{\mathrm{BPSK}}^{2}\left(y;\rho\right)=\left[\cosh\left(2y\rho\right)\right]^{-4}\leq16e^{-8\left|y\right|\rho}$
we find that
\begin{equation}
2\int_{-\infty}^{0}\sqrt{\frac{\rho}{\pi}}e^{-\rho\nu^{2}}\phi_{\mathrm{BPSK}}^{2}\left(-1+\nu;\rho\right)d\nu\leq32e^{8\rho}\int_{-\infty}^{0}\sqrt{\frac{\rho}{\pi}}e^{-\rho\left(\nu-4\right)^{2}}d\nu=32e^{8\rho}Q\left(\sqrt{32\rho}\right)=\bar{C}\left(\rho\right)
\end{equation}
where $Q\left(\cdot\right)$ is the standard error function (\ref{eq:Qfunc_def}).
Moreover, similarly to (\ref{eq:PAM phix2 int lower bound}), for
$m=1$ and $m=M$, we have
\begin{equation}
2\int_{-\infty}^{\infty}\sqrt{\frac{\gamma}{\pi}}e^{-\gamma\nu^{2}}\phi_{X}^{2}\left(x_{m}+\nu;\gamma\right)d\nu\geq-\left(\frac{d}{2}\right)^{4}\mathrm{mmse}_{\mathrm{BPSK}}'\left(\left(\frac{d}{2}\right)^{2}\gamma\right)
\end{equation}
We thus conclude that,
\begin{equation}
\mathrm{mmse}_{d,M\textrm{-PAM}}'\left(\gamma\right)\leq2\frac{M-1}{M}\left(\frac{d}{2}\right)^{4}\left[\mathrm{mmse}_{\mathrm{BPSK}}'\left(\left(\frac{d}{2}\right)^{2}\gamma\right)+\bar{C}\left(\left(\frac{d}{2}\right)^{2}\gamma\right)\right]
\end{equation}

\subsection{\label{sub:PAM dmmse low proof}Lower bound on $\mathrm{mmse}_{d,M\textrm{-PAM}}'\left(\gamma\right)$}

`

We apply the pointwise upper bound of theorem \ref{thm:pointwise PAM bounds}
to obtain (similarly to (\ref{eq:phiX_bpsk_bound})),
\begin{equation}
\frac{\phi_{X}\left(x_{m}+\nu;\gamma\right)}{\left(d/2\right)^{2}}\leq\bar{D}\left(\left(\frac{d}{2}\right)^{2}\gamma\right)+\begin{cases}
\phi_{\mathrm{BPSK}}\left(-1+\frac{\nu}{d/2};\left(\frac{d}{2}\right)^{2}\gamma\right) & 0\leq\nu\leq d\\
1 & \nu\geq d
\end{cases}
\end{equation}
with $\bar{D}\left(\gamma\right)=4\sum_{k=1}^{\infty}\left(k+1\right)^{2}e^{-4\gamma k^{2}}$
and we have used the fact that $\phi_{\mathrm{BPSK}}\left(y;\rho\right)\leq1$.
Squaring this inequality, we have
\begin{equation}
\frac{\phi_{X}^{2}\left(x_{m}+\nu;\gamma\right)}{\left(d/2\right)^{4}}\leq c\left(\left(\frac{d}{2}\right)^{2}\gamma\right)+\begin{cases}
\phi_{\mathrm{BPSK}}^{2}\left(-1+\frac{\nu}{d/2};\left(\frac{d}{2}\right)^{2}\gamma\right) & 0\leq\nu\leq d\\
1 & \nu\geq d
\end{cases}\label{eq:phiX square upper bound}
\end{equation}
where $c\left(\rho\right)=2\bar{D}\left(\rho\right)+\bar{D}^{2}\left(\rho\right)$.
Letting $\rho=\left(d/2\right)^{2}\gamma$, we have for $m<M$,
\begin{alignat}{1}
\left(\frac{d}{2}\right)^{-4}\int_{0}^{\infty}\sqrt{\frac{\gamma}{\pi}}e^{-\gamma\nu^{2}} & \phi_{X}^{2}\left(x_{m}+\nu;\gamma\right)d\nu\leq\nonumber \\
 & \int_{0}^{2}\sqrt{\frac{\rho}{\pi}}e^{-\rho\nu^{2}}\phi_{\mathrm{BPSK}}^{2}\left(-1+\nu;\rho\right)d\nu+\int_{2}^{\infty}\sqrt{\frac{\rho}{\pi}}e^{-\rho\nu^{2}}d\nu+c\left(\rho\right)\leq\nonumber \\
 & \int_{-\infty}^{\infty}\sqrt{\frac{\rho}{\pi}}e^{-\rho\nu^{2}}\phi_{\mathrm{BPSK}}^{2}\left(\nu;\rho\right)d\nu+Q\left(\sqrt{8\rho}\right)+c\left(\rho\right)=\nonumber \\
 & -\frac{1}{2}\mathrm{mmse}_{\mathrm{BPSK}}'\left(\rho\right)+\frac{1}{2}\underline{C}\left(\rho\right)
\end{alignat}
with $\underline{C}\left(\rho\right)=2\left[c\left(\rho\right)+Q\left(\sqrt{8\rho}\right)\right]$
and $Q\left(\cdot\right)$ the standard error function (\ref{eq:Qfunc_def}).
The first transition in the above equation follows from integrating
(\ref{eq:phiX square upper bound}) and scaling the integration variable
by $d/2$ as in (\ref{eq:phiX integral upper bound}). The second
transition is obtained by extending the integration limits of the
first term, and evaluating the integral in the second term. Similarly,
for $m>1$ we have
\begin{equation}
\left(\frac{d}{2}\right)^{-4}\int_{-\infty}^{0}\sqrt{\frac{\gamma}{\pi}}e^{-\gamma\nu^{2}}\phi_{X}^{2}\left(x_{m}+\nu;\gamma\right)d\nu\leq-\frac{1}{2}\mathrm{mmse}_{\mathrm{BPSK}}'\left(\rho\right)+\frac{1}{2}\underline{C}\left(\rho\right)
\end{equation}
and for $m=1,M$ it is simple to show that
\begin{equation}
\left(\frac{d}{2}\right)^{-4}\int_{-\infty}^{\infty}\sqrt{\frac{\gamma}{\pi}}e^{-\gamma\nu^{2}}\phi_{X}^{2}\left(x_{m}+\nu;\gamma\right)d\nu\leq-\frac{1}{2}\mathrm{mmse}_{\mathrm{BPSK}}'\left(\rho\right)+\frac{1}{2}\underline{C}\left(\rho\right)\:,\: m=1,M
\end{equation}
Therefore
\begin{alignat}{1}
\mathrm{mmse}_{d,M\textrm{-PAM}}'\left(\gamma\right) & =-2\sum_{m=1}^{M}\frac{1}{M}\int_{-\infty}^{\infty}\sqrt{\frac{\gamma}{\pi}}e^{-\gamma\nu^{2}}\phi_{X}^{2}\left(x_{m}+\nu;\gamma\right)d\nu\nonumber \\
 & \geq2\frac{M-1}{M}\left(\frac{d}{2}\right)^{4}\left[\mathrm{mmse}_{\mathrm{BPSK}}'\left(\left(\frac{d}{2}\right)^{2}\gamma\right)-\underline{C}\left(\left(\frac{d}{2}\right)^{2}\gamma\right)\right]
\end{alignat}

\subsection{\texttt{\label{sub:PAM mmse up proof}}Upper bound on $\mathrm{mmse}_{d,M\textrm{-PAM}}\left(\gamma\right)$}

The upper bound on $\mathrm{mmse}_{d,M\textrm{-PAM}}\left(\gamma\right)$
may be derived in the same way as the lower bound on $\mathrm{mmse}_{d,M\textrm{-PAM}}'\left(\gamma\right)$.
However, we will take a slightly different approach in order to obtain
a better expression for the slackness term $\bar{B}\left(\gamma\right)$.
Let $\tilde{\emph{s}}\left(y\right)$ be the sub-optimal estimator
for $X$ that assumes $X$ is uniformly distributed on the two nearest
neighbors to $y$ in $\mathcal{X}.$ We have
\begin{alignat}{1}
\mathrm{mmse}_{d,M\textrm{-PAM}}\left(\gamma\right) & =\E[\left(X-\Ebrack{X|Y_{\gamma}}{}\right)^{2}]\nonumber \\
 & \leq\E[\left(X-\tilde{s}\left(Y_{\gamma}\right)\right)^{2}]=\sum_{m=1}^{M}\frac{1}{M}\int_{-\infty}^{\infty}\sqrt{\frac{\gamma}{\pi}}e^{-\gamma\nu^{2}}\left(x_{m}-\tilde{\emph{s}}\left(x_{m}+\nu\right)\right)^{2}d\nu
\end{alignat}
For convenience denote $x_{M+1}\equiv\infty$ and $x_{0}=-\infty$.
We observe that for any $m'\geq m$ and any $x_{m'}\leq\nu\leq x_{m'+1}$,
\begin{equation}
\left(x_{m}-\tilde{\emph{s}}\left(x_{m}+\nu\right)\right)^{2}\leq d^{2}\left(m'-m+1\right)^{2}
\end{equation}
Therefore, for any $m<M$,
\begin{alignat}{1}
\int_{0}^{\infty}\sqrt{\frac{\gamma}{\pi}} & e^{-\gamma\nu^{2}}\left(x_{m}-\tilde{\emph{s}}\left(x_{m}+\nu\right)\right)^{2}d\nu\leq\int_{0}^{d}\sqrt{\frac{\gamma}{\pi}}e^{-\gamma\nu^{2}}\left(x_{m}-\tilde{\emph{s}}\left(x_{m}+\nu\right)\right)^{2}d\nu\nonumber \\
 & +\sum_{m'=m+1}^{M}d^{2}\left(m'-m+1\right)^{2}\int_{x_{m'}-x_{m}}^{x_{m'+1}-x_{m}}\sqrt{\frac{\gamma}{\pi}}e^{-\gamma\nu^{2}}d\nu
\end{alignat}
The first term is clearly upper bounded by $\left(\frac{d}{2}\right)^{2}\mathrm{mmse}_{\mathrm{BPSK}}\left(\left(\frac{d}{2}\right)^{2}\gamma\right)$:
\begin{alignat}{1}
\int_{0}^{d}\sqrt{\frac{\gamma}{\pi}}e^{-\gamma\nu^{2}} & \left(x_{m}-\tilde{\emph{s}}\left(x_{m}+\nu\right)\right)^{2}d\nu=\left(\frac{d}{2}\right)^{2}\int_{0}^{d}\sqrt{\frac{\gamma}{\pi}}e^{-\gamma\nu^{2}}\phi_{\mathrm{BPSK}}\left(-1+\left(\frac{d}{2}\right)^{-1}\nu;\left(\frac{d}{2}\right)^{2}\gamma\right)d\nu\nonumber \\
\leq & \left(\frac{d}{2}\right)^{2}\int_{-\infty}^{\infty}\sqrt{\frac{\left(d/2\right)^{2}\gamma}{\pi}}e^{-\left(d/2\right)^{2}\gamma\nu^{2}}\phi_{\mathrm{BPSK}}\left(-1+\nu;\left(\frac{d}{2}\right)^{2}\gamma\right)d\nu\nonumber \\
= & \left(\frac{d}{2}\right)^{2}\mathrm{mmse}_{\mathrm{BPSK}}\left(\left(\frac{d}{2}\right)^{2}\gamma\right)
\end{alignat}
The second term can be upper bounded as follows
\begin{alignat}{1}
\sum_{m'=m+1}^{M} & d^{2}\left(m'-m+1\right)^{2}\int_{x_{m'}-x_{m}}^{x_{m'+1}-x_{m}}\sqrt{\frac{\gamma}{\pi}}e^{-\gamma\nu^{2}}d\nu\nonumber \\
= & \sum_{k=1}^{M-m-1}d^{2}\left(k+1\right)^{2}\int_{kd}^{\left(k+1\right)d}\sqrt{\frac{\gamma}{\pi}}e^{-\gamma\nu^{2}}d\nu+d^{2}\left(M-m+1\right)^{2}\int_{\left(M-m\right)d}^{\infty}\sqrt{\frac{\gamma}{\pi}}e^{-\gamma\nu^{2}}d\nu\nonumber \\
= & \sum_{k=1}^{M-m-1}d^{2}\left(k+1\right)^{2}\left[Q\left(kd\sqrt{2\gamma}\right)-Q\left(\left(k+1\right)d\sqrt{2\gamma}\right)\right]+d^{2}\left(M-m+1\right)^{2}Q\left(\left(M-m\right)d\sqrt{2\gamma}\right)\nonumber \\
= & 4d^{2}Q\left(\sqrt{2d^{2}\gamma}\right)+d^{2}\sum_{k=2}^{M-m}\left(2k+1\right)Q\left(k\sqrt{2d^{2}\gamma}\right)\nonumber \\
\leq & 4d^{2}Q\left(\sqrt{2d^{2}\gamma}\right)+d^{2}\sum_{k=2}^{\infty}\left(2k+1\right)Q\left(k\sqrt{2d^{2}\gamma}\right)=\left(\frac{d}{2}\right)^{2}\bar{B}\left(\left(\frac{d}{2}\right)^{2}\gamma\right)
\end{alignat}
where $Q\left(\cdot\right)$ is the standard error function (\ref{eq:Qfunc_def}).
This upper bound can be slightly relaxed to obtain a more manageable
expression, using the inequality $\sqrt{2\pi}xQ\left(x\right)\leq e^{-x^{2}/2}$:
\begin{eqnarray}
\bar{B}\left(\rho\right) & \leq & \frac{16}{\sqrt{16\pi\emph{\ensuremath{\rho}}}}e^{-4\emph{\ensuremath{\rho}}}+\frac{4}{\sqrt{16\pi\emph{\ensuremath{\rho}}}}\sum_{k=2}^{\infty}\frac{2k+1}{k}e^{-4k^{2}\emph{\ensuremath{\rho}}}\nonumber \\
 & \leq & \frac{1}{\sqrt{\pi\emph{\ensuremath{\rho}}}}\left(4e^{-4\emph{\ensuremath{\rho}}}+\frac{5}{2}e^{-16\emph{\ensuremath{\rho}}}\sum_{k=0}^{\infty}e^{-4k\left(k+4\right)\emph{\ensuremath{\rho}}}\right)\nonumber \\
 & \leq & \frac{1}{\sqrt{\pi\emph{\ensuremath{\rho}}}}\left(4e^{-4\emph{\ensuremath{\rho}}}+\frac{5}{2}e^{-16\emph{\ensuremath{\rho}}}\sum_{k=0}^{\infty}e^{-20k\emph{\ensuremath{\rho}}}\right)=\frac{1}{2\sqrt{\pi\emph{\ensuremath{\rho}}}}\left(8e^{-4\emph{\ensuremath{\rho}}}+5\frac{e^{-16\emph{\ensuremath{\rho}}}}{1-e^{-20\rho}}\right)
\end{eqnarray}
where we used $\left(2k+1\right)/k\leq5/2$ for every $k\geq2$ and
$4k\left(k+4\right)\geq20k$ for every $k\geq0$. We conclude that,
with $\rho=\left(d/2\right)^{2}\gamma$,
\begin{equation}
\int_{0}^{\infty}\sqrt{\frac{\gamma}{\pi}}e^{-\gamma\nu^{2}}\left(x_{m}-\tilde{\emph{s}}\left(x_{m}+\nu\right)\right)^{2}d\nu\leq\left(\frac{d}{2}\right)^{2}\left[\mathrm{mmse}_{\mathrm{BPSK}}\left(\rho\right)+\bar{B}\left(\rho\right)\right]\:,\:\forall m<M
\end{equation}
and it may similarly be shown that,
\begin{equation}
\int_{-\infty}^{0}\sqrt{\frac{\gamma}{\pi}}e^{-\gamma\nu^{2}}\left(x_{m}-\tilde{\emph{s}}\left(x_{m}+\nu\right)\right)^{2}d\nu\leq\left(\frac{d}{2}\right)^{2}\left[\mathrm{mmse}_{\mathrm{BPSK}}\left(\rho\right)+\bar{B}\left(\rho\right)\right]\:,\:\forall m>1
\end{equation}
It is also simple to show that for $m=1,M$, 
\begin{equation}
\int_{-\infty}^{\infty}\sqrt{\frac{\gamma}{\pi}}e^{-\gamma\nu^{2}}\left(x_{m}-\tilde{\emph{s}}\left(x_{m}+\nu\right)\right)^{2}d\nu\leq\left(\frac{d}{2}\right)^{2}\left[\mathrm{mmse}_{\mathrm{BPSK}}\left(\rho\right)+\bar{B}\left(\rho\right)\right]\:,\:\forall m=1,M
\end{equation}
and so
\begin{alignat}{1}
\mathrm{mmse}_{d,M\textrm{-PAM}}\left(\gamma\right) & \leq\sum_{m=1}^{M}\frac{1}{M}\int_{-\infty}^{\infty}\sqrt{\frac{\gamma}{\pi}}e^{-\gamma\nu^{2}}\left(x_{m}-\tilde{\emph{s}}\left(x_{m}+\nu\right)\right)^{2}d\nu\nonumber \\
 & \leq2\frac{M-1}{M}\left(\frac{d}{2}\right)^{2}\left[\mathrm{mmse}_{\mathrm{BPSK}}\left(\left(\frac{d}{2}\right)^{2}\gamma\right)+\bar{B}\left(\left(\frac{d}{2}\right)^{2}\gamma\right)\right]
\end{alignat}

\section{\label{sec:BPSK bounds proof}Proof of Theorem \ref{thm:BPSK bounds}}

Using $\phi_{\mathrm{BPSK}}(y;\gamma)=1/\cosh^{2}\left(2\gamma y\right)$,
we find that
\begin{eqnarray}
\mathrm{mmse}_{\mathrm{BPSK}}\left(\gamma\right) & = & \sqrt{\frac{\gamma}{\pi}}\int_{-\infty}^{\infty}\phi_{\mathrm{BPSK}}(y;\gamma)\left(\frac{e^{-\gamma\left(y-1\right)^{2}}+e^{-\gamma\left(y+1\right)^{2}}}{2}\right)dy\nonumber \\
 & = & \sqrt{\frac{\gamma}{\pi}}e^{-\gamma}\int_{-\infty}^{\infty}\frac{1}{\cosh\left(2\gamma y\right)}e^{-\gamma y^{2}}dy\nonumber \\
 & = & \frac{1}{\sqrt{\pi\gamma}}e^{-\gamma}\int_{-\infty}^{\infty}\frac{1}{\cosh\left(2z\right)}e^{-\frac{z^{2}}{\gamma}}dz
\end{eqnarray}
and (\ref{eq:BPSK MMSE asymptotic bound}) is readily found by substituting
$1-\frac{z^{2}}{\gamma}\leq e^{-\frac{z^{2}}{\gamma}}\leq1$ and integrating.
Note that by substituting $e^{-z^{2}/\gamma}=\sum_{k=0}^{\infty}\frac{1}{k!}\left(-z^{2}/\gamma\right)^{k}$,
the high-SNR asymptotic expansion of $\mathrm{mmse}_{\mathrm{BPSK}}\left(\gamma\right)$
is obtained. A different change of variables yields the equality,
\begin{equation}
\mathrm{mmse}_{\mathrm{BPSK}}\left(\gamma\right)=\frac{1}{\sqrt{\pi}}e^{-\gamma}\int_{-\infty}^{\infty}\frac{1}{\cosh\left(2\sqrt{\gamma}z\right)}e^{-z^{2}}dz
\end{equation}
and substituting $1\leq\cosh\left(2\sqrt{\gamma}z\right)\leq e^{2\gamma z^{2}}$
yields the bounds in (\ref{eq:BPSK MMSE lowsnr bound}). Since $\mathrm{mmse}_{\mathrm{BPSK}}'\left(\gamma\right)=-2\E[\phi_{\mathrm{BPSK}}^{2}\left(Y_{\gamma};\gamma\right)][Y_{\gamma}]$,
we find the bounds for $\mathrm{mmse}_{\mathrm{BPSK}}'\left(\gamma\right)$
by replacing $\cosh\left(\cdot\right)$ with $\cosh^{3}\left(\cdot\right)$
in the derivations above.

\section{\label{sec:uniform-convex-proof}Proof of Proposition \ref{prop:uniform-convex}}

Let $X$ be a real-valued RV uniformly distributed in $\left[-A/2,A/2\right]$,
and using the notation of Section \ref{sec:mmse bounds} let $Y_{\gamma}=X+\frac{1}{\sqrt{\gamma}}N$
with $N\sim\mathcal{N}(0,\frac{1}{2})$ and independent of $X$. Using
the orthogonality principle and considering the measurement $Y_{\gamma}$
as a suboptimal estimator, we may express the MMSE as
\begin{equation}
\mathrm{mmse}_{X}\left(\gamma\right)=\frac{1}{2\gamma}-\E[\left(Y_{\gamma}-\Ebrack{X|Y_{\gamma}}{}\right)]^{2}
\end{equation}

Straightforward calculation of $\E[\left(Y_{\gamma}-\Ebrack{X|Y_{\gamma}}{}\right)]^{2}$
shows that we may write
\begin{equation}
\mathrm{mmse}_{X}\left(\gamma\right)=\frac{1}{2\gamma}\left(1-\int_{-\infty}^{\infty}g(y;\gamma)dy\right)\label{eq:uniform-mmse-gform}
\end{equation}
with
\begin{equation}
g\left(y;\gamma\right)\triangleq\frac{1}{2\pi A}\frac{\left(e^{-\gamma\left(y-A/2\right)^{2}}-e^{-\gamma\left(y+A/2\right)^{2}}\right)^{2}}{Q\left(\sqrt{2\gamma}\left[y-A/2\right]\right)-Q\left(\sqrt{2\gamma}\left[y+A/2\right]\right)}
\end{equation}
and with the error function $Q\left(\cdot\right)$ defined in (\ref{eq:Qfunc_def}).

Differentiating (\ref{eq:uniform-mmse-gform}), we have 
\begin{equation}
\mathrm{mmse}_{X}'\left(\gamma\right)=-\frac{1}{\gamma}\mathrm{mmse}_{X}\left(\gamma\right)+\frac{1}{2\gamma}\int_{-\infty}^{\infty}\left[h_{1}\left(y;\gamma\right)-h_{2}\left(y;\gamma\right)\right]dy
\end{equation}
where
\begin{equation}
h_{1}\left(y;\gamma\right)\triangleq\frac{1}{\pi A}\frac{\left(\left(y-A/2\right)^{2}e^{-\gamma\left(y-A/2\right)^{2}}-\left(y+A/2\right)^{2}e^{-\gamma\left(y+A/2\right)^{2}}\right)\left(e^{-\gamma\left(y-A/2\right)^{2}}-e^{-\gamma\left(y+A/2\right)^{2}}\right)}{Q\left(\sqrt{2\gamma}\left[y-A/2\right]\right)-Q\left(\sqrt{2\gamma}\left[y+A/2\right]\right)}
\end{equation}
and
\begin{equation}
h_{2}\left(y;\gamma\right)\triangleq\frac{1}{4A\pi\sqrt{\pi\gamma}}\frac{\left(\left(y-A/2\right)e^{-\gamma\left(y-A/2\right)^{2}}-\left(y+A/2\right)e^{-\gamma\left(y+A/2\right)^{2}}\right)\left(e^{-\gamma\left(y-A/2\right)^{2}}-e^{-\gamma\left(y+A/2\right)^{2}}\right)^{2}}{\left(Q\left(\sqrt{2\gamma}\left[y-A/2\right]\right)-Q\left(\sqrt{2\gamma}\left[y+A/2\right]\right)\right)^{2}}
\end{equation}
For $\left|y\right|\leq A/2$ we have $h_{2}\left(y;\gamma\right)\leq0$.
For $y>A/2$ we find that
\begin{eqnarray}
h_{2}\left(y;\gamma\right) & \leq & \frac{1}{4A\pi\sqrt{\pi\gamma}}\frac{\left(y-A/2\right)e^{-3\gamma\left(y-A/2\right)^{2}}}{Q\left(\sqrt{2\gamma}\left[y-A/2\right]\right)^{2}}
\end{eqnarray}
where we have used
\begin{equation}
\frac{e^{-\gamma\left(y-A/2\right)^{2}}-e^{-\gamma\left(y+A/2\right)^{2}}}{Q\left(\sqrt{2\gamma}\left[y-A/2\right]\right)-Q\left(\sqrt{2\gamma}\left[y+A/2\right]\right)}\leq\frac{e^{-\gamma\left(y-A/2\right)^{2}}}{Q\left(\sqrt{2\gamma}\left[y-A/2\right]\right)}
\end{equation}
for every $\gamma$, $A$ and $y$. Integrating, we have
\begin{alignat}{1}
\int_{-\infty}^{\infty}h_{2}\left(y;\gamma\right)dy & =2\int_{0}^{\infty}h_{2}\left(y;\gamma\right)dz\leq2\int_{A/2}^{\infty}h_{2}\left(y;\gamma\right)dz=\frac{c_{2}}{2A\pi\sqrt{\pi}\gamma\sqrt{\gamma}}\label{eq:unif-h2-bound}
\end{alignat}
where
\begin{equation}
c_{2}=\int_{0}^{\infty}\frac{xe^{-3x^{2}}}{Q\left(\sqrt{2}x\right)^{2}}dx\approx10.6\label{eq:unif-c2}
\end{equation}

Turning to $h_{1}$, we find that for $y>0$,
\begin{alignat}{1}
h_{1}\left(y;\gamma\right) & \geq\frac{1}{\pi A}\frac{\left(y-A/2\right)^{2}e^{-2\gamma\left(y-A/2\right)^{2}}-\left[\left(y-A/2\right)^{2}+\left(y+A/2\right)^{2}\right]e^{-\gamma\left(y-A/2\right)^{2}}e^{-\gamma\left(y+A/2\right)^{2}}}{Q\left(\sqrt{2\gamma}\left[y-A/2\right]\right)-Q\left(\sqrt{2\gamma}\left[y+A/2\right]\right)}\nonumber \\
\geq & \frac{1}{\pi A\gamma}\frac{\gamma\left(y-A/2\right)^{2}e^{-2\gamma\left(y-A/2\right)^{2}}}{Q\left(\sqrt{2\gamma}\left[y-A/2\right]\right)}-\frac{1}{\pi A}\frac{\left(A^{2}/2+2y^{2}\right)e^{-2\gamma y^{2}}}{Q\left(\sqrt{2\gamma}\left[y-A/2\right]\right)-Q\left(\sqrt{2\gamma}\left[y+A/2\right]\right)}e^{-A^{2}\gamma/2}
\end{alignat}
 Therefore,
\begin{equation}
\int_{-\infty}^{\infty}h_{1}\left(y;\gamma\right)dy=2\int_{0}^{\infty}h_{1}\left(y;\gamma\right)dz\geq\frac{2c_{1}}{\pi A\gamma\sqrt{\gamma}}-\left[\frac{A}{\pi\sqrt{\gamma}}K_{0}\left(A\right)+\frac{4}{\pi A\gamma\sqrt{\gamma}}K_{2}\left(A\right)\right]e^{-A^{2}\gamma/2}\label{eq:unif-h1-bound}
\end{equation}
with
\begin{equation}
c_{1}=\int_{0}^{\infty}\frac{x^{2}e^{-2x{}^{2}}}{Q\left(\sqrt{2}x\right)}dz\approx2.26\label{eq:unif-c1}
\end{equation}
and
\begin{eqnarray}
K_{i}\left(A\right) & = & \int_{0}^{\infty}\frac{x^{i}e^{-2x{}^{2}}dx}{Q\left(\sqrt{2}\left(x-A/2\right)\right)-Q\left(\sqrt{2}\left(x+A/2\right)\right)}
\end{eqnarray}
Putting the bounds together, and simplifying the exponential term
by assuming $\gamma>1$,
\begin{equation}
\mathrm{mmse}_{X}'\left(\gamma\right)\geq-\frac{1}{\gamma}\mathrm{mmse}_{X}\left(\gamma\right)+\frac{c_{0}}{\pi A\gamma^{2}\sqrt{\gamma}}-k\left(A\right)e^{-A^{2}\gamma/2}\label{eq:unif-dmmse-bound}
\end{equation}
where
\begin{equation}
c_{0}=c_{1}-\frac{1}{4\sqrt{\pi}}c_{2}\approx0.77
\end{equation}
and 
\begin{equation}
k\left(A\right)=\frac{A}{\pi}K_{0}\left(A\right)+\frac{4}{\pi A}K_{2}\left(A\right)
\end{equation}
Consequently,
\begin{equation}
\mathrm{mmse}_{X}\left(\gamma\right)+\left(1+\gamma\right)\mathrm{mmse}_{X}'\left(\gamma\right)\geq\frac{c_{0}}{\pi A\gamma\sqrt{\gamma}}-\frac{1}{2\gamma^{2}}-k\left(A\right)e^{-A^{2}\gamma/2}\label{eq:unif-ddIlog-bound}
\end{equation}
for $\gamma>1$, where we have used (\ref{eq:unif-dmmse-bound}) along
with $\mathrm{mmse}_{X}\left(\gamma\right)\leq1/2\gamma$ which is
true for any input.

Let $X$ be the in-phase or quadrature component of a unit power $\infty$-QAM
input, so that $A=\sqrt{6}$ ($X$ has variance $1/2$), and
\begin{equation}
\mathrm{mmse}_{\infty\textrm{-QAM}}\left(\gamma\right)=2\mathrm{mmse}_{X}\left(\gamma\right)
\end{equation}
Therefore, using (\ref{eq:ddIlog}) and (\ref{eq:unif-ddIlog-bound}),
we find that a sufficient condition for $\Ilog[\infty\textrm{-QAM}]$
to be convex is
\begin{equation}
\frac{c_{0}}{\pi\sqrt{6}\gamma\sqrt{\gamma}}-\frac{1}{2\gamma^{2}}-k\left(\sqrt{6}\right)e^{-3\gamma}\geq0
\end{equation}
As a result, there must exist a value of $\gamma$ above which convexity
holds. Using $k\left(\sqrt{6}\right)\approx0.586$, it is seen that
the above inequality becomes positive for $\gamma>25$, or 14 dB,
and therefore convexity holds above this value. Numerically examining
$\Ilog[\infty\textrm{-QAM}]$ and its derivatives for SNR's below
14 dB, it is seen that the function is concave below $\underline{\gamma}_{0}=8.76\text{ dB}$
and then becomes convex. The above analysis guarantees that $\Ilog[\infty\textrm{-QAM}]$
never becomes concave again at higher SNR's.

We remark that bounds (\ref{eq:unif-h2-bound}) and (\ref{eq:unif-h1-bound})
could have been made tighter by extending the lower integration limit
in (\ref{eq:unif-c2}) and (\ref{eq:unif-c1}) to $-\infty$, at the
cost of adding additional exponential factors.

\section{\label{sec:unif-envelopes-proof}Proof of Proposition \ref{prop:unif-envelopes}}

First, we show that the concave envelope of $\Ilog[\infty\textrm{-QAM}]$
is $\Ihatlog[\infty\textrm{-QAM}]\left(\zeta\right)=\zeta$. Assume
by contradiction that there exists another concave function $\tilde{I}\left(\zeta\right)$
that upper bounds $\Ilog[\infty\textrm{-QAM}]$ and satisfies $\tilde{I}(\zeta_{a})<\zeta_{a}$
for some $\zeta_{c}\geq0$. Since $\tilde{I}\left(0\right)\geq\Ilog[\infty\textrm{-QAM}]\left(0\right)=0$,
we must have $\tilde{I}'(\zeta_{i})<1$ for some $\zeta_{b}\in\left[0,\zeta_{a}\right)$
for $\tilde{I}(\zeta_{a})<\zeta_{a}$ to be possible. By the concavity
of $\tilde{I}$, $\tilde{I}'$ is non-increasing, and hence $\tilde{I}\left(\zeta\right)\leq\tilde{I}(\zeta_{b})+(\zeta-\zeta_{b})\tilde{I}'(\zeta_{b})$
for $\zeta\geq\zeta_{b}$. However, by (\ref{eq:unif_Ilog_lim}) we
clearly have that for any $C\in\mathbb{R}$ and $\alpha<1$, $\Ilog[\infty\textrm{-QAM}]\left(\zeta\right)>C+\alpha\zeta$
for sufficiently high $\zeta$. There must therefore exist $\zeta_{c}\geq0$
such that 
\begin{equation}
\Ilog[\infty\textrm{-QAM}]\left(\zeta_{c}\right)>\tilde{I}(\zeta_{b})+(\zeta_{c}-\zeta_{b})\tilde{I}'(\zeta_{b})\geq\tilde{I}\left(\zeta_{c}\right)
\end{equation}
forming a contradiction. We conclude that the concave envelope satisfies
$\Ihatlog[\infty\textrm{-QAM}]\left(\zeta\right)\geq\zeta$. Clearly,
$\zeta$ is concave and upper bounds $\Ilog[\infty\textrm{-QAM}]$
and therefore $\Ihatlog[\infty\textrm{-QAM}]\left(\zeta\right)=\zeta$. 

For any input distribution, $\dIlog\left(\zeta\right)=\left(1+\gamma\right)\mathrm{\mathrm{mmse}}_{x}(\gamma)\leq1$
. Therefore, $\zeta-\Ilog[\infty\textrm{-QAM}]\left(\zeta\right)$
is an increasing function. Thus, given (\ref{eq:unif_Ilog_lim}) and
the expression for $\Ihatlog[\infty\textrm{-QAM }]$, we may easily
find the maximum difference between it and $\Ilog[\infty\textrm{-QAM}]$,
\begin{equation}
\Delta_{\infty\text{-QAM}}=\sup_{\zeta}\left(\Ihatlog[\infty\textrm{-QAM }]\left(\zeta\right)-\Ilog[\infty\textrm{-QAM}]\left(\zeta\right)\right)=\lim_{\zeta\to\infty}\left(\zeta-\Ilog[\infty\textrm{-QAM}]\left(\zeta\right)\right)=\log\left(\frac{\pi e}{6}\right)
\end{equation}

We now consider the interval $\left[0,\bar{\zeta}\right]$ for some
$\bar{\zeta}>\underline{\zeta}_{0}$. Since the constant function
$\Ilog[\infty\textrm{-QAM}](\bar{\zeta})$ is concave and upper bounds
$\Ilog[\infty\textrm{-QAM}]$ on the interval, we must have $\hat{I}_{\infty\text{-QAM}}^{\log;\left[0,\bar{\zeta}\right]}(\bar{\zeta})\leq\Ilog[\infty\textrm{-QAM}](\bar{\zeta})$.
By definition, $\hat{I}_{\infty\text{-QAM}}^{\log;\left[0,\bar{\zeta}\right]}$
also upper bounds $\Ilog[\infty\textrm{-QAM}]$, and so we must have
$\hat{I}_{\infty\text{-QAM}}^{\log;\left[0,\bar{\zeta}\right]}(\bar{\zeta})=\Ilog[\infty\textrm{-QAM}](\bar{\zeta})$.
However, since $\bar{\zeta}>\underline{\zeta}_{0}$, by Proposition
\ref{prop:uniform-convex} $\Ilog[\infty\textrm{-QAM}]$ is convex
around $\bar{\zeta}$, and therefore $\hat{I}_{\infty\text{-QAM}}^{\log;\left[0,\bar{\zeta}\right]}(\zeta)$
cannot be identical to $\Ilog[\infty\textrm{-QAM}](\zeta)$ in a neighborhood
of $\bar{\zeta}$. Hence, there exists $\underline{\zeta}_{1}$ such
that $\hat{I}_{\infty\text{-QAM}}^{\log;\left[0,\bar{\zeta}\right]}$
is linear on the interval $[\underline{\zeta}_{1},\bar{\zeta}]$ and
that $\hat{I}_{\infty\text{-QAM}}^{\log;\left[0,\bar{\zeta}\right]}(\underline{\zeta}_{1})=\Ilog[\infty\textrm{-QAM}](\underline{\zeta}_{1})$.
By Proposition \ref{prop:uniform-convex}, $\dIlog[\infty\textrm{-QAM}]$
has only a single minimum, located at $\underline{\zeta}_{0}$, below
which $\Ilog[\infty\textrm{-QAM}]$ is concave. As is easily confirmed
from inspection of Figure \ref{fig:unif-ilog}, this implies that
$\underline{\zeta}_{1}<\underline{\zeta}_{0}$ and that $\hat{I}_{\infty\text{-QAM}}^{\log;\left[0,\bar{\zeta}\right]}$
is given by (\ref{eq:unif_concave_envelope_interval}) , since $\hat{I}_{\infty\text{-QAM}}^{\log;\left[0,\bar{\zeta}\right]}$
may be identical to $\Ilog[\infty\textrm{-QAM}]$ in the interval
$[0,\underline{\zeta}_{1}]$, where the latter is concave. Moreover,
$\underline{\zeta}_{1}$ is uniquely determined by (\ref{eq:unif_concave_envelope_interval})
and the condition $\hat{I}_{\infty\text{-QAM}}^{\log;\left[0,\bar{\zeta}\right]}(\bar{\zeta})=\Ilog[\infty\textrm{-QAM}](\bar{\zeta})$.

Since 
\begin{equation}
\frac{d}{d\zeta}\left(\hat{I}_{\infty\text{-QAM}}^{\log;\left[0,\bar{\zeta}\right]}(\zeta)-\Ilog[\infty\textrm{-QAM}](\zeta)\right)=\dIlog[\infty\textrm{-QAM}](\underline{\zeta}_{1})-\dIlog[\infty\textrm{-QAM}](\zeta)
\end{equation}
 for $\zeta\in[\underline{\zeta}_{1},\bar{\zeta}]$, the maximum difference
between $\Ilog[\infty\textrm{-QAM}]$ and its concave envelope on
$\left[0,\bar{\zeta}\right]$ is obtained for $\zeta_{m}$ which satisfies
$\dIlog[\infty\textrm{-QAM}](\zeta_{m})=\dIlog[\infty\textrm{-QAM}](\underline{\zeta}_{1})$
and may therefore be easily found numerically. 

The construction of the convex envelope of $\Ilog[\infty\textrm{-QAM}]$
follows exactly the same lines as the construction of $\hat{I}_{\infty\text{-QAM}}^{\log;\left[0,\bar{\zeta}\right]}$
above. Since the convex function 0 lower bounds $\Ilog[\infty\textrm{-QAM}]$,
we must have $\Ichecklog[\infty\text{-QAM}]\left(0\right)\geq0$.
By definition, $\Ichecklog[\infty\text{-QAM}]$ also lower bounds
$\Ilog[\infty\textrm{-QAM}]$, and so we must have $\Ichecklog[\infty\text{-QAM}]\left(0\right)=0$.
However, by Proposition \ref{prop:uniform-convex} $\Ilog[\infty\textrm{-QAM}]$
is concave around $\zeta=0$, and therefore $\Ichecklog[\infty\text{-QAM}](\zeta)$
cannot be identical to $\Ilog[\infty\textrm{-QAM}](\zeta)$ in a neighborhood
of $0$. Hence, there exists $\tilde{\zeta}_{2}$ such that $\Ichecklog[\infty\text{-QAM}]$
is linear on the interval $[0,\tilde{\zeta}_{2}]$ and that $\Ichecklog[\infty\text{-QAM}](\tilde{\zeta}_{2})=\Ilog[\infty\textrm{-QAM}](\tilde{\zeta}_{2})$.
By Proposition \ref{prop:uniform-convex}, $\dIlog[\infty\textrm{-QAM}]$
has only a single minimum, located at $\underline{\zeta}_{0}$, above
which $\Ilog[\infty\textrm{-QAM}]$ is convex. As is easily confirmed
from inspection of Figure \ref{fig:unif-ilog}, this implies that
$\tilde{\zeta}_{2}>\underline{\zeta}_{0}$ and that $\Ichecklog[\infty\text{-QAM}]$
is given by (\ref{eq:unif_convex_envelope}) , since $\Ichecklog[\infty\text{-QAM}]$
may be identical to $\Ilog[\infty\textrm{-QAM}]$ in the interval
$[\tilde{\zeta}_{2},\infty)$, where the latter is convex. Moreover,
$\tilde{\zeta}_{2}\approx5.52$ {[}bits{]} is uniquely determined
by (\ref{eq:unif_concave_envelope_interval}) and the condition $\Ichecklog[\infty\text{-QAM}](\tilde{\zeta}_{2})=\Ilog[\infty\textrm{-QAM}](\tilde{\zeta}_{2})$.

Since 
\begin{equation}
\frac{d}{d\zeta}\left(\Ilog[\infty\textrm{-QAM}](\zeta)-\Ichecklog[\infty\text{-QAM}](\zeta)\right)=\dIlog[\infty\textrm{-QAM}](\zeta)-\dIlog[\infty\textrm{-QAM}](\tilde{\zeta}_{2})
\end{equation}
 for $\zeta\in[0,\tilde{\zeta}_{2}]$, the maximum difference between
$\Ilog[\infty\textrm{-QAM}]$ and its convex envelope is obtained
for $\tilde{\zeta}_{m}<\tilde{\zeta}_{2}$ which satisfies $\dIlog[\infty\textrm{-QAM}](\zeta_{m})=\dIlog[\infty\textrm{-QAM}](\tilde{\zeta}_{2})$.
Simple numerical computation shows that $\tilde{\zeta}_{m}\approx1.70\text{ [bits]}$
and that
\begin{equation}
\tilde{\Delta}_{\infty\textrm{-QAM}}=\Ilog[\infty\text{-QAM}]\left(\tilde{\zeta}_{m}\right)-\Ichecklog[\infty\text{-QAM}](\tilde{\zeta}_{m})\approx0.0608\text{ [bit]}
\end{equation}

\bibliographystyle{unsrt}
\bibliography{sc-ofdm_paper}

\begin{thebibliography}{10}

\bibitem{proakis1987digital}
J.G. Proakis.
\newblock {\em Digital communications}, volume 1221.
\newblock McGraw-hill, 1987.

\bibitem{forney1972maximum}
G.~Forney~Jr.
\newblock Maximum-likelihood sequence estimation of digital sequences in the
  presence of intersymbol interference.
\newblock {\em Information Theory, IEEE Transactions on}, 18(3):363--378, 1972.

\bibitem{cioffi1995mmse}
J.M. Cioffi, G.P. Dudevoir, M.~Vedat~Eyuboglu, and G.D. Forney~Jr.
\newblock {MMSE} decision-feedback equalizers and coding {I}: {E}qualization
  results.
\newblock {\em Communications, IEEE Transactions on}, 43(10):2582--2594, 1995.

\bibitem{tuchler2002turbo}
Michael Tuchler, Ralf Koetter, and Andrew~C Singer.
\newblock Turbo equalization: principles and new results.
\newblock {\em Communications, IEEE Transactions on}, 50(5):754--767, 2002.

\bibitem{hwang2009ofdm}
Taewon Hwang, Chenyang Yang, Gang Wu, Shaoqian Li, and G~Ye~Li.
\newblock {OFDM} and its wireless applications: a survey.
\newblock {\em Vehicular Technology, IEEE Transactions on}, 58(4):1673--1694,
  2009.

\bibitem{li2013ofdma}
Junyi Li, Xinzhou Wu, and Rajiv Laroia.
\newblock {\em {OFDMA} Mobile Broadband Communications}.
\newblock Cambdige University Press, 2013.

\bibitem{bingham2000adsl}
John~AC Bingham.
\newblock {\em ADSL, VDSL, and multicarrier modulation}.
\newblock Wiley New York, 2000.

\bibitem{van2006wifi}
Richard Van~Nee, VK~Jones, Geert Awater, Allert Van~Zelst, James Gardner, and
  Greg Steele.
\newblock The 802.11n {MIMO-OFDM} standard for wireless {LAN} and beyond.
\newblock {\em Wireless Personal Communications}, 37(3-4):445--453, 2006.

\bibitem{ghosh2005wimax}
Arunabha Ghosh, David~R Wolter, Jeffrey~G Andrews, and Runhua Chen.
\newblock Broadband wireless access with {W}i{M}ax/802.16: current performance
  benchmarks and future potential.
\newblock {\em Communications Magazine, IEEE}, 43(2):129--136, 2005.

\bibitem{reimers1996dvbt}
Ulrich Reimers.
\newblock {DVB-T}: the {COFDM}-based system for terrestrial television.
\newblock 1996.

\bibitem{ghosh2010lte}
Amitava Ghosh, Rapeepat Ratasuk, Bishwarup Mondal, Nitin Mangalvedhe, and Tim
  Thomas.
\newblock {LTE}-advanced: next-generation wireless broadband technology.
\newblock {\em Wireless Communications, IEEE}, 17(3):10--22, 2010.

\bibitem{benvenuto2010single}
Nevio Benvenuto, Rui Dinis, David Falconer, and Stefano Tomasin.
\newblock Single carrier modulation with nonlinear frequency domain
  equalization: an idea whose time has come --- again.
\newblock {\em Proceedings of the IEEE}, 98(1):69--96, 2010.

\bibitem{myung2008single}
Hyung~G Myung and David Goodman.
\newblock {\em Single carrier {FDMA}: a new air interface for long term
  evolution}, volume~8.
\newblock John Wiley \& Sons, 2008.

\bibitem{perahia2011gigabit}
Eldad Perahia and Michelle~X Gong.
\newblock Gigabit wireless {LAN}s: an overview of {IEEE} 802.11 ac and 802.11
  ad.
\newblock {\em ACM SIGMOBILE Mobile Computing and Communications Review},
  15(3):23--33, 2011.

\bibitem{ryan2009channel}
William Ryan and Shu Lin.
\newblock {\em Channel codes: classical and modern}.
\newblock Cambridge University Press, 2009.

\bibitem{nguyen2011spatially}
Phong~S Nguyen, Arvind Yedla, Henry~D Pfister, and Krishna~R Narayanan.
\newblock Spatially-coupled codes and threshold saturation on
  intersymbol-interference channels.
\newblock {\em arXiv preprint arXiv:1107.3253}, 2011.

\bibitem{hirt1988capacity}
W.~Hirt and J.L. Massey.
\newblock Capacity of the discrete-time gaussian channel with intersymbol
  interference.
\newblock {\em Information Theory, IEEE Transactions on}, 34(3):38--38, 1988.

\bibitem{shamai1996intersymbol}
S.~Shamai and R.~Laroia.
\newblock The intersymbol interference channel: Lower bounds on capacity and
  channel precoding loss.
\newblock {\em Information Theory, IEEE Transactions on}, 42(5):1388--1404,
  1996.

\bibitem{guo2005mutual}
D.~Guo, S.~Shamai, and S.~Verd{\'u}.
\newblock Mutual information and minimum mean-square error in {G}aussian
  channels.
\newblock {\em Information Theory, IEEE Transactions on}, 51(4):1261--1282,
  2005.

\bibitem{guo2011estimation}
Dongning Guo, Yihong Wu, Shlomo Shamai, and Sergio Verd{\'u}.
\newblock Estimation in gaussian noise: Properties of the minimum mean-square
  error.
\newblock {\em Information Theory, IEEE Transactions on}, 57(4):2371--2385,
  2011.

\bibitem{payaro2009hessian}
Miquel Payar{\'o} and Daniel~P Palomar.
\newblock Hessian and concavity of mutual information, differential entropy,
  and entropy power in linear vector gaussian channels.
\newblock {\em Information Theory, IEEE Transactions on}, 55(8):3613--3628,
  2009.

\bibitem{wilson1995comparison}
Sarah~Kate Wilson and John~M Cioffi.
\newblock A comparison of a single-carrier system using a {DFE} and a coded
  {OFDM} system in a broadcast {R}ayleigh-fading channel.
\newblock In {\em Information Theory, 1995. Proceedings., 1995 IEEE
  International Symposium on}, page 335. IEEE, 1995.

\bibitem{czylwik1997comparison}
Andreas Czylwik.
\newblock Comparison between adaptive {OFDM} and single carrier modulation with
  frequency domain equalization.
\newblock In {\em Vehicular Technology Conference, 1997, IEEE 47th}, volume~2,
  pages 865--869. IEEE, 1997.

\bibitem{tubbax2001ofdm}
Jan Tubbax, Boris C{\^o}me, Liesbet Van~der Perre, Luc Deneire, Stephane
  Donnay, and Marc Engels.
\newblock {OFDM} versus single carrier with cyclic prefix: a system-based
  comparison.
\newblock In {\em Vehicular Technology Conference, 2001. VTC 2001 Fall. IEEE
  VTS 54th}, volume~2, pages 1115--1119. IEEE, 2001.

\bibitem{wang2004ofdm}
Zhengdao Wang, Xiaoli Ma, and Georgios~B Giannakis.
\newblock {OFDM} or single-carrier block transmissions?
\newblock {\em Communications, IEEE Transactions on}, 52(3):380--394, 2004.

\bibitem{lin2003ber}
Yuan-Pei Lin and See-May Phoong.
\newblock {BER} minimized {OFDM} systems with channel independent precoders.
\newblock {\em Signal Processing, IEEE Transactions on}, 51(9):2369--2380,
  2003.

\bibitem{de2011uncoded}
Amanda de~Paula and Cristiano Panazio.
\newblock An uncoded {BER} comparison between {DFE-SCCP} and {OFDM} using a
  convex analysis framework.
\newblock In {\em Circuits and Systems (ISCAS), 2011 IEEE International
  Symposium on}, pages 2397--2400. IEEE, 2011.

\bibitem{fischer1997equivalence}
Robert~FH Fischer and Johannes~B Huber.
\newblock On the equivalence of single-and multicarrier modulation: A new view.
\newblock In {\em Information Theory. 1997. Proceedings., 1997 IEEE
  International Symposium on}, page 197. IEEE, 1997.

\bibitem{zervos1989optimized}
N~Zervos and Irving Kalet.
\newblock Optimized decision feedback equalization versus optimized orthogonal
  frequency division multiplexing, for high-speed data transmission over the
  local cable network.
\newblock In {\em Communications, 1989. ICC'89, BOSTONICC/89. Conference
  record.'World Prosperity Through Communications', IEEE International
  Conference on}, pages 1080--1085. IEEE, 1989.

\bibitem{benvenuto2002comparison}
Nevio Benvenuto and Stefano Tomasin.
\newblock On the comparison between {OFDM} and single carrier modulation with a
  {DFE} using a frequency-domain feedforward filter.
\newblock {\em Communications, IEEE Transactions on}, 50(6):947--955, 2002.

\bibitem{zhang2012comparison}
Jiaqi Zhang, Yukui Pei, and Ning Ge.
\newblock Comparison of achievable rates of {OFDM} and single carrier
  communication systems.
\newblock {\em Tsinghua Science and Technology}, 17(1):73--77, 2012.

\bibitem{franceschini2008information}
M~Franceschini, R~Pighi, G~Ferrari, and R~Raheli.
\newblock On information theoretic aspects of single-and multi-carrier
  communications.
\newblock In {\em Information Theory and Applications Workshop, 2008}, pages
  94--99. IEEE, 2008.

\bibitem{aue1998comparison}
Volker Aue, Gerhard~P Fettweis, and Reinaldo Valenzuela.
\newblock A comparison of the performance of linearly equalized single carrier
  and coded {OFDM} over frequency selective fading channels using the random
  coding technique.
\newblock In {\em Communications, 1998. ICC 98. Conference Record. 1998 IEEE
  International Conference on}, volume~2, pages 753--757. IEEE, 1998.

\bibitem{de2011comparison}
Amanda de~Paula and Cristiano Panazio.
\newblock A comparison between {OFDM} and single-carrier with cyclic prefix
  using channel coding and frequency-selective block fading channels.
\newblock {\em Journal of Communication and Information Systems.},
  26(1):19--29, 2011.

\bibitem{de2013comparison}
Amanda de~Paula and Cristiano Panazio.
\newblock Comparison of {OFDM} and {SC-DFE} capacities without channel
  knowledge at the transmitter.
\newblock {\em arXiv preprint arXiv:1306.3440}, 2013.

\bibitem{gray2010entropy}
R.M. Gray.
\newblock {\em Entropy and information theory}.
\newblock Springer Verlag, 2010.

\bibitem{cover2012elements}
Thomas~M Cover and Joy~A Thomas.
\newblock {\em Elements of {I}nformation {T}heory}.
\newblock John Wiley \& Sons, 2012.

\bibitem{jeong2012easily}
S.~Jeong and J.~Moon.
\newblock Easily computed lower bounds on the information rate of intersymbol
  interference channels.
\newblock {\em Information Theory, IEEE Transactions on}, 58(2):864--877, 2012.

\bibitem{carmon2013slc}
Y.~Carmon and S.~Shamai.
\newblock Lower bounds and approximations for the information rate of the {ISI}
  channel.
\newblock {\em arXiv preprint arXiv:1401.1480}, 2014.

\bibitem{arnold2006simulation}
Dieter~M Arnold, H-A Loeliger, Pascal~O Vontobel, Aleksandar Kavcic, and Wei
  Zeng.
\newblock Simulation-based computation of information rates for channels with
  memory.
\newblock {\em Information Theory, IEEE Transactions on}, 52(8):3498--3508,
  2006.

\bibitem{pfister2001achievable}
H.D. Pfister, J.B. Soriaga, and P.H. Siegel.
\newblock On the achievable information rates of finite state {ISI} channels.
\newblock In {\em Global Telecommunications Conference, 2001. GLOBECOM'01.
  IEEE}, volume~5, pages 2992--2996. IEEE, 2001.

\bibitem{radosevic2011bounds}
A.~Radosevic, D.~Fertonani, T.M. Duman, J.G. Proakis, and M.~Stojanovic.
\newblock Bounds on the information rate for sparse channels with long memory
  and iud inputs.
\newblock {\em Communications, IEEE Transactions on}, 59(12):3343--3352, 2011.

\bibitem{gray2006toeplitz}
Robert~M Gray.
\newblock {\em Toeplitz and circulant matrices: A review}.
\newblock Now Pub, 2006.

\bibitem{lozano2006optimum}
A.~Lozano, A.M. Tulino, and S.~Verd{\'u}.
\newblock Optimum power allocation for parallel gaussian channels with
  arbitrary input distributions.
\newblock {\em Information Theory, IEEE Transactions on}, 52(7):3033--3051,
  2006.

\bibitem{xiao2011globally}
C.~Xiao, Y.R. Zheng, and Z.~Ding.
\newblock Globally optimal linear precoders for finite alphabet signals over
  complex vector gaussian channels.
\newblock {\em Signal Processing, IEEE Transactions on}, 59(7):3301--3314,
  2011.

\bibitem{boyd2009convex}
Stephen Boyd and Lieven Vandenberghe.
\newblock {\em Convex optimization}.
\newblock Cambridge university press, 2009.

\bibitem{alvarado2013high}
Alex Alvarado, Fredrik Brannstrom, Erik Agrell, and Tobias Koch.
\newblock High-{SNR} asymptotics of mutual information for discrete
  constellations with applications to {BICM}.
\newblock 2013.

\bibitem{smith1971information}
Joel~G Smith.
\newblock The information capacity of amplitude-and variance-constrained sclar
  {G}aussian channels.
\newblock {\em Information and Control}, 18(3):203--219, 1971.

\bibitem{erceg2004ieee}
V~Erceg, L~Schumacher, et~al.
\newblock {TG}n channel models.
\newblock {\em IEEE 802.11-03/940r4}, 2004.

\end{thebibliography}

\end{document}